\documentclass[preprint,preprintnumbers,aps,11pt,superscriptaddress,nofootinbib,tightenlines,floatfix]{revtex4-1}

\usepackage{amsmath,amssymb}
\usepackage{graphicx}
\usepackage{bm}
\usepackage{comment}
\usepackage{color}
\usepackage{subfigure}
\usepackage{array}
\usepackage{multirow}

\usepackage[T1]{fontenc}
\usepackage[latin9]{inputenc}
\usepackage{graphicx}
\usepackage{esint}
\usepackage{hyperref}
\usepackage{atbegshi}
\usepackage{lipsum}
\usepackage{url}

\begin{document}

\hspace{5.2in} \mbox{NT@UW-13-19}

\title{Two-Nucleon Systems in a Finite Volume:
\\
(I) Quantization Conditions}
\author{Ra\'ul Brice\~no{\footnote{\tt briceno@uw.edu}}
}
\affiliation{Department of Physics, University of Washington\\
Box 351560, Seattle, WA 98195, USA}
\author{Zohreh Davoudi{\footnote{\tt davoudi@uw.edu}}
}
\affiliation{Department of Physics, University of Washington\\
Box 351560, Seattle, WA 98195, USA}
\author{Thomas Luu{\footnote{\tt luu5@llnl.gov}}
}
\affiliation{Lawrence Livermore National Laboratory,  Livermore, CA 94550, USA}
\date{\today}

\begin{abstract} 
 The quantization condition for interacting energy eigenvalues of the two-nucleon system in a finite cubic volume is derived in connection to the nucleon-nucleon scattering amplitudes. This condition is derived using an auxiliary (dimer) field formalism that is generalized to arbitrary partial waves in the context of non-relativistic effective field theory.
 The quantization condition presented gives access to the scattering parameters of the two-nucleon systems with arbitrary parity, spin, isospin, angular momentum and center of mass motion, from a lattice QCD calculation of the energy eigenvalues. In particular, as it includes all non-central interactions, such as the two-nucleon tensor force, it makes explicit the dependence of the mixing parameters of  nucleon-nucleon systems calculated from lattice QCD when there is a physical mixing among different partial-waves, e. g. S-D mixing in the deuteron channel.  We provide explicit relations among scattering parameters and their corresponding point group symmetry class eigenenergies with orbital angular momentum $l\le3$, and for center of mass boost vectors of the form $\frac{2\pi}{L}(2n_1,2n_2,2n_3)$, $\frac{2\pi}{L}(2n_1,2n_2,2n_3+1)$ and $\frac{2\pi}{L}(2n_1+1,2n_2+1,2n_3)$. $L$ denotes the special extent of the cubic volume and $n_1,n_2,n_3$ are integers. Our results are valid below inelastic thresholds up to exponential volume corrections that are governed by the pion mass.

\end{abstract}

\maketitle
\tableofcontents
\vfill\eject
\section{Introduction
\label{sec:Intro}
}
Being able to make reliable predictions for few-body and many-body nuclear systems requires truly \textit{ab initio} methods with quantifiable uncertainties. In the two-body sector empirical models are sufficiently precise to provide a reliable estimation of the two-body nuclear force.  However, they do not give much insight into the nature of such systems at extreme energies and densities where experiments are not available or when more exotic nuclear systems involving hyper-nucleons -- such as those in astrophysical environments -- become relevant. The nature of the nuclear forces in connection to the parameters of the Standard Model of particle physics is unknown and further insights into this problem require first-principle calculations that use these fundamental parameters as input. When it comes to nuclear reactions, our knowledge of some of the most fundamental reactions, such as those in energy production chains in stars, diagnostic reactions in terrestrial nuclear reactor facilities or capture processes in neutrino detection experiments, remains limited. In particular matrix elements of electroweak currents between two-body hadronic states are poorly constrained by both model calculations and experiments. So unless precise experimental determinations become available, the only hope to make progress regarding these nuclear reactions is to perform model-independent calculations of  matrix elements of two-body weak currents between hadronic states. Currently the only rigorous method with which one can study such systems from the fundamental theory of strong interactions, quantum chromodynamics (QCD), is lattice QCD (LQCD). Although both analytically and computationally demanding, this approach has been successfully implemented for studying nuclear systems in recent years. With constant developments in formalism and algorithms, as well as ever-increasing computational resources, the precision needed for these calculations will be within reach in the upcoming years.

Lattice QCD (LQCD) calculations are performed in a finite, discretized Euclidean spacetime volume and currently at unphysical light quark masses. Effective field theory methods allow for calculations to be extrapolated to the physical point. Although the Euclidean nature of the calculations imposes limitations on extraction of two-body scattering quantities away from the kinematic threshold in infinite volume \cite{Maiani:1990ca}, \textit{finite volume} (FV) calculations of energy eigenvalues give access to these physical quantities through the so-called \textit{L\"uscher} method \cite{luscher1, luscher2}. This method, which has been widely used to extract scattering phase shifts of two-hadron systems from LQCD (see for example Refs. \cite{Beane:2010hg, Beane:2011xf, Beane:2011iw, Beane:2011sc,  Dudek:2012gj}), not only is applicable to single-channel two-body scattering processes below inelastic threshold, but has also been generalized to multi-coupled channel two-body systems below three-particle production thresholds \cite{He:2005ey, Lage:2009zv, Bernard:2010fp, Hansen:2012tf, Briceno:2012yi, Hansen:2012tf, Li:2012bi}. This latter generalization has been a crucial step in moving toward studies of hadronic resonances from model-independent calculations of the coupled-scattering states directly from the underlying theory of QCD (For a review of recent progress in this direction see Ref. \cite{Mohler:2012nh}.).
 
L\"uscher presented a master formula for the scattering phase shifts of two \textit{scalar} particles in \textit{arbitrary} partial-waves in connection to the FV energy levels of the two-meson system \cite{luscher1, luscher2}. Although this master formula is self-contained and incorporates all the necessary details to be implemented in practice, deducing the relations, or \emph{Quantization Conditions} (QC), among phase shifts in different partial-waves and the energy levels of a specific LQCD calculation requires multiple non-trivial steps. The corresponding procedure is sometimes called the reduction of the L\"uscher formula. The difficulty associated with this procedure is due to the fact that LQCD calculations are performed in a finite periodic cubic volume. As a result, the degeneracy of energy eigenvalues of the system in such calculations is determined according to the irreducible representations (irreps) of the cubic group. Since the phase shifts are characterized according to the irreps of the SO(3) rotational group, the energy eigenvalues of the system in a given irrep of the cubic group  in general depend on the phase shifts of more than one partial-wave channel. Performing  LQCD calculations of energy levels in different irreps of the cubic group would provide multiple QCs depending on different linear combinations of the scattering phase shifts, leading to better constraints on these quantities. Therefore it is necessary to identify all the QCs satisfied by a given scattering parameter in a partial-wave channel. While L\"uscher's original work presents the reduction of the master formula to a QC for the cubic $A_1$ irrep, Ref. \cite{Luu:2011ep} provides the full quantization conditions for the energy eigenvalues of different irreps of the cubic group, in both positive and negative parity sectors for orbital angular momentum $l\leq6$ as well as $l=9$ in the scalar sector. For scattering involving a spin-$\frac{1}{2}$ particle and a scalar particle, the L\"uscher formula can be generalized such that the energy eigenvalues of the meson-baryon system in a given irrep of the double-cover of the cubic group is related to the corresponding phase shifts \cite{Bernard:2008ax}. This generalization has been also presented for nucleon-nucleon scattering\footnote{The L\"uscher formula to study two-nucleon systems were first presented in Ref. \cite{Beane:2003da}, although due to constraining the calculation to the S-wave scattering, the complexity of the two-nucleon systems has not been dealth with. The only previous attempt to address this problem, including the spin, isospin and angular momentum degrees of freedom, is the work by N. Ishizuka \cite{Ishizuka:2009bx}, where the quantization conditions for energy eigenvalues of a two-nucleon system at rest in the positive and negative parity isosinglet channels were obtained for $J\leq 4$.} where due to the the possibility of physical mixing among different partial-wave channels, more complexities arise. This is an important problem as it provides the formalism needed for a first-principle extraction of the S-D mixing parameter in the deuteron channel, and will eventually shed light on the nature of the tensor interaction in nuclear physics.\footnote{For a different approach in studying tensor nuclear force using LQCD calculations, see Ref. \cite{Aoki:2009ji, Murano:2013xxa}. These calculations rely on constructing scheme-dependent potentials from multi-particle Nambu-Bethe-Salpeter wave-functions that are extracted from LQCD.}
 
In this paper, we derive and present the generalization of the L\"uscher formula for nucleon-nucleon scattering valid below the inelastic threshold for all spin and isospin channels in both positive and negative parity sectors. This formula is derived using the auxiliary field (dimer) formalism in the language of a non-relativistic effective field theory (EFT) for nucleon-nucleon interactions. It is well-known that introducing an S-wave dimer field -- that sums all $2\rightarrow2$ interactions non-perturbatively \cite{pionless2, pionless3} -- significantly simplifies the diagrammatic representation of multi-nucleon scattering amplitudes \cite{Bedaque:1997qi, Bedaque:1998mb, Bedaque:1998kg, Bedaque:1998km, Gabbiani:1999yv, Bedaque:1999vb, Bedaque:1999ve, Bedaque:2000ft}. However, to account for scattering in higher angular momentum channels, this dimer field must be generalized to arbitrary partial-waves. This is particularly important when such an auxiliary field is used in constructing a FV formalism for three-body scattering processes. As is pointed out in Ref. \cite{Briceno:2012rv}, the leading systematics of the results presented in Refs. \cite{Briceno:2012rv,Kreuzer:2008bi, Kreuzer:2009jp, Kreuzer:2010ti, Kreuzer:2012sr}, for the relation between three-body scattering amplitude and the FV spectrum of the three-particle system, arises from the FV-induced mixing between S-wave and D-wave scattering modes of the two-particle sub-system.  An S-wave dimer field therefore does not incorporate possible mixings in the FV formalism and will not give rise to a full quantization condition in arbitrary partial-waves in both two-body and three-body systems. To address this defect, we generalize the dimer field to higher partial-waves  and utilize the result to derive the generalized L\"uscher formula for the two-body boosted systems within both scalar and nucleon sectors. The generalization of the dimer formalism will also impact infinite volume three-body EFT calculations since it enables studying scattering in arbitrary partial-wave of two-particle sub-system (in addition to the higher partial-wave scattering of the dimer and the spectator particle that has been previously accounted for in Ref. \cite{Gabbiani:1999yv}).

Performing LQCD for systems with different center of mass (CM) momenta gives access to more energy levels at a given volume and provides additional QCs for the energy eigenvalues of the system in terms of scattering parameters. Boosting the two-particle system however reduces the symmetry of the problem even further and introduces more FV-induced mixings among different partial-waves \cite{Moore:2005dw, Luu:2009}. By investigating the symmetry group of the boosted systems along one and two Cartesian axes as well as that of the unboosted system, we have identified all the QCs satisfied by the phase shifts and mixing parameters in channels with total angular momentum $J\leq4$; ignoring scattering in partial-wave channels with $l\geq4$. Different QCs correspond to different irreps of the cubic ($O$), tetragonal ($D_{4}$) and orthorhombic ($D_{2}$) point groups that represent the symmetry group of systems with CM momentum $\mathbf{P}=0$, $\mathbf{P}=\frac{2\pi}{L}(0,0,1)$ and $\mathbf{P}=\frac{2\pi}{L}(1,1,0)$ respectively, where $L$ denotes the spatial extent of the cubic volume. As will be discussed later, these QCs can be also utilized for boost vectors of the form $\frac{2\pi}{L}(2n_1,2n_2,2n_3)$, $\frac{2\pi}{L}(2n_1,2n_2,2n_3+1)$ and $\frac{2\pi}{L}(2n_1+1,2n_2+1,2n_3)$ and all cubic rotations of these vectors where $n_1,n_2,n_3$ are integers.  Although the master formula presented in this article in the limit of zero CM momentum has been already derived in Ref. \cite{Ishizuka:2009bx} for nucleon-nucleon systems using a relativistic quantum filed theory approach, the full classifications of different QCs for all the spin and isospin channels and for two non-zero CM momenta are being presented and tabulated for the first time in the following sections. These relations make the implementation of the generalized L\"uscher formula for nucleon-nucleon systems straightforward for future LQCD calculations of the nucleon-nucleon (NN) system. In an upcoming paper, we discuss the expectation for energy eigenvalues of the NN systems with physical light-quark masses based on the QCs presented here, and will comment on the implementation of such formalism for studying the NN systems from LQCD in practice \cite{BDL:2013}.

Despite the tight empirical constraints on the two-body nuclear force, the investigation of the two-nucleon sector within LQCD is still warranted.  Understanding the energy-dependence of the scattering phase-shifts of the two-body hadronic states \cite{nnd, Meyer:2012wk, Briceno:2012yi, Bernard:2012bi, Meyer:2013dxa}, for example, is essential to obtaining physical matrix elements of current operators in the two-body sector.  Additionally, as LQCD calculations are currently done at unphysical pion masses, a rigorous study of multi-nucleon systems from LQCD also requires not only the knowledge of two-nucleon phase shifts,  but also their $m_\pi$-dependence \cite{Briceno:2012rv, Kreuzer:2008bi, Kreuzer:2009jp, Kreuzer:2010ti, Kreuzer:2012sr}.   The LQCD determination of the scattering parameters of the nucleon-nucleon systems at unphysical pion masses by itself is an interesting problem as it reveals the dependence of the two-body nuclear force on the masses of quarks in nature. Progress in this direction will have striking impact on our understanding of some of the most fundamental questions regarding the nuclear fine tunings in nature and the anthropic view of the Universe. As discussed in Refs. \cite{Epelbaum:2012iu, Epelbaum:2013wla, Bedaque:2010hr}, the survivability of Carbon-Oxygen based life is related to the variation of the inverse scattering lengths of NN scattering in the isosinglet and isotriplet channels, and a precise LQCD determination of these parameters will put tighter constraints on this quantity. In fact, for the first time LQCD has started addressing the question of the naturalness of the NN scattering length and effective range of the S-wave isosinglet and isotriplet channels, albeit at the SU(3) flavor symmetric point with $m_{\pi}\approx800$ MeV \cite{Beane:2013br}.   To appropriately utilize these LQCD calculations, the FV formalisms and their associated QCs must be determined.

The rest of this paper is structured as follows. In Sec.~\ref{sec: dimer} we present the auxiliary dimer field formalism for arbitrary partial-waves and spin and isospin channels, in both infinite volume and finite volume.  The formalism has been used to come at a master equation for the energy eigenvalues of the NN system in the finite volume in connection to the scattering parameters in all partial-waves and for any non-relativistic CM boost.   The master equation alone is not directly feasible to work with, as it represents a determinant over an infinite dimensional matrix.  The symmetries of the NN system in the infinite volume can, in principle, aid in reducing the master equation to block diagonal form that depends on the total angular momentum $J$ of the system.   In Sec.~\ref{sec: Reduction}, by taking advantage of the symmetries of the cubic, tetragonal and orthorhombic point groups, the strategy to further reduce the master QC to blocks of manageable finite size is discussed.  For systems with total angular momentum $J\le4$, ignoring scattering in partial-wave $l\geq4$, we find the largest block to have dimension $9\times9$.  The mixing of different angular momenta due to finite-volume symmetries is made explicit within this section, as is the reduction of symmetry due to the CM boosting. The invariance of the QCs under a cubic rotation of the boost vectors has been proved in appendix~\ref{app:invariant}. Appendix~\ref{app:red-example} contains an explicit example of the reduction of the master equation to obtain the QCs corresponding to each irrep of the cubic group for the zero CM momentum in the NN positive parity isosinglet channel.  In a subsequent appendix \ref{app:QC}, all such explicit relations are listed for scattering in all spin and isospin channels with $\frac{2\pi}{L}(0,0,0)$, $\frac{2\pi}{L}(0,0,1)$ and $\frac{2\pi}{L}(1,1,0)$ boost vectors that are valid for scattering with partial-waves $l\le3$ below inelastic thresholds. These QCs are also provided as a supplemental material to this article in a mathematica notebook format \cite{BDLsupp}. We recapitulate our findings in sect.~\ref{sec: S&M}.
                               
 \section{Finite Volume Formalism with the Auxiliary Field Method \label{sec: dimer}}
 
 L\"uscher FV formalism was first developed for scalar particles of equal mass with zero CM momentum, using a non-relativistic potential theory which was then generalized to the quantum field theory \cite{luscher1, luscher2}. The generalization to the case of non-zero CM momenta, as well as non-equal masses followed in Refs. \cite{movingframe, sharpe1, Christ:2005gi, Bour:2011ef, Davoudi:2011md, Fu:2011xz, Leskovec:2012gb}. The goal of this section is to extend L\"usher's formula to the case of two nucleons within the context of a non-relativistic effective field theory by using an auxiliary field method. Although there has been many derivations for the L\"uscher formula, the formalism that will be presented here makes the study of two-baryon systems with arbitrary quantum numbers straightforward. Additionally the methodology developed here can be used to generalize the FV formalism presented in Ref. \cite{Briceno:2012rv} to three nucleons with arbitrary partial-waves \cite{Briceno:2013}.
  
 \subsection{Two-boson systems \label{sec: Scalar}}
 
\begin{figure}[t]
\begin{center}
\subfigure[]{
\includegraphics[scale=0.425]{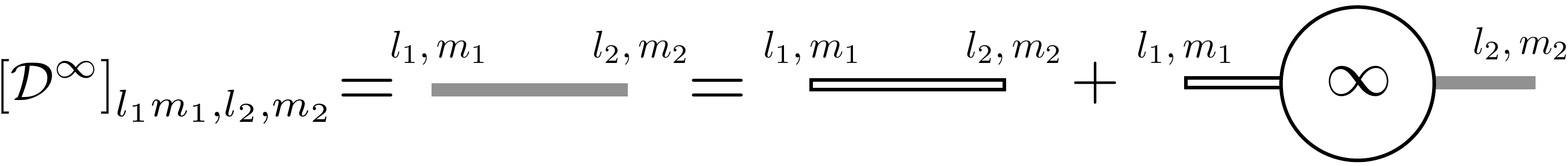}}
\subfigure[]{
\includegraphics[scale=0.425]{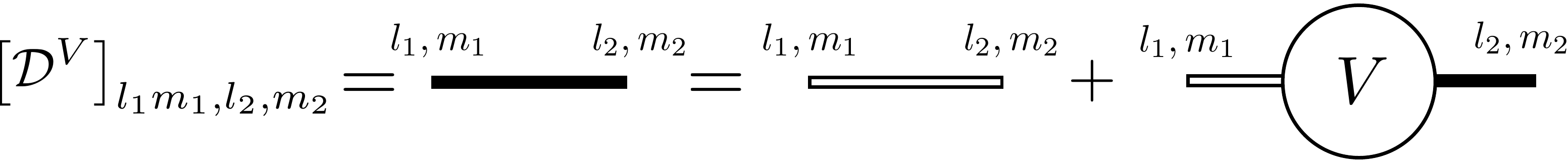}}
\caption{a) Diagrammatic equation satisfied by the matrix elements of the full dimer propagator in a) infinite volume and b) finite volume. The grey (black) band represents the full infinite (finite) volume propagator, $\mathcal{D}^{\infty}$ ($\mathcal{D}^V$), while the double lines represent the bare propagator, $\mathcal{D}^B$.}\label{fig:dimer}
\end{center}
\end{figure}

Consider two identical bosons with mass $M$ that interact in a partial-wave channel $(l,m)$ via a short-range interaction that can be effectively described by derivative couplings to the fields. Let $\phi_{k}$ and  $d_{lm,P}$ denote the interpolating operators that annihilate a boson with non-relativistic (NR) four-momentum $k$, and a dimer (with quantum numbers of two bosons) with NR four-momentum $P$ and angular momentum $(l,m)$, respectively. Then if $P^\mu=(E,\mathbf{P})$ denotes the NR four-momentum of the system, one can write a Galilean-invariant action that describes such system in the infinite volume in terms of a Lagrange density in the momentum space,
\begin{eqnarray}
\label{action}
{S}^{\infty}&=&\int\frac{d^{4}P}{(2\pi)^{4}}\left[\phi_{P}^{\dagger}(E-\frac{\textbf{P}^{2}}{2M})\phi_{P}-\sum_{l,m}d_{lm,P}^{\dagger}\left(E-\frac{\textbf{P}^{2}}{4M}-\Delta_{l}+\sum_{n=2}^{\infty}c_{n,l}(E-\frac{\textbf{P}^{2}}{4M})^{n}\right)d_{lm,P}\right]
\nonumber\\
&~&\qquad \qquad \qquad  -\int\frac{d^{4}P}{(2\pi)^{4}}~\frac{d^{4}k}{(2\pi)^{4}}\sum_{l,m}~\frac{g_{2,l}}{2}\left[d_{lm,P}^{\dagger}~\sqrt{4\pi}~Y_{lm}(\hat{\textbf{k}}^{*})~|\mathbf{k}^{*}|^{l}\phi_{{k}}\phi_{P-{k}}+h.c.\right],
\end{eqnarray}
where ${\textbf{k}}^*=\textbf{k}-\textbf{P}/2$ denotes the relative momentum of two bosons in the interaction term. Note that the interactions between bosons in partial-wave channel $(l,m)$ is mediated by a corresponding dimer field, $d_{lm}$. As is evident, upon integrating out such auxiliary field, one recovers the four-boson interaction term in a Lagrangian with only $\phi$-field degrees of freedom. Since this is a theory of identical bosons, all couplings of the dimer field to a two-boson state with an odd partial-wave vanish. Eq. (\ref{action}) clearly reduces to the S-wave result of Refs. \cite{pionless2, pionless3, Griesshammer:2004pe}. This action can be easily generalized for systems involving distinguishable scalar bosons (e.g. for P-wave scattering see Ref. \cite{Braaten:2011vf}). As usual, the low-energy coefficients (LECs) $\{\Delta_{l},c_{l,n}, g_{2,l}\}$ in the effective Lagrangian must be tuned to reproduce the ERE of the $l^{th}$-partial-wave,
\begin{eqnarray}
k^{*2l+1}\cot\delta_{d}^{(l)}=-\frac{1}{a_l}+\frac{r_{l}k^{*2}}{2}+\sum_{n=2}^\infty\frac{\rho_{n,l}}{n!}~(k^{*2})^{n},
\end{eqnarray}
where $k^*\equiv |\mathbf{k}^*|=\sqrt{ME-\frac{\mathbf{P}^2}{4}}$ is the relative \emph{on-shell} momentum of the bosons in the CM frame. $\delta_d^{(l)}$ is the phase shift in the $l^{th}$-partial-wave, and $\{a_l,r_l, \rho_{n,l}\}$ are the corresponding scattering length, effective range and all higher order shape parameters, respectively. The fully dressed dimer propagator can be obtained by summing up the self-energy bubble diagrams to all orders, Fig. (\ref{fig:dimer}, a), the result of which is the following
\begin{eqnarray}
\mathcal{D}^{\infty}(E,\mathbf{P})=\frac{1}{(\mathcal{D}^{B})^{-1}-I^{\infty}(E,\mathbf{P})},
\label{D-infinity}
\end{eqnarray}
where $\mathcal{D}^B$ denotes the bare dimer propagator, 
\begin{eqnarray}
\left[\mathcal{D}^{B}(E,\mathbf{P})\right]_{l_1m_1,l_2m_2}=\frac{-i~\delta_{l_1l_2}\delta_{m_1m_2}}{E-\frac{\mathbf{P}^2}{4M}-\Delta_{l}+\sum_{n=2}^{\infty}c_{n,l}(E-\frac{\textbf{P}^{2}}{4M})^{n}+i\epsilon},
\label{D-bare}
\end{eqnarray}
and $I^{\infty}$ denotes the value of the bubble diagram evaluated using the power divergence subtraction (PDS) scheme \cite{pds, pds2, pds3},
\begin{eqnarray}
\left[I^{\infty}(E,\mathbf{P})\right]_{l_1m_1,l_2,m_2}=\frac{iM}{8\pi}g_{2,l_1}^2k^{*2l_1}(\mu+ik^*)\delta_{l_1l_2}\delta_{m_1m_2},
\label{I-infinity}
\end{eqnarray}
 where $\mu$ is the renormalization scale. By requiring the full dimer propagator, $\mathcal{D}^{\infty}$, in the infinite volume to reproduce the full scattering amplitude in any given partial-wave,
\begin{eqnarray} 
\mathcal{M}^{\infty}_{l_1m_1,l_2m_2}&=&-[~g~\mathcal{D}^{\infty}(E,\mathbf{P})~g~]_{l_1m_1,l_2m_2}=
\frac{8\pi}{M}~
\frac{1}{k^{*}\cot{\delta^{(l_1)}_d}-ik^{*}}\delta_{l_1l_2}\delta_{m_1m_2},
\end{eqnarray}
one arrives at
\begin{eqnarray}
g_{2,l}^2=\frac{16\pi}{M^2r_{l}} ~\text{for}~l~\text{even}, ~~ \Delta_l=\frac{2}{Mr_l}\left(\frac{1}{a_l}-\mu k^{*2l}\right), ~~ c_{n,l}=\frac{2}{Mr_l}\frac{\rho_{n,l}M^n}{n!}.
\label{g2l}
\end{eqnarray}

In the finite volume, the two-boson system can still be described by the action in Eq. (\ref{action}) except the periodic boundary conditions constrain the momenta to be discretized. In particular, the integral over three-vector momenta in Eq. (\ref{action}) is replaced by a sum over discrete momenta, $P=\frac{2\pi}{L}\mathbf{n}$, where $\mathbf{n}$ is a triplet integer. Then it is straightforward to evaluate the corresponding bubble diagram in the finite volume,\footnote{We keep the temporal extent of the Minkowski space infinite.}
\begin{eqnarray}
\left[I^{V}\right]_{l_1m_1,l_2,m_2}&=&\frac{iM}{8\pi}g_{2,l_1}g_{2,l_2}k^{*l_1+l_2}
\nonumber\\
&~&\times\left[\mu~\delta_{l_1l_2}\delta_{m_1m_2}+\sum_{l,m}\frac{(4\pi)^{3/2}}{k^{*l}}c^{{P}}_{lm}(k^{*2})\int d\Omega~Y^*_{l_1,m_1}Y^*_{l,m}Y_{l_2,m_2}\right],
\label{I-V}
\end{eqnarray}
with
\begin{eqnarray}
\label{clm}
c^{{\mathbf{P}}}_{lm}(x)=\left[\frac{1}{L^3}\sum_{\textbf{q}}-\mathcal{P}\int\frac{d^3\mathbf{q}}{(2\pi)^3}\right]{q}^{*l}\frac{\sqrt{4\pi}Y_{lm}(\hat{\mathbf{q}^*})}{{\mathbf{q}^*}^{2}-x} \ ,
\end{eqnarray}
where $\mathbf{q}^*=\mathbf q- \mathbf P/2$, and $\mathcal{P}$ denotes the principal value of the integral\footnote{When comparing the result presented here with that of Ref. \cite{sharpe1}, note that $c^{\mathbf{P}}_{lm}(x)$ as defined in Eq. (\ref{clm}) is equal to negative $c^{\mathbf{P}}_{lm}(x)$ as defined in Ref. \cite{sharpe1}.}
\footnote{The kinematic function $c_{lm}^\mathbf{P}(k^{*2})$ can be written in terms of the three-dimensional Zeta function, $\mathcal{Z}^\mathbf{d}_{lm}$,
 \begin{eqnarray}
 \nonumber
 	c^\mathbf{P}_{lm}(k^{*2})=\frac{\sqrt{4\pi}}{L^3}\left(\frac{2\pi}{L}\right)^{l-2}\mathcal{Z}^\mathbf{d}_{lm}[1;(k^*L/2\pi)^2],\hspace{1cm} 
\mathcal{Z}^\mathbf{d}_{lm}[s;x^2]=\sum_{\mathbf{n}}\frac{\mathbf{r}^lY_{l,m}(\mathbf{r})}{(r^2-x^2)^s},
\end{eqnarray}
where $\mathbf{\mathbf{r}}=\mathbf{n}-\mathbf{d}/2$ for non-relativistic theories with degenerate masses, $\mathbf{n}$ is an integer triplet, and $\mathbf d$ is the normalized boost vector $\mathbf d=\mathbf{P}L/2\pi$.}. The full dimer propagator, $\mathcal{D}^V$, can then be obtained by summing up the infinite series of bubble diagrams in Fig. (\ref{fig:dimer}, b), where the LEC of the theory are matched with the the physical quantities, Eq. (\ref{g2l}),
\begin{eqnarray}
\mathcal{D}^{V}(E,\mathbf{P})=\frac{1}{(\mathcal{D}^{B})^{-1}-(\mathcal{D}^{B})^{-1}I^{V}(E,\mathbf{P})\mathcal{D}^{B}}.
\label{D-finite}
\end{eqnarray}
Note that, just like $\mathcal{D}^\infty$ in Eq.~(\ref{D-infinity}), $\mathcal{D}^V$ is a matrix in the angular momentum space. The poles of the FV dimer propagator give the spectrum of two-boson system in a finite volume in terms of the scattering parameters. These energy eigenvalues satisfy the following determinant condition
\begin{eqnarray}
\det \left[k^*\cot \delta-\mathcal{F}^{FV}\right]=0,
\label{FullQCboson}
\end{eqnarray}
where both $\cot \delta$ and $\mathcal{F}^{FV}$ are matrices in the angular momentum space,
\begin{eqnarray}
\cot \delta \equiv \cot (\delta_{l_1}) \delta_{l_1l_2}\delta_{m_1m_2},
\label{cot}
\end{eqnarray}
\begin{eqnarray}
\left[\mathcal{F}^{FV}\right]_{l_1m_1,l_2m_2}=\sum_{l,m}\frac{(4\pi)^{3/2}}{k^{*l}}c^{\mathbf{P}}_{lm}(k^{*2})\int d\Omega~Y^*_{l_1,m_1}Y^*_{l,m}Y_{l_2,m_2}.
\label{F}
\end{eqnarray}

In Eq.~(\ref{cot}) the Kronecker deltas that are constraining the $(l,m)$ quantum numbers should not be confused with the phase shift $\delta_{l_1}$. This quantization condition agrees with the non-relativistic limit of the results presented in Refs. \cite{movingframe, sharpe1, Christ:2005gi} for the generalization of the L\"uscher formula to the boosted systems, and upon truncating the angular momentum sum to $l_{max}=0$, reduces to the S-wave result of Ref. \cite{Briceno:2012rv} where an S-wave dimer field is used to derive the L\"uscher formula. This derivation shows that upon incorporating higher partial-waves in the construction of the dimer Lagrangian, as well as accounting for higher order terms in the EFR expansion, all the two-body physics is fully encapsulated in this formalism. As a result the systematic errors of those FV multi-particle calculations that have used an S-wave dimer field up to next-to-leading order in ERE (see Refs. \cite{Briceno:2012rv, Kreuzer:2008bi, Kreuzer:2009jp, Kreuzer:2010ti, Kreuzer:2012sr}), can be easily avoided.

\subsection{Two-nucleon systems \label{sec: Nuclear}}
Due to spin and isospin degrees of freedom, the two-nucleon system exhibits some specific features. In particular, the anti-symmetricity of the two-nucleon state constrains the allowed spin and isospin channels for a given parity state. Additionally, any spin-triplet two-nucleon state is an admixture of two different orbital-angular momentum states. For example, as is well-known, the two-nucleon state in the deuteron channel with $J^{P}=1^{+}$ is an admixture of S-wave and D-wave states. In general, a positive parity two-nucleon state with total angular momentum $J$ is a linear combination of states with\footnote{The $L$ that is introduced here and elsewhere as the partial-wave label of quantities should not be confused with the spatial extent of the lattice $L$ that appears in the definition of the $c_{lm}^{\mathbf{P}}$ functions.}
\begin{eqnarray}
\left(L=J\mp\frac{1}{2}(1-(-1)^J), S=\frac{1}{2}(1-(-1)^J)\right),
\label{positive}
\end{eqnarray}
while in the negative parity sector, the states that are being mixed have\footnote{Note, however, that for a $J$-even state in the first case and a $J$-odd state in the second case, there is only one angular momentum state allowed and no mixing occurs.}
\begin{eqnarray}
\left(L=J\mp\frac{1}{2}(1+(-1)^J), S=\frac{1}{2}(1+(-1)^J)\right).
\label{negative}
\end{eqnarray}
Table (\ref{JP}) shows the allowed spin and angular momentum of NN states in both isosinglet and isotriplet channels with $J\leq4$.
 \begin{center}
\begin{table}[h]
\begin{tabular}{|cc|ccccccccccc|}
\hline
&$J^P$&$0^+$&$0^-$&$1^+$&$1^-$&$2^+$&$2^-$&$3^+$&$3^-$&$4^+$&$4^-$&$$ \\\hline \hline
 &I=0&---&---&\{(0,1),(2,1)\}&(1,0)&(2,1)&---&\{(2,1),(4,1)\}&(3,0)&(4,1)&---&$$\\
&I=1&(0,0)&(1,1)&---&(1,1)& (2,0)&\{(1,1),(3,1)\}&---&(3,1)&(4,0)&\{(3,1),(5,1)\}&$$
\\\hline\hline
\end{tabular}
\caption{The allowed spin and angular momentum states, $(L,S)$, for NN-states with $J\leq4$. Note that depending on the parity of the states, the partial-wave mixing can occur in either the isosinglet or isotriplet channels.}
\label{JP}
\end{table}
\end{center}

In order to write the most general Lagrangian describing nucleon-nucleon scattering in all spin, isospin and angular momentum channels, let us introduce an operator that creates an NN-state with total four-momentum $P$ and the relative momentum ${\textbf{k}}^*={\textbf{k}}-\frac{{\textbf{P}}}{2}$ in an arbitrary partial-wave $(L,M_L)$ in the following way
\begin{eqnarray}
|NN;P,k^*\rangle _{LM_L,SM_S,IM_I}=\mathcal{N}_L\int d\Omega_{\textbf{k}^*}~Y^*_{LM_L}(\hat{\textbf{k}}^{*})k^{*L}\left[N^T_{P-k}~\hat{\mathcal{P}}_{(SM_S,IM_I)}~N_k\right]^\dag|0\rangle,
\end{eqnarray}
where $k^*=\left|\mathbf{k}^*\right|$. $\hat{\mathcal{P}}_{(SM_S,IM_I)}$ is an operator which projects onto a two-nucleon state with spin $(S,M_S)$ and isospin $(I,M_I)$, and $\mathcal{N}_L$ is a normalization factor. By requiring such state to have a non-zero norm, and given the anti-commutating nature of nucleon fields, one can infer that for positive parity states the operator $\hat{\mathcal{P}}_{(SM_S,IM_I)}$ must be necessarily antisymmetric, while for negative parity states it must be symmetric. Since this operator is a direct product of two projection operators in the space of spin and isospin, these requirements can be fulfilled by constructing the corresponding operators using the appropriate combinations of Pauli matrices, $\sigma_j$ ($\tau _j$), that act on the spin (isospin) components of the nucleon field. To proceed with such construction, let us define the following operators
\begin{eqnarray}
\alpha_j^I=\tau_y\tau_j,\hspace{1cm}
\alpha_j^S=\sigma_y\sigma_j,\hspace{1cm}
\beta^I=\tau_y,\hspace{1cm}
\beta^S=\sigma_y. 
\end{eqnarray}
Note that the matrices that are named as $\alpha$ are symmetric while those that are named as $\beta$ are antisymmetric. Superscript $I$ ($S$) implies that the operator is acting on the spin (isospin) space, and index $j=1,2,3$ stands for the Cartesian components of the operators. Alternatively one can form linear combinations of $\alpha^S_j$ ($\alpha^I_j$) that transform as a rank one spherical tensor.\footnote{A Cartesian vector $\mathbf{r}$ can be brought into a spherical vector according to 
\begin{eqnarray}
\label{spherical}
r^{(0)}\equiv r_{z},\hspace{1cm}
r^{(\pm 1)}\equiv\mp \frac{\left(r_{x} \pm ir_{y}\right)}{\sqrt{2}}.
\nonumber
\end{eqnarray}
} Using these matrices, it is straightforward to see that an antisymmetric $\hat{\mathcal{P}}_{(SM_S,IM_I)}$ can have one of the following forms 
\begin{eqnarray}
{\hat{\mathcal{P}}}_{(00,1M_I)} \equiv \frac{\alpha^{(M_I)}_I\otimes \beta_S}{\sqrt{8}},\hspace{.25cm}
{\hat{\mathcal{P}}}_{(1M_S,00)} \equiv \frac{\beta_I\otimes \alpha^{(M_s)}_S}{\sqrt{8}},
\end{eqnarray}
which can project onto two-nucleon states with $\left(S=0,I=1\right)$ and $(S=1,I=0)$ respectively. Note that these are the conventional isotriplet and isosinglet projection operators in the positive parity sector that are used frequently in literature \cite{Savage:1998ae, Chen:1999tn}.
On the other hand, a symmetric $\hat{\mathcal{P}}_{(SM_S,IM_I)}$ can project onto two-nucleon states with $(S=0,I=0)$ and $(S=1,I=1)$ and should have one of the following forms,
\begin{eqnarray}
{\hat{\mathcal{P}}}_{(00,00)} \equiv \frac{\beta_I\otimes \beta_S}{\sqrt{8}},\hspace{.25cm}
{\hat{\mathcal{P}}}_{(1M_S,1M_I)} \equiv \frac{\alpha^{(M_I)}_I\otimes \alpha^{(M_S)}_S}{\sqrt{8}},
\end{eqnarray}
respectively.

As it is the total angular momentum $J$ that is conserved in a two-nucleon scattering process, as opposed to the orbital angular momentum $L$, it is convenient to project a two-nucleon state in the $|LM_L,SM_S\rangle$ basis into a state in the $|JM_J,LS\rangle$ basis using the Clebsch-Gordan coefficients,
\begin{eqnarray} 
|NN;P,k^*\rangle _{JM_J,LS,IM_I}&=&\sum_{M_L,M_S}\langle JM_J|LM_L,SM_S\rangle~ |NN;P,k^*\rangle _{LM_L,SM_S,IM_I}.
\label{NNstate}
\end{eqnarray} 

Note that isospin remains a conserved quantum number up to small isospin breaking effects that we ignore for the nucleon systems.
In order to describe nucleon-nucleon interactions, we introduce an auxiliary dimer filed, similar to the scalar theory.\footnote{The S-wave dimer field in the nuclear sector is commonly referred to as a di-baryon field.} This field, that will be labeled $d^{LS}_{JM_J,IM_I;P}$, has the quantum numbers of two-nucleon states with total angular momentum $(J,M_J)$ and isospin quantum number of ${(I,M_I)}$ with orbital angular momentum $L$ and spin $S$. Now the action corresponding to the Lagrangian density of free nucleon and dimer fields in the momentum space can be written as
\begin{eqnarray} 
S^{\infty}_{kin}&=&\int\frac{d^4P}{(2\pi)^4}\left[N^\dag_P(E-\frac{\textbf{P}^2}{2m})N_P \right .
\nonumber\\
&~& ~~~ \left . -\sum_{\substack{J,M_J, I,M_I}}\sum_{L,S}~\left(d^{LS}_{JM_J,IM_I;P}\right)^\dag\left(E-\frac{\textbf{P}^2}{4m}-\Delta^{LS}_{JI}+\sum_{n=2}^{\infty} c^{LS}_{JI,n}(E-\frac{\textbf{P}^2}{4m})^{n}\right)d^{LS}_{JM_J,IM_I;P}\right].
\nonumber\\
\label{Skin}
\end{eqnarray} 

In order to write the interaction Lagrangian, one should note that, while the total angular momentum, parity, isospin and spin are conserved in a strongly interacting nucleon-nucleon process, the orbital angular momentum can change due to the action of tensor forces in nuclear physics. This is easy to implement in this formalism, as the two-nucleon states that are formed, Eq. (\ref{NNstate}), are compatible with the symmetries of the two-nucleon states. The interacting part of the action that does not mix angular momentum states, ${S}^{\infty}_{int,1}$, can then be written as
\begin{eqnarray} 
{S}^{\infty}_{int,1}&=&-\int\frac{d^4P}{(2\pi)^4}~\frac{d^4k}{(2\pi)^4}
\sum_{\substack{J,M_J, I,M_I}} ~ \sum_{L,M_L,S,M_S}
~{g^{LS}_{JI}}~
\langle JM_J|LM_L,SM_S\rangle
\nonumber\\
&~& \qquad \qquad \qquad \qquad \times \left[\left(d^{LS}_{JM_J,IM_I;P}\right)^\dag~\sqrt{4\pi}~Y_{LM_L}(\hat{\textbf{k}}^*)~{k}^{*L}~N^T_{k}~\hat{\mathcal{P}}_{(SM_S,IM_I)}~N_{P-k}+h.c. \right],
\nonumber\\
\label{S1}
\end{eqnarray} 
where ${g^{LS}_{JI}}$ denotes the coupling of a dimer field to the two-nucleon state with quantum numbers $\{J,I,L,S\}$. Note that the interactions must be azimuthally symmetric and so the reason the couplings are independent of azimuthal quantum numbers. Eqs. (\ref{positive}, \ref{negative}) now guide us to write the most general form of the interacting part of the action that is not diagonal in the angular momentum space, ${S}^{\infty}_{int,2}$, as follows
\begin{eqnarray} 
{S}^{\infty}_{int,2}&=&-\int\frac{d^4P}{(2\pi)^4}\frac{d^4k}{(2\pi)^4}
\sum_{\substack{J,M_J, I,M_I}} ~ \sum_{L,M_L,L',M_L',S,M_S}{h_{JI}} ~ \delta_{I,\frac{1+(-1)^J}{2}}\delta_{S,1}(\delta_{L,J+1}\delta_{L',J-1}+\delta_{L,J-1}\delta_{L',J+1})
\nonumber\\
&~& ~ \times
\langle JM_J|L'M_L',SM_S\rangle~\left[\left(d^{LS}_{JM_J,IM_I;P}\right)^\dag~\sqrt{4\pi}~Y_{L'M_{L}'}(\hat{\textbf{k}}^*)~{k}^{*{L'}}~N^T_{k}~\hat{\mathcal{P}}_{(SM_S,IM_I)}~N_{P-{k}}+h.c.\right].
\nonumber\\
\label{S2}
\end{eqnarray} 
Note that in this interacting term, spin, isospin and the initial and final angular momenta are all fixed for any given total angular momentum $J$. As a result we have only specified the $(JI)$ quantum numbers corresponding to coupling $h$. As in the scalar case, all the LECs of this effective Lagrangian, $\{\Delta^{LS}_{JI},c^{LS}_{JI,n}, g^{LS}_{JI}, h_{JI}\}$, can be tuned to reproduce the \textit{low-energy} expansion of the scattering amplitudes in the $J^{th}$ angular momentum channel with a given spin and isospin. As discussed in Sec. \ref{sec: Scalar}), in the scalar sector the LECs can be easily determined in terms of the ERE parameters and the renormalization scale. For coupled-channel systems, obtaining the LECs in terms of the scattering parameters requires solving a set of coupled equations. The tuning of the LECs is only an intermediate step in obtaining the relationship between the FV spectrum and the scattering amplitude, which can be easily circumvented by introducing the Bethe-Salpeter kernel.  

\begin{figure}
\begin{centering}
\includegraphics[scale=0.425]{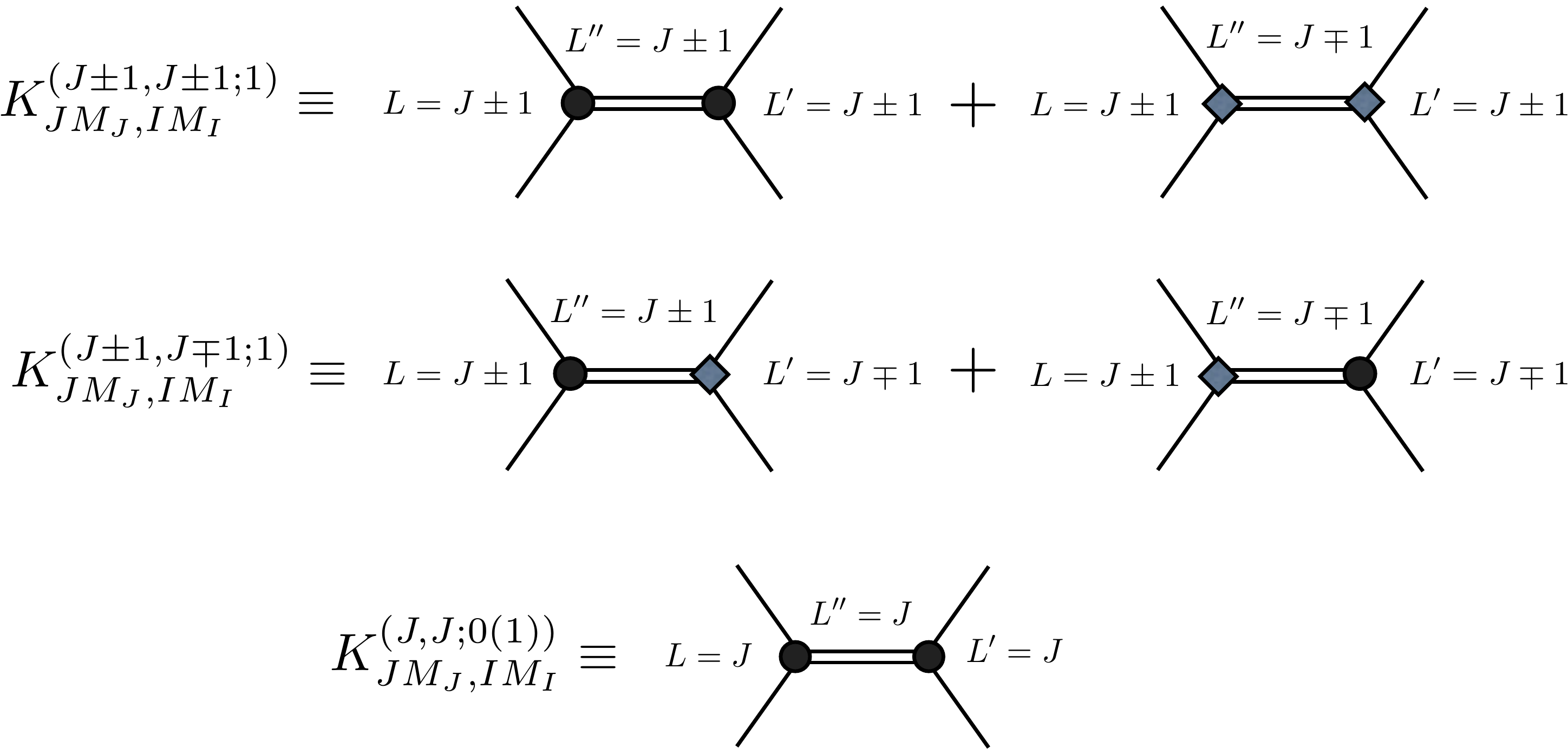}
\par
\caption{The leading $2\rightarrow2$ transition amplitudes in the sector with total angular momentum $J$, Eq. (\ref{KernelJ}). The superscripts in the kernels denote the initial angular momentum, $L$, final angular momentum, $L'$ and the conserved spin of the channels, $S$, respectively. The black dot represents the interaction vertex that conserves the partial-wave of the channel, and whose strength is parametrized by the coupling $g^{LS}_{JI}$, Eq. (\ref{S1}). The grey diamond denotes the vertex that mixes partial-waves, and whose strength is given by $h_{JI}$, Eq. (\ref{S2}). the double lines are the bare propagators corresponding to a dimer field with angular momentum $L''$.}\label{fig: Kernels}
\end{centering}
\end{figure}

Let us encapsulate the leading $2\rightarrow2$ transition amplitude between a two-nucleon state with $(JM_J,IM_I,LS)$ quantum numbers and a two-nucleon state with $(JM_J,IM_I,L'S')$ quantum numbers in the Bethe-Salpeter kernel, $K$. Note that since total angular momentum, spin and isospin are conserved in each $2\rightarrow 2$ transition, the kernel can be fully specified by $K_{JM_J;IM_I}^{(LL';S)}$. Since $J$ is conserved, the full kernel in the space of total angular momentum can be expressed as a block-diagonal matrix. In fact, it is straightforward to see that for each $J$-sector, the corresponding sub-block of the full matrix has the following form
\begin{eqnarray}
\left(\begin{array}{cccc}
K_{JM_{J};IM_{I}}^{(J-1,J-1;1)} & 0 & 0 & K_{JM_{J};IM_{I}}^{(J-1,J+1;1)}\\
0 & K_{JM_{J};IM_{I}}^{(J,J;0)} & 0 & 0\\
0 & 0 & K_{JM_{J};I'M_{I'}}^{(J,J;1)} & 0\\
K_{JM_{J};IM_{I}}^{(J+1,J-1;1)} & 0 & 0 & K_{JM_{J};IM_{I}}^{(J+1,J+1;1)}
\end{array}\right).
\label{KernelJ}
\end{eqnarray}
Note that for any given $J$, $I$, $L$ and $S$, there are $(2J+1)^2\times(2I+1)^2$ elements accounting for different values of $M_J$ and $M_I$ quantum numbers. Note also that the value of the isospin is fixed for each transition kernel. Explicitly, one finds that $I=\frac{1+(-1)^J}{2}$ and $I'=\frac{1+(-1)^{J+1}}{2}$.\footnote{Note that there is no $(I=0,S=0)$ channel for scattering in an even $J$ sector. Also there is no $(I=1,S=0)$ channel for scattering in an odd $J$ sector.} For the special case of $J=0$, the corresponding sub-sector is 
\begin{eqnarray}
\left(\begin{array}{cc}
K_{00;1M_I}^{(0,0;0)} & 0\\
0 & K_{00;1M_{I}}^{(1,1;1)}
\end{array}\right).
\label{KernelJ0}
\end{eqnarray}
These kernels, that correspond to leading transitions in all spin and isospin channels, are depicted in Fig. \ref{fig: Kernels}. Although one can read off the Feynman rules corresponding to these kernels from the  Lagrangian, Eqs. (\ref{Skin}, \ref{S1}, \ref{S2}), the FV energy eigenvalues can be determined without having to reference to the explicit form of these kernels, as will become evident shortly.

The scattering amplitude can be calculated by summing up all the $2\rightarrow2$ diagrams which can be obtained by any number of insertions of the transition kernels and the two-particle propagator loops. It can be easily seen that the infinite-volume two-particle loops, $\mathcal{G}^{\infty}$, are diagonal in total angular momentum, spin, isospin and orbital angular momentum. It is easy to show that $\mathcal{G}^{\infty}=2~I^{\infty}$, where $I^{\infty}$ is the infinite-volume loop for two identical bosons, Eq. (\ref{I-infinity}), hence the overall factor of two.
As a result, the scattering amplitude can be expressed as
\begin{eqnarray}
\mathcal{M}^{\infty}=-\mathcal{K}\frac{1}{1-\mathcal{G}^{\infty}\mathcal{K}},
\end{eqnarray}
where $\mathcal{K}$ is a matrix whose $J^{th}$-sub-block is given by Eq. (\ref{KernelJ}). Since $\mathcal{G}^{\infty}$ is diagonal, the $J^{th}$-sub-block of the infinite-volume scattering amplitude reads 
\begin{eqnarray}
\left(\begin{array}{cccc}
\mathcal{M}_{JM_{J};IM_{I}}^{(J-1,J-1;1)} & 0 & 0 & \mathcal{M}_{JM_{J};IM_{I}}^{(J-1,J+1;1)}\\
0 & \mathcal{M}_{JM_{J};IM_{I}}^{(J,J;0)} & 0 & 0\\
0 & 0 & \mathcal{M}_{JM_{J};I'M_{I'}}^{(J,J;1)} & 0\\
\mathcal{M}_{JM_{J};IM_{I}}^{(J+1,J-1;1)} & 0 & 0 & \mathcal{M}_{JM_{J};IM_{I}}^{(J+1,J+1;1)}
\end{array}\right),
\label{amplitude}
\end{eqnarray}
for any non-zero $J$ and
\begin{eqnarray}
\left(\begin{array}{cc}
\mathcal{M}_{00;1M_I}^{(0,0;0)} & 0\\
0 & \mathcal{M}_{00;1M_{I}}^{(1,1;1)}
\end{array}\right),
\label{amplitude}
\end{eqnarray}
for $J=0$. As is conventional, the scattering amplitude in channels with no partial-wave mixing can be parametrized by a scattering phase shift, $\delta_{JI}^{LS}$, according to
\begin{eqnarray}
\mathcal{M}_{JM_{J};IM_{I}}^{(JJ;S)}=\frac{4\pi}{Mk^*}\frac{e^{2i\delta_{JI}^{LS}}-1}{2i}\delta_{L,J}=\frac{4\pi}{Mk^*}\frac{1}{\cot{\delta_{JI}^{LS}}-i}\delta_{L,J},
\label{M-single}
\end{eqnarray}
while in channels where there is a mixing between the partial-waves, it can be characterized by two phase-shifts and one mixing angle, $\bar{\epsilon}_J$, \cite{Smatrix},
\begin{eqnarray}
\mathcal{M}_{JM_{J};IM_{I}}^{(J\pm1,J\pm1;S)}=\frac{4\pi}{Mk^*}\frac{\cos{2\bar{\epsilon}_J}e^{2i\delta_{JI}^{LS}}-1}{2i}\delta_{L,J\pm1},
\label{M-coupled1}
\\
\mathcal{M}_{JM_{J};IM_{I}}^{(J\pm1,J\mp1;S)}=\frac{4\pi}{Mk^*}\sin{2\bar{\epsilon}_J}\frac{e^{i(\delta_{JI}^{LS}+\delta_{JI}^{L'S})}}{2}\delta_{L,J\pm1}\delta_{L',J\mp1}.
\label{M-coupled2}
\end{eqnarray}
These relations are independent of $M_J$ and $M_I$ as the scatterings are azimuthally symmetric. We emphasize again that Kronecker deltas used to specify the $L$ quantum numbers should not be confused with the phase shifts. Note that for each $J$ sector, there is only one mixing parameter and as result no further labeling other than the $J$ label is necessary for $\bar{\epsilon}_J$.

The FV kernels are equal to the infinite-volume kernels (up to exponentially suppressed terms in volume below the pion production threshold \footnote{These corrections have been previously calculated for $\pi^+\pi^+$ \cite{Bedaque:2006yi} and $NN$ \cite{Sato:2007ms} S-wave scattering.}), and in particular the $J^{th}$-sub-block of such kernel is given by Eq. (\ref{KernelJ}). As in the scalar case, the only difference between the finite volume and infinite volume shows up in the s-channel bubble diagrams, where the two particles running in the loops can go on-shell and give rise to power-law volume corrections. It is straightforward to show that the two-nucleon propagator in the finite volume, $\mathcal{G}^V$, can be written as
\begin{eqnarray}
\mathcal{G}^V=\mathcal{G}^{\infty}+\delta\mathcal{G}^V,
\label{M-infinity}
\end{eqnarray}
where $\delta\mathcal{G}^V$ is a matrix in the $(JM_J,IM_I,LS)$ basis whose matrix elements are given by
\begin{eqnarray}
&& \left[\delta\mathcal{G}^V\right]_{JM_J,IM_I,LS;J'M_J',I'M_I',L'S'}=\frac{iMk^*}{4\pi}\delta_{II'}\delta_{M_IM_I'}\delta_{SS'}\left[\delta_{JJ'}\delta_{M_JM_J'}\delta_{LL'} +i\sum_{l,m}\frac{(4\pi)^{3/2}}{k^{*l+1}}c_{lm}^{\mathbf{P}}(k^{*2}) \right.
\nonumber\\
&& \qquad \qquad \qquad \qquad ~ \left .  \times \sum_{M_L,M_L',M_S}\langle JM_J|LM_L,SM_S\rangle \langle L'M_L',SM_S|J'M_J'\rangle \int d\Omega~Y^*_{L,M_L}Y^*_{l,m}Y_{L',M_L'}\right],
\nonumber\\
\label{deltaG}
\end{eqnarray}
and, as is evident, is \emph{neither} diagonal in the $J$-basis nor in the $L$-basis. The kinematic function $c_{lm}^{\mathbf{P}}(k^{*2})$ is defined in Eq. (\ref{clm}) and is evaluated at the on-shell relative momentum of two nucleons in the CM frame. The full FV two-nucleon scattering amplitude can be evaluated by summing up all $2\rightarrow2$ FV diagrams,\footnote{We note that the notion of a FV scattering amplitude is merely a mathematical concept as there is no asymptotic state by which one could define the scattering amplitude in a finite volume.   In principle one should look at the pole locations of the two-body correlation function to gain access to physically relevant states. However, it can be easily shown that both correlation function and the so-called FV scattering amplitude have the same pole structure, so we use the latter to make the presentation simpler.}
\begin{eqnarray}
\mathcal{M}^{V}=-\mathcal{K}\frac{1}{1-\mathcal{G}^{V}\mathcal{K}}=\frac{1}{(\mathcal{M}^{\infty})^{-1}+\delta\mathcal{G}^{V}},
\label{M-V}
\end{eqnarray}
where in the second equality the kernel is eliminated in favor of $\mathcal{M}^{\infty}$ and $\mathcal{G}^{\infty}$ using Eq. (\ref{M-infinity}). The energy eigenvalues of the two-nucleon system arise from the poles of $\mathcal{M}^V$ which satisfy the following determinant condition
\begin{eqnarray}
\det\left[{(\mathcal{M}^{\infty})^{-1}+\delta\mathcal{G}^{V}}\right]=0.
\label{NNQC}
\end{eqnarray}
This quantization condition clearly reduces to Eq. (\ref{FullQCboson}) for two-boson systems when setting $S=0$ \footnote{The symmetry factor in both scattering amplitude and the FV function will cancel out in the determinant condition, leaving the FV QC, Eq. \ref{FullQCboson}, insensitive to the distinguishability of the particles.}, and is in agreement with the result of Ref. \cite{Bernard:2008ax} for meson-baryon scattering after setting $S=1/2$. This result also extends the result of Ref. \cite{Ishizuka:2009bx} for two-nucleon systems to moving frames.
\footnote{Although both $(\mathcal{M}^{\infty})^{-1}$ and $\delta\mathcal{G}^{V}$ have imaginary parts, for each partial-wave, there occurs an intricate cancellation of their imaginary parts.  One could, on the other hand, work with the kernel $\mathcal{K}$ and finite-volume propagator $\mathcal{G}^V$, both manifestly real expressions, to define the QC.  From Eq. (\ref{M-V}) it is evident that the finite volume QC can be equivalently written as $\det[\mathcal{K}^{-1}-\mathcal{G}^{V}]=0$,
which has yet no reference to the scattering amplitude. However Eq. (\ref{NNQC}) is comprised of renormalization-scale \emph{independent} quantities, whereas $\mathcal{K}$ and $\mathcal{G}^V$ are both renormalization-scale \emph{dependent} (only their difference is scale independent).  For this reason and to make connection with infinite-volume observables easier, we utilize Eq. \ref{NNQC}. Since the cancellation of imaginary terms is not trivial in channels with partial-wave mixing, we explicitly show how this cancellation occurs in appendix \ref{app:red-example}.}

It is important to note that in deriving this result we have only assumed that the FV kernels are exponentially close to their infinite volume counterparts. Therefore the result obtained is valid for energies up to the inelastic threshold. In the nuclear sector this corresponds to the pion production threshold, $E^*=m_\pi$, which is well above the t-channel cut defined by $E^*_{cut}\equiv m_\pi^2/4m_N$ ($\sim 5$~MeV at the physical point). For energies above $E^*_{cut}$, the dimer formalism written in Eqs.~(\ref{Skin}-\ref{S2}) will get corrections from coupling of nucleons to pions and the LECs appearing in the action will get $m_\pi$-dependent corrections \cite{Weinberg:1990rz, Jenkins:1990jv, pds, pds2, Beane:2001bc}. All such corrections due to the dynamical pions above the t-channel cut, including pion exchange diagrams, can be still embedded in the interacting kernels in both infinite volume and finite volume. Below the pion production threshold, these corrections in the FV kernels are still exponentially close to their infinite volume counterparts, making the results presented in this section valid beyond the t-channel cut.

The determinant of the QC is defined in the basis of $(JM_J,IM_I,LS)$ quantum numbers and is over an infinite dimensional matrix. To be practical, this determinant should be truncated in the space of total angular momentum and orbital angular momentum. Such truncation is justified since in the low-momentum limit the scattering phase shift of higher partial-waves $L$  scales as $k^{*2L+1}$. In the next section, by truncating the partial-waves to $L\leq3$, we unfold this determinant condition further, and present the reduction of this master formula to separate QCs for energy eigenvalues in different irreps of the corresponding symmetry group of the two-nucleon system. The first trivial reduction in the QC clearly takes place among different spin-isospin channels. In particular, it is straightforward to see that the QC in Eq. (\ref{NNQC}) does not mix $(S=0,I=1)$, $(S=1,I=0)$, $(S=0,I=0)$ and $(S=1,I=1)$ sectors, and automatically breaks into four independent determinant conditions that correspond to different spin-isospin sectors,
\begin{eqnarray}
\textrm{Det}\left[{(\mathcal{M}^{\infty})^{-1}+\delta\mathcal{G}^{V}}\right]=\prod_{I=0}^{1}\prod_{S=0}^{1} \det \left[(\mathcal{M}^{\infty}_{(I,S)})^{-1}+\delta \mathcal{G}^{V}_{(I,S)} \right]=0 .
\label{NNQC-IS}
\end{eqnarray}
This is due to the fact that each J-sub block of the scattering amplitude matrix can be separated into three independent sectors as following
\begin{eqnarray}
\mathcal{M}^{\infty}_{(I,1)}\equiv\left(\begin{array}{ccc}
\mathcal{M}_{J;I}^{(J-1,J-1;1)} &  & \mathcal{M}_{J;I}^{(J-1,J+1;1)}\\
\\
\mathcal{M}_{J;I}^{(J+1,J-1;1)} &  & \mathcal{M}_{J;I}^{(J+1,J+1;1)}
\end{array}\right), ~\mathcal{M}^{\infty}_{(I,0)}\equiv\begin{array}{c}
\mathcal{M}_{J;I}^{(J,J;0)}\end{array}, ~\mathcal{M}^{\infty}_{(I',1)}\equiv
\begin{array}{c}
\mathcal{M}_{J;I'}^{(J,J;1)}\end{array},
\nonumber\\
\label{amplitude-IS}
\end{eqnarray}
where $I$ and $I'$ are defined after Eq. (\ref{KernelJ}). Since the $M_J$ and $M_I$ indices are being suppressed, one should keep in mind that each block is still a $(2J+1)^2\times(2I+1)^2$ diagonal matrix. If $J$ is even, these amplitudes describe scattering in the negative parity isotriplet, positive parity isotriplet and positive parity isosinglet channels, respectively. For an odd $J$, these amplitudes correspond to scattering in the positive parity isosinglet, negative parity isosinglet and negative parity isotriplet channels, respectively.
Due to the reduced symmetry of the FV, $\delta \mathcal{G}^V$ has off-diagonal terms in the basis of total angular momentum $J$. So although the QC in Eq. (\ref{NNQC}) fully breaks down in the $(I,S)$-basis, it remains coupled in the $(J,L)$-basis. In order to further reduce the determinant conditions in Eq. (\ref{NNQC-IS}), the symmetries of the FV functions must be studied in more detail. This will be the topic of the next section, Sec. \ref{sec: Reduction}.

\section{Symmetry Considerations and Quantization Conditions \label{sec: Reduction}}
Lattice QCD calculations are performed in cubic volumes with periodic boundary conditions on the fields in spatial directions. As a result, the energy eigenstates of the two-particle systems at rest transform according to various irreps of the cubic group, depending on the interpolating operators that are used. Although it is convenient to think of the determinant condition, Eq. (\ref{NNQC}), as a determinant in the $J$-basis, one should expect that for zero CM momentum, this equation splits into $5$ independent quantization conditions corresponding to the $5$ irreps of the cubic group (see table (\ref{groups})). Furthermore, the degeneracy of the energy eigenvalues will reflect the dimension of the corresponding irrep. In general, the FV matrix $\delta \mathcal{G}^{V}$, Eq. (\ref{deltaG}), although being sparse, mixes states corresponding to different irreps of the cubic group. As a result, at least a partial block-diagonalization of this matrix is necessary to unfold different irreps that are present due to the decomposition of a given total angular momentum $J$. When the two-particle system is boosted, the symmetry group of the system is no longer cubic, and the reduction of the determinant condition, Eq. (\ref{NNQC}), takes place according to the irreps of the corresponding point group, table. (\ref{groups}). In the following section, this reduction procedure and the method of block diagonalization will be briefly discussed. In particular we aim to obtain all the QC satisfied by the phase shifts and mixing parameters of the NN system of channels with $l\leq3$. We constrain this study to the CM momenta $\mathbf{d}=(0,0,0)$, $\mathbf{d}=(0,0,1)$ and $\mathbf{d}=(1,1,0)$, where $\mathbf{d}=\frac{L\mathbf{P}}{2\pi}$, which  provides $49$ independent QCs satisfied by different scattering parameters in these channels. As mentioned earlier, these boost vectors correspond to cubic ($O$), tetragonal ($D_4$) and orthorhombic ($D_2$) point groups, respectively.\footnote{Of course with the inclusion of the inversion operation, the relevant symmetry groups are $O_h$, $D_{4h}$ and $D_{2h}$ with twice as number of elements as that of $O$, $D_{4}$ and $D_{2}$ groups, respectively. Consequently the corresponding irreps of these groups will have definite parities. Since we have separated the NN system into four different spin-isospin channels with definite parities, it suffices to constrain the symmetry groups of the calculation to $O$, $D_4$ and $D_2$. However the QCs presented in the last appendix of this paper should be realized according to the assigned irrep of the symmetry group with the specified parity of the channel.}
\begin{center}
\begin{table} [h]
\label{tab:param1}
\begin{tabular}{|ccccc|}
\hline
$\hspace{.3cm}\mathbf{d}\hspace{.3cm}$&$\hspace{.3cm}$point group$\hspace{.3cm}$&$\hspace{.3cm}$classification$\hspace{.3cm}$&$\hspace{.3cm}N_{\text{elements}}\hspace{.3cm}$&irreps (dimension) \\\hline \hline
$~(0,0,0)~$&${O}$&cubic&$24$&$~A_1(1),A_2(1),E(2),T_1(3),T_2(3)~$\\
$~(0,0,1)~$&$D_{4}$&tetragonal&$8$&$~A_1(1),A_2(1),E(2),B_1(1),B_2(1)~$\\ 
$~(1,1,0)~$&$D_{2}$&orthorhombic&$4$&$~A(1),B_1(1),B_2(1),B_3(1)~$
\\\hline\hline
\end{tabular}
\caption{The classification of the point groups corresponding to the symmetry groups of the FV calculations with different boost vectors. The forth column shows the number of elements of each group.}
\label{groups}
\end{table}
\end{center}

In order to calculate matrix elements of the FV matrix $\delta \mathcal{G}^{V}$, one can take advantage of the symmetries of the $c_{lm}^{\mathbf{P}}$ functions as defined in Eq. (\ref{clm}). The relations between non-zero $c_{lm}^{\mathbf{P}}$s for any given angular momentum $l$ can be easily deduced from the transformation properties of these functions under symmetry operations of the corresponding point groups
\begin{eqnarray}
c^{\mathbf{P}}_{lm}=\sum_{m'=-l}^{l}\mathcal{D}^{(l)}_{mm'}(R_{\mathcal{X}})~c^{\mathbf{P}}_{lm'},
\label{clm-trans}
\end{eqnarray}
where $R_{\mathcal{X}}$ is the rotation matrix corresponding to each symmetry operation $\mathcal{X}$ of the group, and $\mathcal{D}^{(l)}_{mm'}$ denotes the matrix elements of the Wigner $\mathcal{D}$-matrix \cite{luscher2}. Beside these transformations, one can see that $c^\mathbf{P}_{lm}$s are invariant under inversion as can be easily verified from Eq. (\ref{clm}) for an arbitrary boost, and as a result all $c^\mathbf{P}_{lm}$s with an odd $l$ vanish.\footnote{For systems with non-equal masses, this is no longer true when the system is boosted. Since the parity is broken for such systems, even and odd partial-waves mix with each other in the QCs, see Refs. \cite{Bour:2011ef, Davoudi:2011md, Fu:2011xz, Leskovec:2012gb}.} Table (\ref{nonzero-clm}) contains all such relations for non-vanishing $c^\mathbf{P}_{lm}$s up to $l=6$ for $\mathbf{d}=(0,0,0)$, $\mathbf{d}=(0,0,1)$ and $\mathbf{d}=(1,1,0)$ boost vectors.\footnote{A closer look at nonrelativistic $c_{lm}^{\mathbf{P}}$ functions shows that all $c_{2,\pm2}^{\mathbf{(1,1,0)}}$ and $c_{4,\pm2}^{\mathbf{(1,1,0)}}$ vanish. This extra symmetry of NR systems significantly simplifies the QCs presented in appendix C for boost vector $(1,1,0)$.}
\begin{center}
\begin{table} [h]
\label{tab:param3}
\begin{tabular}{|c|c|c|}
\hline 
\textbf{d}=(0,0,0) & \textbf{d}=(0,0,1) & \textbf{d}=(1,1,0)\tabularnewline
\hline 
\hline 
$~c_{00}^{P}~$ & $~c_{00}^{P}~$ & $~c_{00}^{P}~$\tabularnewline
$~c_{40}^{P}~$ & $~c_{20}^{P}~$ & $~c_{20}^{P}~$\tabularnewline
$~c_{44}^{P}=c_{4,-4}^{P}=\sqrt{\frac{5}{14}}c_{40}^{P}~$ & $~c_{40}^{P}~$ & $~c_{22}^{P}=-c_{2,-2}^{P}~$\tabularnewline
$~c_{60}^{P}~$ & $~c_{44}^{P}=c_{4,-4}^{P}~$ & $~c_{40}^{P}~$\tabularnewline
$~c_{64}^{P}=c_{6,-4}^{P}=-\sqrt{\frac{7}{2}}c_{60}^{P}~$ & $~c_{60}^{P}~$ & $~c_{42}^{P}=-c_{4,-2}^{P}~$\tabularnewline
 & $~c_{64}^{P}=c_{6,-4}^{P}~$ & $~c_{44}^{P}=c_{4,-4}^{P}~$\tabularnewline
 &  & $~c_{60}^{P}~$\tabularnewline
 &  & $~c_{62}^{P}=-c_{6,-2}^{P}~$\tabularnewline
 &  & $~c_{64}^{P}=c_{6,-4}^{P}~$\tabularnewline
\hline 
\end{tabular}
\caption{The nonzero $c_{lm}^P$s up to $l=6$ for three different boost vectors $\mathbf{d}$.}
\label{nonzero-clm}
\end{table}
\par\end{center}

An important point regarding the $c^\mathbf{P}_{lm}$ functions is that they explicitly depend on the direction of the boost vector. In other words, $c^\mathbf{P}_{lm}$s that correspond to different boost vectors with the same magnitude $|\mathbf{d}|=n$  are not equal. As a result the corresponding set of non-zero $c^\mathbf{P}_{lm}$s as well as the relations among them, for permutations of the components of $(0,0,1)$ and $(1,1,0)$ boost vectors are different from those that are listed in Table (\ref{nonzero-clm}). Although this difference in general results in different $\delta \mathcal{G}^{V}$ matrices, e. g. for $(1,0,0)$, $(0,1,0)$ and $(0,0,1)$ boost vectors, as is shown in appendix \ref{app:invariant}, the master equation (\ref{NNQC}) is invariant under a  $\mathbf{P}\rightarrow \mathbf{P}'$ transformation when $\mathbf{P}$ and $\mathbf{P}'$ are related by a cubic rotation and  $|\mathbf{P}|=|\mathbf{P}'|$. The reason is that there exists a unitary transformation that relates $\delta \mathcal{G}^{V,\mathbf{P}}$ to $\delta \mathcal{G}^{V,\mathbf{P}'}$, leaving the determinant condition invariant. 

Since the relations among $c^\mathbf{P}_{lm}$ are simpler when one assumes boost vectors that discriminate the z-axis relative to the other two Cartesian axes, we will present the QCs corresponding to $\mathbf{d}=(0,0,1)$ and $\mathbf{d}=(1,1,0)$ boost vectors only. The lattice practitioner can still use the QCs presented to extract the scattering parameters of the NN system from the energy eigenvalues of lattice calculations with other permutations of these boost vectors. It is however crucial to input the boost vectors that are specified in this paper when calculating the $c^\mathbf{P}_{lm}$ functions in the QCs (instead of the boost vectors that are used in the lattice calculation). In order to increase statistics and the precision of results, one should perform the lattice calculation with all possible boost vectors of a given magnitude that belong to the same $A_1$ irrep of the cubic group,\footnote{Note that in higher momentum shells, there occurs multiple $A_1$ irreps of the cubic group. This indicates that there are classes of momentum vector that do not transform into each other via a symmetry operation of the cubic group, e. g. $(2,2,1)$ and $(0,0,3)$ vectors in the $\mathbf{n}^2=9$ shell. However, as is discussed, another property of the $c_{lm}^{\mathbf{P}}$ functions for non-relativistic degenerate masses indicates that the value of the FV function is the same for these two boost vectors as they are both of the form $(2n_1,2n_2,2n_3+1)$ with $n_i \in \mathbb{Z}$.} and use the average energy eigenvalues in the QCs presented to determine the scattering parameters; keeping in mind that $c^\mathbf{P}_{lm}$ functions have to be evaluated at the boost vectors considered in this paper.

The other fact that should be pointed out is that due to the symmetries of the $c^\mathbf{P}_{lm}$ function for equal masses, the system at rest with $\mathbf{d}=(0,0,0)$ exhibits the same symmetry transformation as that of the $(2n_1,2n_2,2n_3)$ boost where $n_1,n_2,n_3$ are integers. Similarly, the symmetry group of the calculations with $(0,0,1)$ ($(1,1,0)$) boost is the same as that of $(2n_1,2n_2,2n_3+1)$ ($(2n_1+1,2n_2+1,2n_3)$) boosts. As a result, the quantization conditions presented in appendix \ref{app:QC} can be used with these boost vectors as well. It is worth mentioning that for relativistic two-particle systems with degenerate masses, the above statement is no longer true. This is due to the fact that the boost vector dependence of the relativistic $c_{lm}^{\mathbf{P}}$ function is different from that of the NR counterpart, leading to more distinct point group symmetries for different boosts \cite{movingframe, sharpe1, Christ:2005gi}.

Back to our main goal, we aim to break the master equations (\ref{NNQC-IS}) into separate QCs corresponding to each irrep of the symmetry group of the problem. In fact, from the transformation law of the $\delta \mathcal{G}^{V}$ function under a symmetry operation of the group, 
\begin{eqnarray}
\left[\delta\mathcal{G}^V\right]_{JM_J,LS;J'M_J',L'S}=\sum_{\bar{M}_J=-J}^{J}\sum_{\bar{M}_J'=-J'}^{J'}\mathcal{D}^{(J)}_{M_J,\bar{M}_J}(R_{\mathcal{X}})\left[\delta\mathcal{G}^V\right]_{J\bar{M}_J,LS;J'\bar{M}_J',L'S}\mathcal{D}^{(J')}_{\bar{M}_J',M_J'}(R_{\mathcal{X}}^{-1}),
\nonumber\\
\label{dG-trans}
\end{eqnarray}
one can deduce that there is a unitary transformation which brings the matrix $\delta \mathcal{G}^{V}$ to a block-diagonal form. Note that we have suppressed the isospin quantum numbers as $\delta\mathcal{G}^V$ is diagonal in the isospin basis. Each of these blocks then can be identified by a given irrep of the symmetry group of the problem. Such transformation eventually breaks the determinant conditions (\ref{NNQC-IS}) to separate determinant conditions corresponding to each irrep of the point group of the system. Explicitly in each spin and isospin sector,
\begin{eqnarray}
\det \left[(\mathcal{M}^{\infty}_{(I,S)})^{-1}+\delta \mathcal{G}^{V}_{(I,S)} \right]=\prod_{\Gamma^i}\det\left[(\mathcal{M}^{\infty-1}_{(I,S)})_{\Gamma^i}+\delta \mathcal{G}^{V,\Gamma^i}_{(I,S)} \right]^{N(\Gamma^i)}=0 .
\label{NNQC-irrep}
\end{eqnarray}
where $\Gamma^i$ denotes each irrep of the corresponding group and $N(\Gamma^i)$ is the dimensionality of each irrep. The dimensionality of each of these smaller determinant conditions is  given by the multiplicity of each irrep in the decomposition of angular momentum channels that are being included in the scattering problem. As is seen in  appendix \ref{app:QC}, although from the master quantization condition, for some of the NN channels with $J\leq4$ and $l\leq 3$, one has to deal with a determinant of $30 \times 30$ matrices, upon such reduction of the master equation, one arrives at QCs that require taking the determinant of at most $9 \times 9$ matrices. We demonstrate this procedure in more detail for one example in appendix \ref{app:red-example}. For the rest of the channels and boosts only the final result of such reduction will be presented (see appendix \ref{app:QC}).\footnote{Although these QCs are the main results of this paper, to achieve a better presentation of these long equations, we have tabulated them in an appendix.} It is also shown in appendix \ref{app:red-example} that the QC in Eq. (\ref{NNQC}) is real despite the fact that both $\delta \mathcal{G}^V$ and $\mathcal{M}^{\infty}$ are complex quantities.

\section{Summary and Conclusion \label{sec: S&M}}

The auxiliary field formalism has been extended to arbitrary partial-waves in both the scalar and nuclear sectors in infinite and finite volumes. Such a formalism can be used to derive a master equation that relates the FV two-nucleon energies and the scattering parameters of the two-nucleon systems with arbitrary spin, isospin and angular momentum. This master equation, Eq.~(\ref{NNQC}), that is the extension of the L\"uscher formalism to two-nucleon systems,  is valid up to inelastic thresholds and is general for any non-relativistic CM boost. 

The QC is a determinant over an infinite-dimensional matrix in the basis of angular momentum, and in practice it is necessary to truncate the number of partial-waves that contribute to the scattering. By taking advantage of symmetries within the problem, we show how the master equation can be reduced to finite-size blocks that relate particular partial-wave channels (and their mixing) to different spin-isospin channels and different irreps of the corresponding point group of the system. By truncating the matrices at $J\leq 4$ and $l\leq 3$, this procedure requires block-diagonalizing matrices as large as $30\times30$. The resulting QCs are determinant conditions involving matrices that are at most $9\times9$, and are therefore practical to be used in future LQCD calculations of NN systems. We have provided one explicit example of this reduction for the scattering in the positive parity isosinglet channel for zero CM momentum in appendix \ref{app:red-example}.  All other QCs for different CM boosts, parity, isospin, spin, and angular momentum are enumerated in appendix \ref{app:QC}. Having studied the zero CM boost as well as $(0,0,1)$ and $(1,1,0)$ boosts, we arrive at $49$ independent QCs for four different spin and isospin channels giving access to all 16 phase shifts and mixing parameters in these channels. Table \ref{scatt-param} summarizes all such  scattering parameters and the corresponding equations that give access to each parameter as presented in this paper. As a supplemental material to this article, these QCs are also provided in a mathematica notebook format to facilitate their use in future LQCD calculations, see Ref. \cite{BDLsupp}.  Given the fact that NN-systems couple different partial-waves, in order to reliably extract scattering parameters from LQCD calculations, these calculations must be necessarily performed  in multiple boosts and various irreps of the corresponding symmetry group.

\begin{table}
\begin{centering}
\begin{tabular}{|c||c|c|c|c|c|}
\hline 
J & $0$ & $1$ & $2$ & $3$ & $4$\tabularnewline
\hline 
\hline 
\multirow{4}{*}{$I=0,\; S=1$} & - & $\delta_{1,S},\delta_{1,D},\epsilon_{1,SD}$ & $\delta_{2,D}$ & $\delta_{3,D}$ & -\tabularnewline
\cline{2-6} 
 &  & Eqs.  & Eqs.  & Eqs. & \tabularnewline
 &  & \ref{I000T1}, \ref{I001A2}, \ref{I001E} & \ref{I000E}, \ref{I000T2}, \ref{I001A1}, & \ref{I000A2}, \ref{I000T1}, \ref{I000T2}, \ref{I001B1} & \tabularnewline
 &  & \ref{I110B1}, \ref{I110B2}, \ref{I110B3} & \ref{I001B1}, \ref{I001B2}, \ref{I001E} & \ref{I001B2}, \ref{I001A2}, \ref{I001E}, \ref{I110B1} & \tabularnewline
  &  &  & \ref{I110A} & \ref{I110B2}, \ref{I110B3}, \ref{I110A} & \tabularnewline
\hline
\multirow{5}{*}{$I=1,\; S=0$} & $\delta_{0,S}$ & - & $\delta_{2,D}$ & - & -\tabularnewline
\cline{2-6}
 & Eqs. &  & Eqs.  &  & \tabularnewline
 & \ref{II000A1}, \ref{II001A1}, \ref{II110A} &  & \ref{II000E}, \ref{II000T2}, \ref{II001A1}, &  & \tabularnewline
 &  &  & \ref{II001B1}, \ref{II001B2}, \ref{II001E}, &  & \tabularnewline
 &  &  &  \ref{II110B1}, \ref{II110B2}, \ref{II110B3}, &  & \tabularnewline
 &  &  &   \ref{II110A} &  & \tabularnewline
\hline 
\multirow{5}{*}{$I=0,\; S=0$} & - & $\delta_{1,P}$ & - & $\delta_{3,F}$ & - \tabularnewline
\cline{2-6}
 &  & Eqs.  &  & Eqs.  &  \tabularnewline
 &  & \ref{III000T1}, \ref{III001A2}, \ref{III001E} &  & \ref{III000T1}, \ref{III000A}, \ref{III000T2}, \ref{III001A2}, & \tabularnewline
 &  & \ref{III110B2}, \ref{III110B3}, \ref{III110B1} &  & \ref{III001B1}, \ref{III001B2}, \ref{III001E}, \ref{III110B2}, & \tabularnewline
 &  &  &  & \ref{III110B3}, \ref{III110A}, \ref{III110B1} & \tabularnewline
\hline 
\multirow{5}{*}{$I=1,\; S=1$} & $\delta_{0,P}$ & $\delta_{1,P}$ & $\delta_{2,P},\delta_{2,F},\epsilon_{2,PF}$ & $\delta_{3,F}$ & $\delta_{4,F}$ \tabularnewline
\cline{2-6} 
 & Eqs.  & Eqs.  & Eqs.  & Eqs.  & Eqs.  \tabularnewline
 & \ref{IV000A1}, \ref{IV001A1}, \ref{IV110A} & \ref{IV000T1}, \ref{IV001A2}, \ref{IV001E} & \ref{IV000T2}, \ref{IV000E}, \ref{IV001A1}, & \ref{IV000T1}, \ref{IV000T2}, \ref{IV000A2}, & \ref{IV000T1}, \ref{IV000T2}, \ref{IV000A1}, \tabularnewline
 &  & \ref{IV110A}, \ref{IV110B1}, \ref{IV110B2} & \ref{IV001B1}, \ref{IV001B2}, \ref{IV001E}, & \ref{IV001A2}, \ref{IV001B1}, \ref{IV001B2}, & \ref{IV001A1}, \ref{IV001A2}, \ref{IV001B1}, \tabularnewline
 &  & \ref{IV110B3} & \ref{IV110B1}, \ref{IV110B2}, \ref{IV110B3} & \ref{IV001E}, \ref{IV110A}, \ref{IV110B1}, & \ref{IV001B2}, \ref{IV000E}, \ref{IV001E}, \ref{IV110A},  \tabularnewline
&  &  &  &  \ref{IV110A}, \ref{IV110B2}, \ref{IV110B3} & \ref{IV110B1}, \ref{IV110B2},
\ref{IV110B3}
\tabularnewline
\hline
\end{tabular}
\par\end{centering}

\caption{The scattering parameters that can be determined from the QCs presented in appendix \ref{app:QC} for all four different spin-isospin channels. The reference to the relevant equations in extracting each parameter is given in the table. These equations are assumed to be used in Eq. (\ref{QC-simplified}). The subscript in each parameter denotes the total $J$ as well as the partial-wave of the channel the parameter corresponds to. Alternatively one can use the supplemental material to this article which contains the list of all quantization conditions in a mathematica notebook format \cite{BDLsupp}.}
\label{scatt-param}
\end{table}

As is extensively discussed in this paper, although the QCs are tabulated only for $(0,0,0)$, $(0,0,1)$ and $(1,1,0)$ boosts, they can be used for a more general set of boost vectors. Since the symmetry group of the two equal-mass problem depends on the evenness and oddness of the components of the boost vectors, the QCs presented in appendix \ref{app:QC} for different irreps of the corresponding symmetry groups, can be equally used for the $(2n_1,2n_2,2n_3)$, $(2n_1,2n_2,2n_3+1)$ and $(2n_1+1,2n_2+1,2n_3)$ boost vectors $\forall~n_i\in \mathbb{Z}$. The other generality of the QCs with regard to the boost vectors is that upon a cubic rotations of the CM boost vectors, the QCs remain unchanged although the FV functions will be different. As a result, a lattice practitioner may use an average of the energy levels extracted from the NN correlation functions for all different boosts with the same magnitude, belonging to the same $A_1$ irrep of the cubic group, in the QCs of appendix \ref{app:QC} to improve the statistics of the calculation. However, it should be noted that in evaluating the FV functions ($c_{lm}^{\mathbf{P}}s$) only the boost vectors considered in this paper must be used. In summary, the lattice practitioner can extract the desired scattering parameters of the NN-system by performing the following steps:

\begin{enumerate}
\item For a given irrep $\Gamma$, evaluate the $NN$ correlation function with all possible boost vectors with magnitude ${d}$ that are related to each other via a cubic rotation, $\{C_{NN}^{\Gamma,\textbf{d}_1},\ldots,C_{NN}^{\Gamma,\textbf{d}_{N_d}}\}$.
\item Average the value of the correlation functions over all boost vectors used in the previous step, $C_{NN}^{\Gamma,{d}}=\sum_{i}^{N_d}C_{NN}^{\Gamma,\textbf{d}_i}/N_d$.
\item Obtain the non-relativistic finite volume energy, $E_{NR}^\Gamma=E_{NN}^\Gamma-2m_N$, from the asymptotic behavior of the correlation function and therefore obtain the value of the relative momentum $k^*$ from $k^*=\sqrt{M_NE-(\pi\mathbf{d})^2/L^2}$.
\item Determine scattering parameters from the QCs in appendix \ref{app:QC}:

\begin{enumerate}
\item Use $\mathbf{d}=(0,0,0)$ if $\mathbf{d}$ is a permutation of $(2n_1,2n_2,2n_3)$,
\item Use $\mathbf{d}=(0,0,1)$ if $\mathbf{d}$ is a permutation of $(2n_1,2n_2,2n_3+1)$,
\item Use $\mathbf{d}=(1,1,0)$ if $\mathbf{d}$ is a permutation of $(2n_1+1,2n_2+1,2n_3)$.
\end{enumerate}
\end{enumerate}

In a following paper \cite{Briceno:2013}, we will use the QCs derived here, coupled with empirical two-nucleon phase shifts, to provide estimates for the energy levels expected to be seen in future LQCD calculations at the physical pion mass. This will allow for an estimation of the precision that is needed for such calculations. Clearly the impact of this formalism on our understanding of the nature of nuclear forces depends upon best implementing this formalism in the upcoming LQCD calculations of the NN systems.

\subsection*{Acknowledgment}
We would like to thank Martin J. Savage for valuable discussions and for his feedback on the first manuscript of this paper. RB and ZD also would like to thank Richard J. Furnstahl for useful conversations. RB and ZD were supported
in part by the DOE grant DE-FG02-97ER41014. The work of TL was performed
under the auspices of the U.S. Department of Energy by Lawrence Livermore National
Laboratory under Contract DE-AC52-07NA27344. Part of this work were planned as a result of discussions among the authors during the INT workshop on ``\textit{Nuclear reactions from lattice QCD}'' in the Institute for Nuclear Theory at the University of Washington in March 2013.

\appendix


\section{Quantization Conditions under $\mathbf{P}\rightarrow \mathbf{P}'$ transformation when $\mathbf{P}$ and $\mathbf{P}'$ are related by a cubic rotation and $|\mathbf{P}|=|\mathbf{P}'|$ \label{app:invariant}}
We aim to show that the master QC, Eq. (\ref{NNQC}), is invariant under a $\mathbf{P}\rightarrow \mathbf{P}'$ transformation where $\mathbf{P}$ and $\mathbf{P}'$ are two boost vectors that are related by a cubic rotation. Denoting such rotation by $R$, it is straightforward to show that
\begin{eqnarray}
c^{\mathbf{P}'}_{lm}=\sum_{m'=-l}^{l}\mathcal{D}^{(l)}_{mm'}(R)~c^{\mathbf{P}}_{lm'}.
\label{clm-trans-P}
\end{eqnarray}
Note that for $\mathbf{P}=0$ this relation reduces to Eq. (\ref{clm-trans}), while for a general non-zero boost vector, it only holds if the rotation $R$ corresponds to the symmetry operation of the cube. For example, such a transformation can take the $c^{\mathbf{P}}_{lm}$ function evaluated with $\textbf{d}=(0,0,1)$ to a $c^{\mathbf{P}'}_{lm}$ evaluated with $\textbf{d}=(1,0,0)$. To proceed let us rewrite the $\delta \mathcal{G}^{V}$ matrix elements as given in Eq. (\ref{deltaG}) in terms of the matrix elements of the $\mathcal{F}^{FV}$ that is defined in Eq. (\ref{F}) for the scalar sector,
\begin{eqnarray}
&& \left[\delta\mathcal{G}^{V,\mathbf{P}}\right]_{JM_J,IM_I,LS;J'M_J',I'M_I',L'S'}=\frac{iM}{4\pi}\delta_{II'}\delta_{M_IM_I'}\delta_{SS'}\times
\nonumber\\
&& \qquad \qquad \qquad ~ \times \left[k^*\delta_{JJ'}\delta_{M_JM_J'}\delta_{LL'} +i\sum_{M_L,M_L',M_S}\langle JM_J|LM_L,SM_S\rangle \langle L'M_L',SM_S|J'M_J'\rangle \mathcal{F}^{FV,\mathbf{P}}_{LM_L,L'M_L'} \right].
\nonumber\\
\label{deltaG-F}
\end{eqnarray}
Superscript $\mathbf{P}$ on $\delta\mathcal{G}^{V}$ and $\mathcal{F}^{FV}$ reflects the fact that they depend on both magnitude and direction of the boost vector. Now given the transformation of $c^{\mathbf{P}}_{lm}$ under a cubic rotation of the boost vector, Eq. (\ref{clm-trans-P}), one can write $\mathcal{F}^{FV,\mathbf{P}'}$ as following
\begin{eqnarray}
\left[\mathcal{F}^{FV,\mathbf{P}'}\right]_{LM_L,L'M_L'}&=&\sum_{l,m}\sum_{m'=-l}^{l}\mathcal{D}^{(l)}_{mm'}(R)\frac{(4\pi)^{3/2}}{k^{*l}}c^{\mathbf{P}}_{lm'}(k^{*2})\int d\Omega~Y^*_{L,M_L}Y^*_{l,m}Y_{L',M_L'}
\nonumber\\
\nonumber\\
&=& \sum_{\bar{M}_L=-L}^{L}\sum_{\bar{M}_L'=-L'}^{L'}\mathcal{D}^{(L)}_{\bar{M}_LM_L}(R^{-1})\left[\mathcal{F}^{FV,\mathbf{P}}\right]_{L\bar{M}_L,L'\bar{M}_L'}
\mathcal{D}^{(L')}_{M_L'\bar{M}_L'}(R),
\label{F-trans}
\end{eqnarray}
where in the last equality we have used the fact that under rotation
\begin{eqnarray}
\sum_{M'=-L}^{L}\mathcal{D}^{(L)}_{MM'}(R)~Y_{LM'}(\hat{\mathbf{r}})=Y_{LM}(R\hat{\mathbf{r}}).
\label{deltaG}
\end{eqnarray}

Now one can obtain the relation between $\delta\mathcal{G}^{V,\mathbf{P}'}$ and $\delta\mathcal{G}^{V,\mathbf{P}}$ using Eqs. (\ref{deltaG-F}, \ref{F-trans}),
\begin{eqnarray}
\left[\delta\mathcal{G}^{V,\mathbf{P}'}\right]_{JM_J,L;J'M_J',L'}&=&\frac{iM}{4\pi}\times \left[k^*\delta_{JJ'}\delta_{M_JM_J'}\delta_{LL'} +i \sum_{M_L,M_L',M_S}\langle JM_J|LM_L,SM_S\rangle \langle L'M_L',SM_S|J'M_J'\rangle\right.
\nonumber\\
&& \qquad \qquad \qquad  \left. \times  \sum_{\bar{M}_L=-L}^{L}\sum_{\bar{M}_L'=-L'}^{L'}\mathcal{D}^{(L)}_{\bar{M}_LM_L}(R^{-1})\left[\mathcal{F}^{FV,\mathbf{P}}\right]_{L\bar{M}_L,L'\bar{M}_L'}
\mathcal{D}^{(L')}_{M_L'\bar{M}_L'}(R) \right],
\nonumber\\
\label{G-trans1}
\end{eqnarray}
where we have suppressed spin and isospin indices for the sake of compactness. Using the fact that
\begin{eqnarray}
\langle JM_J|LM_L,SM_S\rangle=\sum_{\widetilde{M}_J=-J}^{J}\sum_{\widetilde{M}_L=-L}^{L}\sum_{\widetilde{M}_S=-S}^{S}\mathcal{D}^{(J)}_{M_J\widetilde{M}_J}(R^{-1})\mathcal{D}^{(L)}_{M_L\widetilde{M}_L}(R)\mathcal{D}^{(S)}_{M_S\widetilde{M}_S}(R)
\langle J\widetilde{M}_J|L\widetilde{M}_L,S\widetilde{M}_S\rangle,
\nonumber\\
\label{CG-trans}
\end{eqnarray}
and given that Wigner $\mathcal{D}$-matrices are unitary, one easily arrives at
\begin{eqnarray}
\left[\delta\mathcal{G}^{V,\mathbf{P}'}\right]_{JM_J,L;J'M_J',L'}=\sum_{\bar{M}_J=-J}^{J}\sum_{\bar{M}_J'=-J'}^{J'}\mathcal{D}^{(J)}_{\bar{M}_JM_J}(R^{-1})\left[\delta\mathcal{G}^{V,\mathbf{P}}\right]_{J\bar{M}_J,L;J'\bar{M}_J',L'}\mathcal{D}^{(J')}_{M_J'\bar{M}_J'}(R),
\nonumber\\
\label{G-trans2}
\end{eqnarray}
or in the matrix notation, $\delta\mathcal{G}^{V,\mathbf{P}'}=\mathcal{D}^*(R)\delta\mathcal{G}^{V,\mathbf{P}}\mathcal{D}^T(R)$. Given that the scattering amplitude is diagonal in the $|J,M_J\rangle$ basis, and that the quantization condition Eq. (\ref{NNQC}) is a determinant condition, one obtains
\begin{eqnarray}
\det\left[{(\mathcal{M}^{\infty})^{-1}+\delta\mathcal{G}^{V,\mathbf{P}'}}\right]&=&\det\left[\mathcal{D}^*(R)\left({(\mathcal{M}^{\infty})^{-1}+\delta\mathcal{G}^{V,\mathbf{P}}}\right)\mathcal{D}^T(R)\right]
\nonumber\\
&=&\det\left[{(\mathcal{M}^{\infty})^{-1}+\delta\mathcal{G}^{V,\mathbf{P}}}\right]=0.
\label{QC-trans}
\end{eqnarray}
As one would expect, although the FV functions are in general different for different boosts with the same magnitude within a given $A_1$ irrep of the cubic group, the spectrum does not depend on the choice of the direction of the boost vector. As discussed in Sec. \ref{sec: Reduction}, in order to extract the scattering parameters of NN systems from the QCs presented in this paper, one needs to use the specific boost vectors that are studied in this paper. However, the value of energy eigenvalues can be taken from the LQCD calculations that are performed with any other boost vector that is a cubic rotation of the boost vectors presented here.

\section{Reduction Procedure for Positive Parity Isosinglet Channel with $\mathbf{P}=\mathbf{0}$ \label{app:red-example}}
\noindent Consider the NN system in the positive parity isosinglet channel where the ground state in the infinite volume is known to be a shallow bound state, the deuteron, whose wave-function is an admixture of both S-wave and D-wave. In order to obtain the phase shifts and mixing parameter in this channel from the energy eigenvalues of the two-nucleon system at rest from a LQCD calculation, one must first construct sources and sinks that transform according to a given irrep of the cubic group, e.g. $T_1$ when $\textbf{P}=0$. The extracted energies then needs to be put in the determinant condition for this channel in the corresponding irrep of the cubic group, Eq. (\ref{NNQC-irrep}), and subsequently solve for the scattering parameters. If one assumes the contributions from scattering channels with $J>4$ and $l\geq4$ to be negligible, the scattering amplitude matrix in the LHS of Eq. (\ref{NNQC-irrep}) can be written as
\begin{eqnarray}
\mathcal{M}^{\infty}_{(0,1)}=\left(\begin{array}{cccc}
\mathcal{M}_{1;0}^{(0,0;1)} & \mathcal{M}_{1;0}^{(0,2;1)} & 0 & 0\\
\mathcal{M}_{1;0}^{(2,0;1)} & \mathcal{M}_{1;0}^{(2,2;1)} & 0 & 0\\
0 & 0 & \mathcal{M}_{2;0}^{(2,2;1)} & 0\\
0 & 0 & 0 & \mathcal{M}_{3;0}^{(2,2;1)}
\end{array}\right),
\end{eqnarray}
where each element, $ \mathcal{M}_{J;I}^{(L,L';S)}$, is a diagonal $(2J+1)^2\times(2I+1)^2$-matrix. Note that this is an $18\times18$ matrix which is parametrized by two phase shifts and a mixing angle in the $J=1$ channel, and two D-wave phase shifts in the $J=2$ and $J=3$ channels. Although there is a mixing between D-wave and G-wave channels in the $J=3$ sector, due to the assumption of a negligible $G$-wave scattering, the scattering amplitude in this channel is truncated to the D-wave.

The elements of the FV matrix $\delta \mathcal{G}^V$ in the LHS of Eq. (\ref{NNQC-irrep}) for this channel can be evaluated from Eq. (\ref{deltaG}). The result reads
\begin{eqnarray}
\delta \mathcal{G}^V_{(0,1)}=\left(\begin{array}{cccc}
\delta\mathcal{G}{}_{1,1;0}^{V,(0,0;1)} & \delta\mathcal{G}_{1,1;0}^{V,(0,2;1)} & \delta\mathcal{G}_{12;0}^{V,(0,2;1)} & \delta\mathcal{G}_{1,3;0}^{V,(0,2;1)}\\
\\
\delta\mathcal{G}_{1,1;0}^{V,(2,0;1)} & \delta\mathcal{G}_{1,1;0}^{V,(2,2;1)} & \delta\mathcal{G}_{1,2;0}^{V,(2,2;1)} & \delta\mathcal{G}_{1,3;0}^{V,(2,2;1)}\\
\\
\delta\mathcal{G}_{2,1;0}^{V,(2,0;1)} & \delta\mathcal{G}_{2,1;0}^{V,(2,2;1)} & \delta\mathcal{G}_{2,2;0}^{V,(2,2;1)} & \delta\mathcal{G}_{2,3;0}^{V,(2,2;1)}\\
\\
\delta\mathcal{G}_{3,1;0}^{V,(2,0;1)} & \delta\mathcal{G}_{3,1;0}^{V,(2,2;1)} & \delta\mathcal{G}_{3,2;0}^{V,(2,2;1)} & \delta\mathcal{G}_{3,3;0}^{V,(2,2;1)}
\end{array}\right),
\end{eqnarray}
where each element still represents a matrix $\delta\mathcal{G}_{J,J';I}^{V,(L,L';S)}$
in the $|J,M_J\rangle$ basis and whose explicit forms are as following\footnote{We will drop the superscript $\mathbf{P}$ on the $c_{lm}$s in this example as they are evaluated for $\mathbf{P}=0$.}
\begin{eqnarray}
\delta\mathcal{G}_{1,1;0}^{V,(0,0;1)}&=&\delta\mathcal{G}_{1,1;0}^{V,(2,2;1)}=M(-c_{00}+\frac{i k^*}{4 \pi })~\mathbf{I}_3,
\end{eqnarray}
\begin{eqnarray}
 \delta\mathcal{G}_{1,3;0}^{V,(2,2;1)}&=&\left[\delta\mathcal{G}_{3,1;0}^{V,(2,2;1)}\right]^T=\frac{M}{k^{*4}}c_{40}\left(
\begin{array}{ccccccc}
 0 & 0 & -\frac{3  }{7} & 0 & 0 & 0 & -\frac{\sqrt{15}}{7} \\
 0 & 0 & 0 & \frac{2 \sqrt{6}}{7} & 0 & 0 & 0 \\
 -\frac{\sqrt{15}}{7} & 0 & 0 & 0 & -\frac{3}{7} & 0 & 0
\end{array}
\right),
\end{eqnarray}
\begin{eqnarray}
\delta \mathcal{G}^{V,(2,2;1)}_{(2,2;0)}&=& M(-c_{00}+\frac{i k^*}{4 \pi})~\mathbf{I}_5+\frac{M}{k^{*4}}c_{40}\left(\begin{array}{ccccc}
 \frac{2 }{21} & 0 & 0 & 0 & \frac{10 }{21} \\
 0 &-\frac{8 }{21} & 0 & 0 & 0 \\
 0 & 0 & \frac{4 }{7} & 0 & 0 \\
 0 & 0 & 0 &-\frac{8 }{21} & 0 \\
 \frac{10 }{21} & 0 & 0 & 0 & \frac{2 }{21}
\end{array}\right),
\end{eqnarray}
\begin{eqnarray}
\delta \mathcal{G}^{V,(2,2;1)}_{(2,3;0)}&=&\left[\delta \mathcal{G}^{V,(2,2;1)}_{(3,2;0)}\right]^T=\frac{M}{k^{*4}}c_{40}\left(
\begin{array}{ccccccc}
 0 & \frac{5 \sqrt{2}}{21} & 0 & 0 & 0 & \frac{5 \sqrt{2}}{21} & 0 \\
 0 & 0 & -\frac{5 \sqrt{5}}{21} & 0 & 0 & 0 & \frac{5}{7 \sqrt{3}} \\
 0 & 0 & 0 & 0 & 0 & 0 & 0 \\
 -\frac{5}{7 \sqrt{3}} & 0 & 0 & 0 & \frac{5 \sqrt{5}}{21} & 0 & 0 \\
 0 & -\frac{5 \sqrt{2}}{21} & 0 & 0 & 0 & -\frac{5 \sqrt{2}}{21} & 0
\end{array}
\right),
\end{eqnarray}
\begin{eqnarray}
\delta \mathcal{G}^{V,(2,2;1)}_{(3,3;0)}&=&M(-c_{00}+\frac{i k^*}{4 \pi})~\mathbf{I}_7-
\frac{M}{k^{*4}}c_{40}\left(\begin{array}{ccccccc}
 \frac{1}{7} & 0 & 0 & 0 & \frac{\sqrt{\frac{5}{3}}}{7} & 0 & 0 \\
 0 & -\frac{1}{3} & 0 & 0 & 0 & \frac{5}{21} & 0 \\
 0 & 0 & \frac{1}{21} & 0 & 0 & 0 & \frac{\sqrt{\frac{5}{3}}}{7} \\
 0 & 0 & 0 & \frac{2}{7} & 0 & 0 & 0 \\
 \frac{\sqrt{\frac{5}{3}}}{7} & 0 & 0 & 0 & \frac{1}{21} & 0 & 0 \\
 0 & \frac{5}{21} & 0 & 0 & 0 & -\frac{1}{3} & 0 \\
 0 & 0 & \frac{\sqrt{\frac{5}{3}}}{7} & 0 & 0 & 0 & \frac{1}{7}
\end{array}\right),
\nonumber\\
\end{eqnarray}
where $\mathbf{I}_n$ is the $n \times n$ identity matrix, and the rest of the blocks are zero. As is suggested in Ref. \cite{Luu:2011ep}, a unitary matrix, that can bring the $\delta \mathcal{G}^V$ matrix into a block-diagonalized form, can be found by diagonalizing the blocks that are located on the diagonal of the $\delta \mathcal{G}^V$ matrix, $\delta \mathcal{G}^{V,(L,L';1)}_{(J,J;0)}$. It turns out that when there are multiple occurrences of a given irrep in each angular momentum $J$ (see table \ref{irreps}), the procedure of block diagonalization becomes more cumbersome, and a systematic procedure must be taken which is based on the knowledge of the basis functions corresponding to each occurrence of any given irrep. Such basis functions for the irreps of the point groups considered in this paper are tabulated in table \ref{irreps}. These basis functions correspond to each occurrence of the irreps in the decomposition of the angular momentum states into the irreps of the $O$, $D_4$ and $D_2$ point groups up to $J=4$. For this channel, however, such a unitary matrix can be found easily based on the method described in Ref. \cite{Luu:2011ep}. One finds
\begin{eqnarray}
S=\left(\begin{array}{ccc}
S_{11} & 0 & 0\\
0 & S_{22} & 0\\
0 & 0 & S_{33}
\end{array}\right),
\end{eqnarray}
where the zero elements denote sub-blocks of appropriate dimension with all elements equal to zero, and the non-trivial blocks are the following matrices
\begin{eqnarray}
S_{11}=\mathbf{I}_6,
~S_{22}=\left(\begin{array}{ccccc}
0 & 0 & 0 & 1 & 0\\
0 & 1 & 0 & 0 & 0\\
0 & 0 & 1 & 0 & 0\\
-\frac{1}{\sqrt{2}} & 0 & 0 & 0 & \frac{1}{\sqrt{2}}\\
\frac{1}{\sqrt{2}} & 0 & 0 & 0 & \frac{1}{\sqrt{2}}
\end{array}\right), 
~S_{33}=\left(\begin{array}{ccccccc}
0 & 0 & \sqrt{\frac{3}{8}} & 0 & 0 & 0 & \sqrt{\frac{5}{8}}\\
\sqrt{\frac{5}{8}} & 0 & 0 & 0 & \sqrt{\frac{3}{8}} & 0 & 0\\
0 & 0 & 0 & 1 & 0 & 0 & 0\\
0 & 0 & -\sqrt{\frac{5}{8}} & 0 & 0 & 0 & \sqrt{\frac{3}{8}}\\
0 & \frac{1}{\sqrt{2}} & 0 & 0 & 0 & \frac{1}{\sqrt{2}} & 0\\
-\sqrt{\frac{3}{8}} & 0 & 0 & 0 & \sqrt{\frac{5}{8}} & 0 & 0\\
0 & -\frac{1}{\sqrt{2}} & 0 & 0 & 0 & \frac{1}{\sqrt{2}} & 0
\end{array}\right).
\nonumber\\
\end{eqnarray}
The resultant partially block-diagonalized matrix can then be obtained by,
\begin{eqnarray}
&& S[(\mathcal{M}^{\infty}_{(0,1)})^{-1}+\delta \mathcal{G}^{V}_{(0,1)}]S^T=
\nonumber\\
&& \qquad ~~~~ \left(\begin{array}{cccccccccccccccccc}
x_{1} & 0 & 0 & y_{1} & 0 & 0 & 0 & 0 & 0 & 0 & 0 & 0 & 0 & 0 & 0 & 0 & 0 & 0\\
0 & x_{1} & 0 & 0 & y_{1} & 0 & 0 & 0 & 0 & 0 & 0 & 0 & 0 & 0 & 0 & 0 & 0 & 0\\
0 & 0 & x_{1} & 0 & 0 & y_{1} & 0 & 0 & 0 & 0 & 0 & 0 & 0 & 0 & 0 & 0 & 0 & 0\\
y_{1} & 0 & 0 & x_{2} & 0 & 0 & 0 & 0 & 0 & 0 & 0 & -y_{2} & 0 & 0 & 0 & 0 & 0 & 0\\
0 & y_{1} & 0 & 0 & x_{2} & 0 & 0 & 0 & 0 & 0 & 0 & 0 & 0 & y_{2} & 0 & 0 & 0 & 0\\
0 & 0 & y_{1} & 0 & 0 & x_{2} & 0 & 0 & 0 & 0 & 0 & 0 & -y_{2} & 0 & 0 & 0 & 0 & 0\\
0 & 0 & 0 & 0 & 0 & 0 & x_{3} & 0 & 0 & 0 & 0 & 0 & 0 & 0 & 0 & 0 & y_{3} & 0\\
0 & 0 & 0 & 0 & 0 & 0 & 0 & x_{3} & 0 & 0 & 0 & 0 & 0 & 0 & y_{3} & 0 & 0 & 0\\
0 & 0 & 0 & 0 & 0 & 0 & 0 & 0 & x_{4} & 0 & 0 & 0 & 0 & 0 & 0 & 0 & 0 & 0\\
0 & 0 & 0 & 0 & 0 & 0 & 0 & 0 & 0 & x_{3} & 0 & 0 & 0 & 0 & 0 & -y_{3} & 0 & 0\\
0 & 0 & 0 & 0 & 0 & 0 & 0 & 0 & 0 & 0 & x_{4} & 0 & 0 & 0 & 0 & 0 & 0 & 0\\
0 & 0 & 0 & -y_{2} & 0 & 0 & 0 & 0 & 0 & 0 & 0 & x_{5} & 0 & 0 & 0 & 0 & 0 & 0\\
0 & 0 & 0 & 0 & 0 & -y_{2} & 0 & 0 & 0 & 0 & 0 & 0 & x_{5} & 0 & 0 & 0 & 0 & 0\\
0 & 0 & 0 & 0 & y_{2} & 0 & 0 & 0 & 0 & 0 & 0 & 0 & 0 & x_{5} & 0 & 0 & 0 & 0\\
0 & 0 & 0 & 0 & 0 & 0 & 0 & y_{3} & 0 & 0 & 0 & 0 & 0 & 0 & x_{6} & 0 & 0 & 0\\
0 & 0 & 0 & 0 & 0 & 0 & 0 & 0 & 0 & -y_{3} & 0 & 0 & 0 & 0 & 0 & x_{6} & 0 & 0\\
0 & 0 & 0 & 0 & 0 & 0 & y_{3} & 0 & 0 & 0 & 0 & 0 & 0 & 0 & 0 & 0 & x_{6} & 0\\
0 & 0 & 0 & 0 & 0 & 0 & 0 & 0 & 0 & 0 & 0 & 0 & 0 & 0 & 0 & 0 & 0 & x_{7}
\end{array}\right),
\label{BD-form}
\end{eqnarray}
where
\begin{eqnarray}
x_{1}&=&-M c_{00}+\frac{iMk^*}{4\pi}+\frac{\mathcal{M}_{1;0}^{(2,2;1)}}{\det(\mathcal{M}^{SD})},~
x_{2}=-M c_{00}+\frac{iMk^*}{4\pi}+\frac{\mathcal{M}_{1;0}^{(0,0;1)}}{\det(\mathcal{M}^{SD})},
\nonumber\\
x_{3}&=&-M c_{00}-\frac{8}{21}\frac{M}{k^{*4}}c_{40}+\frac{iMk^*}{4\pi}+\frac{1}{\mathcal{M}^{(22;1)}_{2;0}}, ~
x_{4}=-M c_{00}+\frac{4}{7}\frac{M}{k^{*4}}c_{40}+\frac{iMk^*}{4\pi}+\frac{1}{\mathcal{M}^{(22;1)}_{2;0}},
\nonumber\\
x_{5}&=&-M c_{00}-\frac{2}{7}\frac{M}{k^{*4}}c_{40}+\frac{iMk^*}{4\pi}+\frac{1}{\mathcal{M}^{(22;1)}_{3;0}},~
x_{6}=-M c_{00}+\frac{2}{21}\frac{M}{k^{*4}}c_{40}+\frac{iMk^*}{4\pi}+\frac{1}{\mathcal{M}^{(22;1)}_{3;0}},
\nonumber\\
x_{7}&=&-M c_{00}+\frac{4}{7}\frac{M}{k^{*4}}c_{40}+\frac{iMk^*}{4\pi}+\frac{1}{\mathcal{M}^{(22;1)}_{3;0}},
y_{1}=-\frac{\mathcal{M}_{1;0}^{(0,2;1)}}{\det(\mathcal{M}^{SD})},
y_{2}=\frac{2\sqrt{6}}{7}\frac{M}{k^{*4}}c_{40},
y_{3}=\frac{10\sqrt{2}}{21}\frac{M}{k^{*4}}c_{40},
\nonumber\\
\end{eqnarray}
and $\det(\mathcal{M}^{SD})$ in these relations denotes the determinant of the $J=1$ sub-block of the scattering amplitude, $\det(\mathcal{M}^{SD})=\mathcal{M}_{1;0}^{(0,0;1)} \mathcal{M}_{1;0}^{(2,2;1)}- (\mathcal{M}_{1;0}^{(0,2;1)})^2$. This matrix can now clearly be broken to 4 independent blocks corresponding to 4 irreps of the cubic group. The degeneracy of the diagonal elements of this matrix, as well as the coupling between different rows and columns, indicate which irrep of the cubic group each block corresponds to. According to table \ref{irreps}, the one-dimensional irrep $A_2$ only occurs in the decomposition of $J=3$ angular momentum. As is seen from Eq. (\ref{BD-form}), the element $x_7$ belongs to the $J=3$ sector and has a one-fold degeneracy. Also it does not mix with other angular momentum channels, therefore it must correspond to the $A_2$ irrep. So the one-dimensional QC corresponding to the $A_2$ irrep is
\begin{eqnarray}
A_2: ~ \frac{1}{\mathcal{M}^{(22;1)}_{3;0}}-M c_{00}+\frac{4}{7}\frac{M}{k^{*4}}c_{40}+\frac{iMk^*}{4\pi}=0.
\label{A2}
\end{eqnarray}

The QC corresponding to the two-dimensional irrep $E$ can be also deduced easily as it only has overlap with the $J=2$ channel. Clearly the element corresponding to this irrep is $x_4$ with two-fold degeneracy and the corresponding QC reads
\begin{eqnarray}
E: ~ \frac{1}{\mathcal{M}^{(22;1)}_{2;0}}-M c_{00}+\frac{4}{7}\frac{M}{k^{*4}}c_{40}+\frac{iMk^*}{4\pi}=0.
\label{E}
\end{eqnarray}

The three-dimensional irrep $T_2$ appears in the decomposition of both $J=2$ and $J=3$ angular momentum, and as is seen from Eq. (\ref{BD-form}) mixes the $x_3$, $x_6$ and $y_3$ elements through the following QC
\begin{eqnarray}
T_2: ~ \det\left(\begin{array}{cc}
\frac{1}{\mathcal{M}^{(22;1)}_{2;0}}-M c_{00}-\frac{8}{21}\frac{M}{k^{*4}}c_{40}+\frac{iMk^*}{4\pi} & \frac{10\sqrt{2}}{21}\frac{M}{k^{*4}}c_{40} \\
\frac{10\sqrt{2}}{21}\frac{M}{k^{*4}}c_{40} & \frac{1}{\mathcal{M}^{(22;1)}_{3;0}}-M c_{00}+\frac{2}{21}\frac{M}{k^{*4}}c_{40}+\frac{iMk^*}{4\pi}
\end{array}\right)=0.
\nonumber\\
\label{T2}
\end{eqnarray}
As is clear, the energy eigenvalues in this irrep have a three-fold degeneracy (there are three copies of this QCs) that is consistent with the dimensionality of the irrep. The remaining irrep is $T_1$ which is a three-dimensional irrep and contribute to both $J=1$ and $J=3$ channels. As there are two $J=1$ sectors corresponding to S-wave and D-wave scatterings, the QC must be the determinant of a $3 \times 3$ matrix. This is in fact the case by looking closely at the partially block-diagonalized matrix in Eq. (\ref{BD-form}). One finds explicitly
\begin{eqnarray}
T_1: &~& \det\left(\begin{array}{ccc}
\frac{\mathcal{M}_{1;0}^{(2,2;1)}}{\det(\mathcal{M}^{SD})}-M c_{00}+\frac{iMk^*}{4\pi} & -\frac{\mathcal{M}_{1;0}^{(0,2;1)}}{\det(\mathcal{M}^{SD})} & 0\\
-\frac{\mathcal{M}_{1;0}^{(0,2;1)}}{\det(\mathcal{M}^{SD})} & \frac{\mathcal{M}_{1;0}^{(0,0;1)}}{\det(\mathcal{M}^{SD})}-M c_{00}+\frac{iMk^*}{4\pi} & -\frac{2\sqrt{6}}{7}\frac{M}{k^{*4}}c_{40}\\
0 & -\frac{2\sqrt{6}}{7}\frac{M}{k^{*4}}c_{40} & \frac{1}{\mathcal{M}^{(22;1)}_{3;0}}-M c_{00}-\frac{2}{7}\frac{M}{k^{*4}}c_{40}+\frac{iMk^*}{4\pi}
\end{array}\right)
\nonumber\\
&& \qquad \qquad \qquad \qquad \qquad \qquad
\qquad \qquad \qquad \qquad \qquad \qquad 
\qquad \qquad \qquad \qquad \qquad \qquad \qquad ~ =0.
\nonumber\\
\label{T1}
\end{eqnarray}
Again there is a three-fold degeneracy for the energy-eigenvalues as there are three copies of this QC for this irrep. This is an important QC as it gives access to the mixing angle between S- and D-partial-waves. Note that the QC for $A_2$ irrep, Eq. (\ref{A2}), by its own determines the phase-shift in the $J=3$ channel, which can then be used in Eq. (\ref{T1}) for the $T_1$ irrep to determine the phase-shifts and mixing angle in the $J=1$ channel. Eq. (\ref{E}) for the $E$ irrep gives access to the phase shift in the $J=2$ channel, and finally Eq. (\ref{T2}) provides another relation for the phase-shifts in the $J=2$ and $J=3$ channels. In practice, one needs multiple energy levels in order to be able to reliably extract these parameters from the QCs presented. This is specially a challenging task when it comes to the determination of the scattering parameters in the channels with physical mixing, e. g. S-D mixing, since there are at least three unknown parameters to be determined from the QC, e. g. see Eq. (\ref{T1}). By doing the LQCD calculations of the boosted two-nucleon system, one will attain more energy levels that will correspond to another set of QCs. These QCs then provide a set of equations that the same scattering parameters satisfy, and therefore better constraints can be put on these quantities. Without going into the detail of the reduction procedure that leads to such QCs for boosted systems, as well as QCs for the other three spin-isospin channels, we tabulate these QCs in the next appendix. Before presenting the rest of QCs though, let us show that the QCs are all real conditions.

\begin{center}
\begin{table}
\label{tab:param2}
\begin{tabular}{|c|c|c|cc|}
\hline
&$O$&${D}_{4}$&${D}_{2}$&\\
$\hspace{.3cm}J\hspace{.3cm}$&
$\hspace{.3cm}\Gamma:~$basis functions$\hspace{.3cm}$&
$\hspace{.3cm}\Gamma:~$basis functions$\hspace{.3cm}$&
$\hspace{.3cm}\Gamma:~$basis functions$\hspace{.3cm}$&
 \\\hline \hline
0&${A}_1:Y_{00}$&${A}_1:Y_{00}$&${A}:Y_{00}$& 
\\\hline
1&${T}_1:(Y_{11},Y_{10},Y_{1-1})$&${A}_2:Y_{10}$&${B}_1:Y_{10}$& \\
& &$E:\left(\overline{\mathcal{Y}}_{11},\widetilde{\mathcal{Y}}_{11}\right)$&${B}_2:\overline{\mathcal{Y}}_{11},~{B}_3:\widetilde{\mathcal{Y}}_{11}$& 
\\\hline
2&$E:\left(Y_{20},\overline{\mathcal{Y}}_{22}\right)$&
$A_1:Y_{20}$&$A:Y_{20}$&
\\
&$T_2:\left({\widetilde{\mathcal{Y}}_{22}},{\overline{\mathcal{Y}}_{21}},{\widetilde{\mathcal{Y}}_{21}}\right)$
&$E:\left(\overline{\mathcal{Y}}_{21},\widetilde{\mathcal{Y}}_{21}\right)$
&$B_3:\overline{\mathcal{Y}}_{21},~B_2:\widetilde{\mathcal{Y}}_{21}$&
\\
&&$B_1:\overline{\mathcal{Y}}_{22},~B_2:\widetilde{\mathcal{Y}}_{22}$
&$A:\overline{\mathcal{Y}}_{22},~B_1:\widetilde{\mathcal{Y}}_{22}$
&\\
\hline
3&$A_2:~\widetilde{\mathcal{Y}}_{32}$
&$A_2:~{{Y}}_{30}$
&$B_1:~{{Y}}_{30}$
&
\\
&$T_1:~\left(Y_{30},\widetilde{\mathcal{Y}}_{33}-\frac{3}{\sqrt{15}}\widetilde{\mathcal{Y}}_{31},
\overline{\mathcal{Y}}_{33}+\frac{3}{\sqrt{15}}\overline{\mathcal{Y}}_{31}\right)$
&$E:~\left(\widetilde{\mathcal{Y}}_{31},\overline{\mathcal{Y}}_{31}\right)$
&$B_2:~\overline{\mathcal{Y}}_{31},~B_3:~\widetilde{\mathcal{Y}}_{31}$
&\\
&$T_2:~\left(\overline{\mathcal{Y}}_{32},
\overline{\mathcal{Y}}_{31}-\sqrt{\frac{3}{{5}}}\overline{\mathcal{Y}}_{31},
\widetilde{\mathcal{Y}}_{31}+\sqrt{\frac{3}{{5}}}\widetilde{\mathcal{Y}}_{31}\right)$
&$B_2:~\overline{\mathcal{Y}}_{32},~B_1:~\widetilde{\mathcal{Y}}_{32}$
&$B_1:~\overline{\mathcal{Y}}_{32},~A:~\widetilde{\mathcal{Y}}_{32}$
&\\
&
&$E:~\left(\widetilde{\mathcal{Y}}_{33},\overline{\mathcal{Y}}_{33}\right)$&
$B_2:~\overline{\mathcal{Y}}_{33},~B_3:~\widetilde{\mathcal{Y}}_{33}$&
\\\hline
4&
$A_1:~Y_{40}+\sqrt{\frac{5}{14}}\overline{\mathcal{Y}}_{44}$
&
$A_1:~Y_{40}$
&
$A:~Y_{40}$&\\
&$E:\left(\overline{\mathcal{Y}}_{42},
Y_{40}-\sqrt{\frac{7}{10}}\overline{\mathcal{Y}}_{44}
\right)$&
$E:~\left(\overline{\mathcal{Y}}_{41},\widetilde{\mathcal{Y}}_{41}\right)$&
$B_3:~\overline{\mathcal{Y}}_{41},~B_2:~\widetilde{\mathcal{Y}}_{41}$&
\\
&$T_1:~\left(\widetilde{\mathcal{Y}}_{44},
\overline{\mathcal{Y}}_{43}+\sqrt{7}\overline{\mathcal{Y}}_{41},
\widetilde{\mathcal{Y}}_{43}+\sqrt{{7}}\widetilde{\mathcal{Y}}_{41}
\right)$&
$B_1:~\overline{\mathcal{Y}}_{42},~B_2:~\widetilde{\mathcal{Y}}_{42}$&
$A:~\overline{\mathcal{Y}}_{42},~B_1:~\widetilde{\mathcal{Y}}_{42}$&
\\
&$T_2:~\left(\widetilde{\mathcal{Y}}_{42},
\overline{\mathcal{Y}}_{43}-\sqrt{\frac{1}{7}}\overline{\mathcal{Y}}_{41},
\widetilde{\mathcal{Y}}_{43}+\sqrt{\frac{1}{7}}\widetilde{\mathcal{Y}}_{41}
\right)$&$E:~\left(\overline{\mathcal{Y}}_{43},\widetilde{\mathcal{Y}}_{43}\right)$&
$B_3:~\overline{\mathcal{Y}}_{43},~B_2:~\widetilde{\mathcal{Y}}_{43}$&\\
&&$A_1:~\overline{\mathcal{Y}}_{44},~A_2:~\widetilde{\mathcal{Y}}_{44}$
&$A:~\overline{\mathcal{Y}}_{44},~B_1:~\widetilde{\mathcal{Y}}_{44}$&
\\
\hline
\end{tabular}
\caption{The decomposition of the irreps of the rotational group up to $J=4$ in terms of the irreps of the cubic ($O$), tetragonal ($D_4$) and orthorhombic ($D_2$) groups, see Refs. \cite{luscher2, Feng:2004ua, Dresselhaus}. The corresponding basis functions  of each irrep are also given in the table, where $\overline{\mathcal{Y}}_{lm}\equiv{Y}_{lm}+{Y}_{l-m}$ and $\widetilde{\mathcal{Y}}_{lm}\equiv{Y}_{lm}-{Y}_{l-m}$. These basis functions become useful in reduction of the full determinant condition, Eq. (\ref{NNQC}) into separate QCs corresponding to each irrep of the point group considered, see Sec. \ref{sec: Reduction}.
}
\label{irreps}
\end{table}
\end{center}

As all the $c_{lm}^{\mathbf{P}}$ functions are real, the only imaginary part of the FV matrix $\delta \mathcal{G}^V$ shows up in the diagonal elements of this matrix.\footnote{This is not always the case as for example, the $\delta \mathcal{G}^V$ matrix for the $\mathbf{d}=(1,1,0)$ boost contains off-diagonal complex elements as well. For all of those case, it can be checked that although the elements of the matrix $(\mathcal{M}^{\infty})^{-1}+\delta\mathcal{G}^{V}$ are complex, the determinant of the matrix remains real, see QCs in appendix \ref{app:QC}.} For the angular momentum channels $J$ where there is no coupling between different partial-waves, the inverse scattering amplitude matrix has only diagonal elements, whose imaginary part exactly cancels that of the $\delta \mathcal{G}^V$ matrix, see Eq. (\ref{M-single}). Explicitly,
\begin{eqnarray}
\Im [(\mathcal{M}^{LL;S}_{JM_J;IM_I})^{-1}+\delta \mathcal{G}^{V,(LL;S)}_{JM_J,JM_J;IM_I}]=-\frac{iMk^*}{4\pi}+\frac{iMk^*}{4\pi}=0.
\label{Im-diagonal}
\end{eqnarray}
For the angular momentum channels where there are off-diagonal terms due to the partial-wave mixing, one can still write the inverse of the scattering amplitude in that sector, Eqs. (\ref{M-coupled1}, \ref{M-coupled2}), as following
\begin{eqnarray}
(\mathcal{M}^{LL';1})^{-1}&=&
\left( \begin{array}{cc}
-\frac{Mk^*}{4\pi}
\frac{\cos{2\epsilon}~\sin({\delta '-\delta})
+\sin({\delta '+\delta})}{\cos({\delta '+\delta})-\cos({\delta '-\delta})\cos({2\epsilon})}-\frac{iMk^*}{4\pi}&
\frac{Mk^*}{2\pi}
\frac{\cos(\epsilon)\sin(\epsilon)}{\cos(\delta '+\delta)-\cos(\delta '-\delta)\cos(2\epsilon)}
\\
\frac{Mk^*}{2\pi}
\frac{\cos({\epsilon})\sin({\epsilon})}{\cos({\delta '+\delta})-\cos({\delta '-\delta})\cos({2\epsilon})}
&
-\frac{Mk^*}{4\pi}
\frac{\cos(2\epsilon)~\sin({\delta-\delta '})
+\sin(\delta '+\delta)}{\cos(\delta '+\delta)-\cos(\delta '-\delta)\cos(2\epsilon)}-\frac{iMk^*}{4\pi}
\\
\end{array} \right),
\nonumber\\
\label{Minverse-coupled}
\end{eqnarray}
where $L=J \pm 1$ ($L'=J \mp 1$) and $\delta$ ($\delta '$) denotes the phase shift corresponding to the $L$ ($L'$) partial-wave. As is seen the off-diagonal elements of this matrix are real. Given that the FV function $\delta \mathcal{G}^V$ has real off-diagonal terms, these terms in the QC lead to a real off-diagonal element. For the diagonal elements, the imaginary part of the inverse scattering amplitude is isolated and has the same form as the imaginary part of the $\delta \mathcal{G}^V$ matrix, so a similar cancellation as that given in Eq. (\ref{Im-diagonal}) occurs in this case as well.

\newpage

\section{List of Quantization Conditions~\footnote{The mathematica notebook that contains all the quantization conditions presented in this section is also available online through \cite{BDLsupp}.}}
\label{app:QC}
In order to make the presentation of the QCs clear and compact, we will introduce a simpler notation in this section as following. Let us introduce a new FV function $\mathcal{F}^{(\Gamma),{\textbf{P}}}$ that is projected to a particular irrep of the point group of the problem, $\Gamma_i$,
\begin{eqnarray}
\mathcal{F}^{(\Gamma_i),{\textbf{P}}}(k^{*2};L)& \equiv &-\left[\delta\mathcal{G}^{V}(k^{*2};L)-\frac{iM k^*}{4\pi}\right]_{\Gamma_i}
\nonumber\\
&=&{M}\sum_{l,m}\frac{1}{k^{*l}}~{\mathbb{F}
}_{lm}^{(\Gamma_i),{\textbf{P}}}~{c_{lm}^{\textbf{P}}(k^{*2};L)},
\label{def-F}
\end{eqnarray}
where the volume dependence of the FV functions has been made explicit, while the reference to each $(I,S)$ channel is implicit. In this form, all the detail of the corresponding projected FV functions are embedded in purely numerical matrices, ${\mathbb{F}
}_{lm}^{(\Gamma_i),{\textbf{P}}}$. Similarly, the projection of the inverse of the scattering amplitude in each spin-isospin channel unto a particular irreducible representation, ${\mathbb{M}}^{(\Gamma_i)}$, is defined as
\begin{eqnarray}
{\mathbb{M}}_{(I,S)}^{(\Gamma_i)}\equiv\left(\mathcal{M}^{\infty-1}_{(I,S)}\right)_{\Gamma_i}.
\label{def-M}
\end{eqnarray}
With this notation, the quantization condition for the irreducible representation $\Gamma_i$ can be simply written as
\begin{eqnarray}
\det\left({\mathbb{M}}_{(I,S)}^{(\Gamma_i)}+\frac{iMk^{*}}{4\pi }-\mathcal{F}^{(\Gamma_i),{\textbf{P}}}_{(I,S)}\right)=0.
\label{QC-simplified}
\end{eqnarray}
Since we aim to present the QCs for each spin-isospin channel in separate subsections, the $(I,S)$ subscripts can be dropped in the following presentation. Although the $(I,S)$ index of the scattering amplitudes is assumed implicitly, one should keep the $(J,L)$ quantum numbers of the elements of the scattering amplitude matrix explicit. In order to simplify the notation, however, the diagonal elements of the scattering amplitude in the $L$-basis in each spin-isospin channel will be denoted by
\begin{eqnarray}
\mathcal{M}_{J,L} \equiv \mathcal{M}^{(LL;S)}_{JM_J;IM_{I}},
\end{eqnarray}
while the off-diagonal elements will be defined as
\begin{eqnarray}
\mathcal{M}_{J,LL'} \equiv \mathcal{M}^{(LL';S)}_{JM_J;IM_{I}}.
\end{eqnarray}
The determinant of the $2\times2$ sub-sector that presents the mixing between partial-waves in the $J$ sector is denoted by $\det \mathcal{M}_J$. Explicitly,
\begin{eqnarray}
\det\mathcal{M}_{J}=\det \left( \begin{array}{cc}
\mathcal{M}_{J,L}&\mathcal{M}_{J,LL'}\\
\mathcal{M}_{J,L'L}&\mathcal{M}_{J,L}\\
\end{array} \right)\delta_{L,J-1}\delta_{L',J+1}.
\end{eqnarray}
Instead of using numerical values for the partial-wave $L$, we have used the  conventional spectroscopic notations for $L=0,1,2,3$ as S, P, D and F waves, respectively.

\subsection{Positive parity isosinglet channel}
The scattering amplitude matrix in this channel, after truncating the scatterings at $J=4$ and $L=3$, reads
\begin{eqnarray}
\mathcal{M}_{(0,1)}^{\infty}=\left(\begin{array}{cccc}
\mathcal{M}_{1,S} & \mathcal{M}_{1,SD} & 0 & 0\\
\mathcal{M}_{1,DS} & \mathcal{M}_{1;D} & 0 & 0\\
0 & 0 & \mathcal{M}_{2,D} & 0\\
0 & 0 & 0 & \mathcal{M}_{3,D}
\end{array}\right).
\end{eqnarray}
As is clear, each element is still a $(2J+1)^2$ matrix due to the $M_J$ quantum number. As a result, the truncated scattering amplitude is a $16\times16$ matrix that will be used in the master QC for this spin-isospin channel. In the following, the elements of matrices $\mathbb{F}$ and $\mathbb{M}$ as defined above, Eqs. (\ref{def-F}, \ref{def-M}), will be given for this channel for different irreps of the cubic, tetragonal and orthorhombic point groups.
\subsubsection{$\mathbf{d}=(0,0,0)$}
\begin{align}
& E: \hspace{.5cm}
\mathbb{F}_{00}^{(E)}=1, \hspace{.5cm}
\mathbb{F}_{40}^{(E)}=-4/7,\hspace{.5cm}
{\mathbb{M}}^{(E)}=\mathcal{M}_{2,D}^{-1}. \label{I000E}
\\
& A_2: \hspace{.5cm}
\mathbb{F}_{00}^{(A_2)}= 1, \hspace{.5cm}
\mathbb{F}_{40}^{(A_2)}= -4/7, \hspace{.5cm}
{\mathbb{M}}^{(A_2)}=
\mathcal{M}_{3,D}^{-1}. \label{I000A2}
\\
& T_1: \hspace{.5cm}
\mathbb{F}_{00}^{(T_1)}=\textbf{I}_{3},\hspace{.5cm}
\mathbb{F}_{40}^{(T_1)}=
\left(
\begin{array}{ccc}
 0 & 0 & 0 \\
 0 & 0 & \frac{2 \sqrt{6}}{7} \\
 0 & \frac{2 \sqrt{6}}{7} & \frac{2}{7} \\
\end{array}
\right),\hspace{.5cm}
{\mathbb{M}}^{(T_1)}=\left(
\begin{array}{ccc}
 \frac{\mathcal{M}_{1,D}}{{\det\mathcal{M}_{1}}} & -\frac{\mathcal{M}_{1,SD}}{\det\mathcal{M}_{1}} & 0 \\
 -\frac{\mathcal{M}_{1,SD}}{\det\mathcal{M}_{1}} & \frac{\mathcal{M}_{1,S}}{\det\mathcal{M}_{1}} & 0 \\
 0 & 0 &\mathcal{M}_{3,D}^{-1} \\
\end{array}
\right). \label{I000T1}
\\
& T_2: \hspace{.5cm}
\mathbb{F}_{00}^{(T_2)}=\textbf{I}_{2},\hspace{.5cm}
\mathbb{F}_{40}^{(T_2)}=\left(
\begin{array}{cc}
 \frac{8}{21} & -\frac{10 \sqrt{2}}{21} \\
 -\frac{10 \sqrt{2}}{21} & -\frac{2}{21} \\
\end{array}
\right),\hspace{.5cm}
{\mathbb{M}}^{(T_2)}=
\left(
\begin{array}{ccc}
\mathcal{M}_{2,D}^{-1}&0  \\
0 &\mathcal{M}_{3,D}^{-1} \\ 
\end{array}
\right). \label{I000T2}
\end{align}
%
\subsubsection{$\mathbf{d}=(0,0,1)$}
\begin{align}
& A_1:\hspace{.5cm}\mathbb{F}_{00}^{(A_1)}= 1,\hspace{.5cm}
\mathbb{F}_{20}^{(A_1)}=\frac{\sqrt{5}}{7},\hspace{.5cm}
\mathbb{F}_{40}^{(A_1)}= -4/7,\hspace{.5cm}
{\mathbb{M}}^{(A_1)}=
\mathcal{M}_{2,D}^{-1}.
\label{I001A1}
\\
& B_1:\hspace{.5cm}
\mathbb{F}_{00}^{(B_1)}=\textbf{I}_{2},\hspace{.5cm}
\mathbb{F}_{20}^{(B_1)}=\left(
\begin{array}{cc}
 -\frac{\sqrt{5}}{7} & -\frac{\sqrt{10}}{7} \\
 -\frac{\sqrt{10}}{7} & 0 \\
\end{array}
\right),\hspace{.5cm}
\mathbb{F}_{40}^{(B_1)}=
\left(
\begin{array}{cc}
 -\frac{2}{21} & \frac{5 \sqrt{2}}{21} \\
 \frac{5 \sqrt{2}}{21} & -\frac{1}{3} \\
\end{array}
\right),
\nonumber\\
&\hspace{1.2cm}
\mathbb{F}_{44}^{(B_1)}=\frac{1}{3}
\left(
\begin{array}{cc}
 - 2 \sqrt{\frac{10}{7}} & -2 \sqrt{\frac{5}{7}}  \\
 -2 \sqrt{\frac{5}{7}} & -{\sqrt{\frac{10}{7}}} \\
\end{array}
\right),\hspace{.5cm}
{\mathbb{M}}^{(B_1)}=
\left(
\begin{array}{ccc}
\mathcal{M}_{2,D}^{-1}&0  \\
0 &\mathcal{M}_{3,D}^{-1} \\ 
\end{array}
\right).
\label{I001B1}
\\
& B_2:\hspace{.5cm}
\mathbb{F}_{00}^{(B_2)}=\textbf{I}_{2},\hspace{.5cm}
\mathbb{F}_{20}^{(B_2)}=\left(
\begin{array}{cc}
 -\frac{\sqrt{5}}{7} & -\frac{\sqrt{10}}{7} \\
 -\frac{\sqrt{10}}{7} & 0 \\
\end{array}
\right),\hspace{.5cm}
\mathbb{F}_{40}^{(B_2)}=
\left(
\begin{array}{cc}
 -\frac{2}{21} & \frac{5 \sqrt{2}}{21} \\
 \frac{5 \sqrt{2}}{21} & -\frac{1}{3} \\
\end{array}
\right),
\nonumber\\
&\hspace{1.25cm}
\mathbb{F}_{44}^{(B_2)}=\frac{1}{3}
\left(
\begin{array}{cc}
  2 \sqrt{\frac{10}{7}} & 2 \sqrt{\frac{5}{7}}  \\
 2 \sqrt{\frac{5}{7}} & {\sqrt{\frac{10}{7}}} \\
\end{array}
\right),\hspace{.5cm}
{\mathbb{M}}^{(B_2)}=
\left(
\begin{array}{ccc}
\mathcal{M}_{2,D}^{-1}&0  \\
0 &\mathcal{M}_{3,D}^{-1} \\ 
\end{array}
\right).
\label{I001B2}
\end{align}
\begin{align}
& A_2:\hspace{.5cm}
\mathbb{F}_{00}^{(A_2)}=\textbf{I}_{3},\hspace{.5cm}
\mathbb{F}_{20}^{(A_2)}=
\left(
\begin{array}{ccc}
 \frac{2}{\sqrt{5}} & 0 & -\frac{9}{7 \sqrt{5}} \\
 0 & -\frac{1}{\sqrt{5}} & \frac{6 }{7}\sqrt{\frac{2}{5}} \\
 -\frac{9}{7 \sqrt{5}} & \frac{6 }{7}\sqrt{\frac{2}{5}} & \frac{8}{7 \sqrt{5}} \\
\end{array}
\right),\hspace{.5cm}
\mathbb{F}_{40}^{(A_2)}=\left(
\begin{array}{ccc}
 0 & 0 & -\frac{4}{7} \\
 0 & 0 & -\frac{2 \sqrt{2}}{7}  \\
 -\frac{4}{7} & -\frac{2 \sqrt{2}}{7} & \frac{2}{7} \\
\end{array}
\right),
\nonumber\\
&\hspace{1cm}
{\mathbb{M}}^{(A_2)}= \left(
\begin{array}{ccc}
 \frac{2 \mathcal{M}_{1,S}+2 \sqrt{2}\mathcal{M}_{1,SD}+\mathcal{M}_{1,D}}{3 \det\mathcal{M}_{1}} & \frac{\sqrt{2} \mathcal{M}_{1,S}-\mathcal{M}_{1,SD}-\sqrt{2}\mathcal{M}_{1,D}}{3 \det\mathcal{M}_{1}} & 0 \\
 \frac{\sqrt{2} \mathcal{M}_{1,S}-\mathcal{M}_{1,SD}-\sqrt{2}\mathcal{M}_{1,D}}{3 \det\mathcal{M}_{1}} & \frac{\mathcal{M}_{1,S}-2 \sqrt{2}\mathcal{M}_{1,SD}+2\mathcal{M}_{1,D}}{3 \det\mathcal{M}_{1}} & 0 \\
 0 & 0 & \mathcal{M}_{3,D}^{-1}  \\
\end{array}
\right).
\label{I001A2}
\\
& E:\hspace{.5cm}\mathbb{F}_{00}^{(E)}=\textbf{I}_{4},\hspace{.5cm}
\mathbb{F}_{20}^{(E)}=
\left(
\begin{array}{ccccc}
 \frac{1}{2 \sqrt{5}} & 0 & -\frac{\sqrt{3}}{2} & 0 & \frac{4 \sqrt{\frac{3}{5}}}{7} \\
 0 & -\frac{1}{\sqrt{5}} & 0 & 0 & -\frac{3}{7}   \sqrt{\frac{6}{5}} \\
 -\frac{\sqrt{3}}{2} & 0 & \frac{\sqrt{5}}{14} & 0 & -\frac{2}{7} \\
 0 & 0 & 0 & -\frac{2 }{7} \sqrt{5}  & 0 \\
 \frac{4 \sqrt{\frac{3}{5}}}{7} & -\frac{3}{7}   \sqrt{\frac{6}{5}} & -\frac{2}{7} & 0 & \frac{6}{7 \sqrt{5}} \\
\end{array}
\right),
\nonumber\\
&\hspace{1cm}\mathbb{F}_{40}^{(E)}=\left(
\begin{array}{ccccc}
 0 & 0 & 0 & 0 & \frac{\sqrt{3}}{7} \\
 0 & 0 & 0 & 0 & \frac{\sqrt{6}}{7} \\
 0 & 0 & \frac{8}{21} & 0 & -\frac{5 \sqrt{5}}{21} \\
 0 & 0 & 0 & \frac{1}{7} & 0 \\
 \frac{\sqrt{3}}{7} & \frac{\sqrt{6}}{7} & -\frac{5 \sqrt{5}}{21} & 0 & \frac{1}{21}
\end{array}
\right),
\hspace{.5cm}
\mathbb{F}_{44}^{(E)}=
\left(
\begin{array}{ccccc}
 0 & 0 & 0 & \sqrt{\frac{2}{7}} & 0 \\
 0 & 0 & 0 & \frac{2}{\sqrt{7}} & 0 \\
 0 & 0 & 0 & \sqrt{\frac{10}{21}} & 0 \\
 \sqrt{\frac{2}{7}} & \frac{2}{\sqrt{7}} & \sqrt{\frac{10}{21}} & 0 & \sqrt{\frac{2}{21}} \\
 0 & 0 & 0 & \sqrt{\frac{2}{21}} & 0
\end{array}
\right),
\nonumber\\
&\hspace{1cm}{\mathbb{M}}^{(E)}=
 \left(
\begin{array}{ccccc}
 \frac{\mathcal{M}_{1,S}-2 \sqrt{2}\mathcal{M}_{1,SD}+2\mathcal{M}_{1,D}}{3 \det\mathcal{M}_{1}} & \frac{\sqrt{2} \mathcal{M}_{1,S}-\mathcal{M}_{1,SD}-\sqrt{2}\mathcal{M}_{1,D}}{3 \det\mathcal{M}_{1}} & 0 & 0 & 0 \\
 \frac{\sqrt{2} \mathcal{M}_{1,S}-\mathcal{M}_{1,SD}-\sqrt{2}\mathcal{M}_{1,D}}{3 \det\mathcal{M}_{1}} & \frac{2 \mathcal{M}_{1,S}+2 \sqrt{2}\mathcal{M}_{1,SD}+\mathcal{M}_{1,D}}{3 \det\mathcal{M}_{1}} & 0 & 0 & 0 \\
 0 & 0 & \mathcal{M}_{2,D}^{-1} & 0 & 0 \\
 0 & 0 & 0 & \mathcal{M}_{3,D}^{-1} & 0 \\
 0 & 0 & 0 & 0 & \mathcal{M}_{3,D}^{-1} \\
\end{array}
\right).
\label{I001E}
\end{align}
%
\subsubsection{$\mathbf{d}=(1,1,0)$}
\begin{align}
& B_1: \hspace{.5cm}
\mathbb{F}_{00}^{(B_1)}=\textbf{I}_{5},\hspace{.5cm}
\mathbb{F}_{20}^{(B_1)}=\left(
\begin{array}{ccccc}
 \frac{2}{\sqrt{5}} & 0 & 0 & -\frac{9}{7 \sqrt{5}} & 0 \\
 0 & -\frac{1}{\sqrt{5}} & 0 & \frac{6}{7}\sqrt{\frac{2}{5}} & 0 \\
 0 & 0 & -\frac{\sqrt{5}}{7} & 0 & -\frac{\sqrt{10}}{7} \\
 -\frac{9}{7 \sqrt{5}} & \frac{6}{7}\sqrt{\frac{2}{5}} & 0 & \frac{8}{7 \sqrt{5}} & 0 \\
 0 & 0 & -\frac{\sqrt{10}}{7} & 0 & 0 \\
\end{array}
\right),
\nonumber\\
&\hspace{1.2cm}
\mathbb{F}_{22}^{(B_1)}=\left(
\begin{array}{ccccc}
 0 & 0 & 0 & 0 & -\frac{3 \sqrt{2}}{7}   \\
 0 & 0 & \sqrt{2} & 0 & \frac{4}{7} \\
 0 & -\sqrt{2} & 0 & \frac{2}{7} & 0 \\
 0 & 0 & -\frac{2}{7} & 0 & -\frac{4 \sqrt{2}}{7}   \\
 \frac{3 \sqrt{2}}{7} & -\frac{4}{7} & 0 & \frac{4 \sqrt{2}}{7} & 0 \\
\end{array}
\right)
,\hspace{.5cm}
\mathbb{F}_{40}^{(B_1)}=\left(
\begin{array}{ccccc}
 0 & 0 & 0 & -\frac{4}{7} & 0 \\
 0 & 0 & 0 & -\frac{2 \sqrt{2}}{7}   & 0 \\
 0 & 0 & -\frac{2}{21} & 0 & \frac{5 \sqrt{2}}{21} \\
 -\frac{4}{7} & -\frac{2 \sqrt{2}}{7}   & 0 & \frac{2}{7} & 0 \\
 0 & 0 & \frac{5 \sqrt{2}}{21} & 0 & -\frac{1}{3} \\
\end{array}
\right),
\nonumber
\end{align}
\begin{align}
&\hspace{1.2cm}\mathbb{F}_{42}^{(B_1)}=\left(
\begin{array}{ccccc}
 0 & 0 & 0 & 0 & -\frac{2 \sqrt{6}}{7}   \\
 0 & 0 & 0 & 0 & -\frac{2 \sqrt{3}}{7}   \\
 0 & 0 & 0 & -\frac{10}{7 \sqrt{3}} & 0 \\
 0 & 0 & \frac{10}{7 \sqrt{3}} & 0 & -\frac{\sqrt{\frac{2}{3}}}{7} \\
 \frac{2 \sqrt{6}}{7} & \frac{2 \sqrt{3}}{7} & 0 & \frac{\sqrt{\frac{2}{3}}}{7} & 0 \\
\end{array}
\right),\hspace{.5cm}
\mathbb{F}_{44}^{(B_1)}=\left(
\begin{array}{ccccc}
 0 & 0 & 0 & 0 & 0 \\
 0 & 0 & 0 & 0 & 0 \\
 0 & 0 & -\frac{2}{3}\sqrt{\frac{10}{7}} & 0 & -\frac{2}{3}  \sqrt{\frac{5}{7}} \\
 0 & 0 & 0 & 0 & 0 \\
 0 & 0 & -\frac{2}{3}  \sqrt{\frac{5}{7}} & 0 & -\frac{\sqrt{\frac{10}{7}}}{3} \\
\end{array}
\right),
\nonumber\\
&\hspace{1cm}{\mathbb{M}}^{(B_1)}=
\left(
\begin{array}{ccccc}
 \frac{2 \mathcal{M}_{1,S}+2 \sqrt{2}\mathcal{M}_{1,SD}+\mathcal{M}_{1,D}}{3\det\mathcal{M}_{1}} & \frac{\sqrt{2} \mathcal{M}_{1,S}-\mathcal{M}_{1,SD}-\sqrt{2}\mathcal{M}_{1,D}}{3 \det\mathcal{M}_{1}} & 0 & 0 & 0 \\
 \frac{\sqrt{2} \mathcal{M}_{1,S}-\mathcal{M}_{1,SD}-\sqrt{2}\mathcal{M}_{1,D}}{3 \det\mathcal{M}_{1}} & \frac{\mathcal{M}_{1,S}-2 \sqrt{2}\mathcal{M}_{1,SD}+2\mathcal{M}_{1,D}}{3 \det\mathcal{M}_{1}} & 0 & 0 & 0 \\
 0 & 0 & \mathcal{M}_{2,D}^{-1} & 0 & 0 \\
 0 & 0 & 0 & \mathcal{M}_{3,D}^{-1} & 0 \\
 0 & 0 & 0 & 0 & \mathcal{M}_{3,D}^{-1} \\
\end{array}
\right).
\label{I110B1}
\\
& B_2:\hspace{.5cm}\mathbb{F}_{00}^{(B_2)}=\textbf{I}_{5},\hspace{.5cm}
\mathbb{F}_{20}^{(B_2)}=\left(
\begin{array}{ccccc}
 -\frac{1}{\sqrt{5}} & 0 & 0 & 0 & -\frac{3}{7}  \sqrt{\frac{6}{5}} \\
 0 & \frac{1}{2 \sqrt{5}} & -\frac{\sqrt{3}}{2} & 0 & \frac{4 \sqrt{\frac{3}{5}}}{7} \\
 0 & -\frac{\sqrt{3}}{2} & \frac{\sqrt{5}}{14} & 0 & -\frac{2}{7} \\
 0 & 0 & 0 & -\frac{2\sqrt{5}}{7}   & 0 \\
 -\frac{3}{7}  \sqrt{\frac{6}{5}} & \frac{4 \sqrt{\frac{3}{5}}}{7} & -\frac{2}{7} & 0 & \frac{6}{7 \sqrt{5}} \\
\end{array}
\right),
\nonumber\\
&\hspace{1cm}\mathbb{F}_{22}^{(B_2)}=\left(
\begin{array}{ccccc}
 -i \sqrt{\frac{6}{5}} & 0 & 0 & -\frac{3 \sqrt{3}}{7}  & -\frac{3 i}{7 \sqrt{5}} \\
 0 & i \sqrt{\frac{3}{10}} & \frac{i}{\sqrt{2}} & \frac{2 \sqrt{6}}{7} & \frac{2}{7} i \sqrt{\frac{2}{5}} \\
 0 & \frac{i}{\sqrt{2}} & \frac{i}{7}  \sqrt{\frac{15}{2}} & \frac{\sqrt{10}}{7} & -\frac{i \sqrt{6}}{7}  \\
 \frac{3 \sqrt{3}}{7} & -\frac{2 \sqrt{6}}{7}   & -\frac{\sqrt{10}}{7} & 0 & \frac{2 \sqrt{2}}{7} \\
 -\frac{3 i}{7 \sqrt{5}} & \frac{2}{7} i \sqrt{\frac{2}{5}} & -\frac{i \sqrt{6}}{7}  & -\frac{2 \sqrt{2}}{7}   & \frac{-4i}{7}   \sqrt{\frac{6}{5}} \\
\end{array}
\right),\hspace{.5cm}
\mathbb{F}_{40}^{(B_2)}=\left(
\begin{array}{ccccc}
 0 & 0 & 0 & 0 & \frac{\sqrt{6}}{7} \\
 0 & 0 & 0 & 0 & \frac{\sqrt{3}}{7} \\
 0 & 0 & \frac{8}{21} & 0 & -\frac{5 \sqrt{5}}{21}  \\
 0 & 0 & 0 & \frac{1}{7} & 0 \\
 \frac{\sqrt{6}}{7} & \frac{\sqrt{3}}{7} & -\frac{5 \sqrt{5}}{21}  & 0 & \frac{1}{21} \\
\end{array}
\right),
\nonumber\\
&\hspace{1cm}\mathbb{F}_{42}^{(B_2)}=\left(
\begin{array}{ccccc}
 0 & 0 & 0 & \frac{1}{7} & \frac{i \sqrt{15}}{7} \\
 0 & 0 & 0 & \frac{1}{7 \sqrt{2}} & \frac{i}{7}  \sqrt{\frac{15}{2}} \\
 0 & 0 & -i\frac{4  \sqrt{10}}{21}  & -\frac{\sqrt{\frac{15}{2}}}{7} & -\frac{5 i}{21 \sqrt{2}} \\
 -\frac{1}{7} & -\frac{1}{7 \sqrt{2}} & \frac{\sqrt{\frac{15}{2}}}{7} & 0 & -\frac{\sqrt{6}}{7} \\
 \frac{i \sqrt{15}}{7} & \frac{i}{7}  \sqrt{\frac{15}{2}} & -\frac{5 i}{21 \sqrt{2}} & \frac{\sqrt{6}}{7} & - i \frac{2\sqrt{10}}{21}  \\
\end{array}
\right)
,\hspace{.5cm}
\mathbb{F}_{44}^{(B_2)}=\left(
\begin{array}{ccccc}
 0 & 0 & 0 & \frac{2 i}{\sqrt{7}} & 0 \\
 0 & 0 & 0 & i \sqrt{\frac{2}{7}} & 0 \\
 0 & 0 & 0 & i \sqrt{\frac{10}{21}} & 0 \\
 -\frac{2 i}{\sqrt{7}} & -i \sqrt{\frac{2}{7}} & -i \sqrt{\frac{10}{21}} & 0 & -i \sqrt{\frac{2}{21}} \\
 0 & 0 & 0 & i \sqrt{\frac{2}{21}} & 0 \\
\end{array}
\right),
\nonumber\\
&\hspace{1cm}{\mathbb{M}}^{(B_2)}=\left(
\begin{array}{ccccc}
 \frac{2 \mathcal{M}_{1,S}+2 \sqrt{2}\mathcal{M}_{1,SD}+\mathcal{M}_{1,D}}{3\det\mathcal{M}_{1}} & \frac{\sqrt{2} \mathcal{M}_{1,S}-\mathcal{M}_{1,SD}-\sqrt{2}\mathcal{M}_{1,D}}{3 \det\mathcal{M}_{1}} & 0 & 0 & 0 \\
 \frac{\sqrt{2} \mathcal{M}_{1,S}-\mathcal{M}_{1,SD}-\sqrt{2}\mathcal{M}_{1,D}}{3 \det\mathcal{M}_{1}} & \frac{\mathcal{M}_{1,S}-2 \sqrt{2}\mathcal{M}_{1,SD}+2\mathcal{M}_{1,D}}{3 \det\mathcal{M}_{1}} & 0 & 0 & 0 \\
 0 & 0 & \mathcal{M}_{2,D}^{-1} & 0 & 0 \\
 0 & 0 & 0 & \mathcal{M}_{3,D}^{-1} & 0 \\
 0 & 0 & 0 & 0 & \mathcal{M}_{3,D}^{-1} \\
\end{array}
\right).
\label{I110B2}
\end{align}
\begin{align}
&B_3:\hspace{.5cm}
\mathbb{F}_{00}^{(B_3)}=\textbf{I}_{5},\hspace{.5cm}
\mathbb{F}_{20}^{(B_3)}=\left(
\begin{array}{ccccc}
 \frac{1}{2 \sqrt{5}} & 0 & -\frac{\sqrt{3}}{2} & \frac{4 \sqrt{\frac{3}{5}}}{7} & 0 \\
 0 & -\frac{1}{\sqrt{5}} & 0 & -\frac{3}{7} \sqrt{\frac{6}{5}} & 0 \\
 -\frac{\sqrt{3}}{2} & 0 & \frac{\sqrt{5}}{14} & -\frac{2}{7} & 0 \\
 \frac{4 \sqrt{\frac{3}{5}}}{7} & -\frac{3}{7} \sqrt{\frac{6}{5}} & -\frac{2}{7} & \frac{6}{7 \sqrt{5}} & 0 \\
 0 & 0 & 0 & 0 & -\frac{2 \sqrt{5}}{7}  \\
\end{array}
\right),
\nonumber\\
&\hspace{1cm}
\mathbb{F}_{22}^{(B_3)}=\left(
\begin{array}{ccccc}
 -i \sqrt{\frac{3}{10}} & 0 & -\frac{i}{\sqrt{2}} & \frac{-2i}{7}   \sqrt{\frac{2}{5}} & \frac{2 \sqrt{6}}{7} \\
 0 & i \sqrt{\frac{6}{5}} & 0 & \frac{3 i}{7 \sqrt{5}} & -\frac{3 \sqrt{3}}{7}   \\
 -\frac{i}{\sqrt{2}} & 0 & -\frac{i}{7}  \sqrt{\frac{15}{2}} & \frac{i \sqrt{6}}{7} & \frac{\sqrt{10}}{7} \\
 \frac{-2i}{7}   \sqrt{\frac{2}{5}} & \frac{3 i}{7 \sqrt{5}} & \frac{i \sqrt{6}}{7} & \frac{4}{7} i \sqrt{\frac{6}{5}} & -\frac{2 \sqrt{2}}{7}   \\
 -\frac{2 \sqrt{6}}{7}   & \frac{3 \sqrt{3}}{7} & -\frac{\sqrt{10}}{7} & \frac{2 \sqrt{2}}{7} & 0 \\
\end{array}
\right)
,\hspace{.5cm}
\mathbb{F}_{40}^{(B_3)}=\left(
\begin{array}{ccccc}
 0 & 0 & 0 & \frac{\sqrt{3}}{7} & 0 \\
 0 & 0 & 0 & \frac{\sqrt{6}}{7} & 0 \\
 0 & 0 & \frac{8}{21} & -\frac{5 \sqrt{5}}{21}  & 0 \\
 \frac{\sqrt{3}}{7} & \frac{\sqrt{6}}{7} & -\frac{5 \sqrt{5}}{21}  & \frac{1}{21} & 0 \\
 0 & 0 & 0 & 0 & \frac{1}{7} \\
\end{array}
\right),
\nonumber\\
&\hspace{1cm}\mathbb{F}_{42}^{(B_3)}=\left(
\begin{array}{ccccc}
 0 & 0 & 0 & -\frac{i}{7}  \sqrt{\frac{15}{2}} & \frac{1}{7 \sqrt{2}} \\
 0 & 0 & 0 & -\frac{i \sqrt{15}}{7}   & \frac{1}{7} \\
 0 & 0 & \frac{4 i \sqrt{10}}{21} & \frac{5 i}{21 \sqrt{2}} & -\frac{\sqrt{\frac{15}{2}}}{7} \\
 -\frac{i}{7}  \sqrt{\frac{15}{2}} & -\frac{i \sqrt{15}}{7}   & \frac{5 i}{21 \sqrt{2}} & \frac{2 i \sqrt{10}}{21} & \frac{\sqrt{6}}{7} \\
 -\frac{1}{7 \sqrt{2}} & -\frac{1}{7} & \frac{\sqrt{\frac{15}{2}}}{7} & -\frac{\sqrt{6}}{7} & 0 \\
\end{array}
\right)
,\hspace{.5cm}
\mathbb{F}_{44}^{(B_3)}=\left(
\begin{array}{ccccc}
 0 & 0 & 0 & 0 & -i \sqrt{\frac{2}{7}} \\
 0 & 0 & 0 & 0 & -\frac{2 i}{\sqrt{7}} \\
 0 & 0 & 0 & 0 & -i \sqrt{\frac{10}{21}} \\
 0 & 0 & 0 & 0 & -i \sqrt{\frac{2}{21}} \\
 i \sqrt{\frac{2}{7}} & \frac{2 i}{\sqrt{7}} & i \sqrt{\frac{10}{21}} & i \sqrt{\frac{2}{21}} & 0 \\
\end{array}
\right),
\nonumber\\
&\hspace{1cm}{\mathbb{M}}^{(B_3)}=
\left(
\begin{array}{ccccc}
 \frac{\mathcal{M}_{1,S}-2 \sqrt{2}\mathcal{M}_{1,SD}+2\mathcal{M}_{1,D}}{3 \det\mathcal{M}_{1}} & \frac{\sqrt{2} \mathcal{M}_{1,S}-\mathcal{M}_{1,SD}-\sqrt{2}\mathcal{M}_{1,D}}{3 \det\mathcal{M}_{1}} & 0 & 0 & 0 \\
 \frac{\sqrt{2} \mathcal{M}_{1,S}-\mathcal{M}_{1,SD}-\sqrt{2}\mathcal{M}_{1,D}}{3 \det\mathcal{M}_{1}} & \frac{2 \mathcal{M}_{1,S}+2 \sqrt{2}\mathcal{M}_{1,SD}+\mathcal{M}_{1,D}}{3 \det\mathcal{M}_{1}} & 0 & 0 & 0 \\
 0 & 0 & \mathcal{M}_{2,D}^{-1} & 0 & 0 \\
 0 & 0 & 0 & \mathcal{M}_{3,D}^{-1} & 0 \\
 0 & 0 & 0 & 0 & \mathcal{M}_{3,D}^{-1} \\
\end{array}
\right).
\label{I110B3}
\end{align}

\begin{align}
& A: \hspace{.5cm}\mathbb{F}_{00}^{(A)}=\textbf{I}_{3},\hspace{.5cm}
\mathbb{F}_{20}^{(A)}=\left(
\begin{array}{ccc}
 \frac{\sqrt{5}}{7} & 0 & 0 \\
 0 & -\frac{\sqrt{5}}{7} & -\frac{\sqrt{10}}{7} \\
 0 & -\frac{\sqrt{10}}{7} & 0 \\
\end{array}
\right),\hspace{.5cm}
\mathbb{F}_{22}^{(A)}=\left(
\begin{array}{ccc}
 0 & -\frac{\sqrt{10}}{7} & \frac{2 \sqrt{5}}{7} \\
 \frac{\sqrt{10}}{7} & 0 & 0 \\
 -\frac{2 \sqrt{5}}{7}  & 0 & 0 \\
\end{array}
\right),
\nonumber\\
&\hspace{1cm}\mathbb{F}_{40}^{(A)}=\left(
\begin{array}{ccc}
 -\frac{4}{7} & 0 & 0 \\
 0 & -\frac{2}{21} & \frac{5 \sqrt{2}}{21} \\
 0 & \frac{5 \sqrt{2}}{21} & -\frac{1}{3} \\
\end{array}
\right),\hspace{.5cm}
\nonumber
\mathbb{F}_{42}^{(A)}=\left(
\begin{array}{ccc}
 0 & -\frac{2 \sqrt{\frac{10}{3}}}{7}   & \frac{4 \sqrt{\frac{5}{3}}}{7} \\
 \frac{2 \sqrt{\frac{10}{3}}}{7} & 0 & 0 \\
 -\frac{4}{7} \sqrt{\frac{5}{3}} & 0 & 0 \\
\end{array}
\right),
\nonumber\\
&\hspace{1cm}
\mathbb{F}_{44}^{(A)}=\left(
\begin{array}{ccc}
 0 & 0 & 0 \\
 0 & \frac{2 \sqrt{\frac{10}{7}}}{3} & \frac{2 \sqrt{\frac{5}{7}}}{3} \\
 0 & \frac{2 \sqrt{\frac{5}{7}}}{3} & \frac{\sqrt{\frac{10}{7}}}{3} \\
\end{array}
\right),\hspace{.5cm}
{\mathbb{M}}^{(A)}=\left(
\begin{array}{ccc}
 \mathcal{M}_{2,D}^{-1} & 0 & 0 \\
 0 & \mathcal{M}_{2,D}^{-1} & 0 \\
 0 & 0 & \mathcal{M}_{3,D}^{-1} \\
\end{array}
\right).
\label{I110A}
\end{align}
%

\subsection{Positive parity isotriplet channel}
The scattering amplitude matrix for this channel is only a $6\times6$ matrix as following
\begin{eqnarray}
\mathcal{M}^{\infty}_{(1,0)} = \left( \begin{array}{cccccccccc}
\mathcal{M}_{0,S}&0\\
0&\mathcal{M}_{2,D}\\
\end{array} \right),
\end{eqnarray}
where each element is still a diagonal matrix in the $M_J$ basis. The QC in Eq. (\ref{QC-simplified}) for each irrep of the corresponding point group should be understood with the matrices that are given below.

\subsubsection{$\mathbf{d}=(0,0,0)$}
\begin{align}
&A_1:\hspace{.5cm}\mathbb{F}_{00}^{(A_1)}=1,\hspace{.5cm}
{\mathbb{M}}^{(A_1)}=\mathcal{M}_{0,S}^{-1}.
\label{II000A1}
\\
&E:\hspace{.5cm}
\mathbb{F}_{00}^{(E)}=1,\hspace{.5cm}
\mathbb{F}_{40}^{(E)}=\frac{6}{7},\hspace{.5cm}
{\mathbb{M}}^{(E)}=\mathcal{M}_{2,D}^{-1}.
\label{II000E}
\\
&T_2:\hspace{.5cm}
\mathbb{F}_{00}^{(T_2)}=1,\hspace{.5cm}
\mathbb{F}_{40}^{(T_2)}=-\frac{4}{7},\hspace{.5cm}
{\mathbb{M}}^{(T_2)}=\mathcal{M}_{2,D}^{-1}.
\label{II000T2}
\end{align}
%
\subsubsection{$\mathbf{d}=(0,0,1)$}
\begin{align}
&A_1:\hspace{.5cm}\mathbb{F}_{00}^{(A_1)}=\textbf{I}_{2},\hspace{.5cm}
\mathbb{F}_{20}^{(A_1)}=\left(
\begin{array}{cc}
 0 & 1 \\
 1 & \frac{2 \sqrt{5}}{7} \\
\end{array}
\right),\hspace{.5cm}
\mathbb{F}_{40}^{(A_1)}=\left(
\begin{array}{cc}
 0 & 0 \\
 0 & \frac{6}{7} \\
\end{array}
\right),
\hspace{.5cm} {\mathbb{M}}^{(A_1)}=\left(
\begin{array}{cc}
 \mathcal{M}_{0,S}^{-1} & 0 \\
 0 & \mathcal{M}_{2,D}^{-1} \\
\end{array}
\right).
\label{II001A1}
\\
&B_1:\hspace{.5cm}\mathbb{F}_{00}^{(B_1)}=1,\hspace{.5cm}
\mathbb{F}_{20}^{(B_1)}=-\frac{2 \sqrt{5}}{7} ,\hspace{.5cm}
\mathbb{F}_{40}^{(B_1)}=\frac{1}{7},\hspace{.5cm}
\mathbb{F}_{44}^{(B_1)}=\sqrt{\frac{10}{7}},\hspace{.5cm}
{\mathbb{M}}^{(B_1)}=\mathcal{M}_{2,D}^{-1}.
\label{II001B1}
\\
&B_1:\hspace{.5cm}\mathbb{F}_{00}^{(B_1)}=1,\hspace{.5cm}
\mathbb{F}_{20}^{(B_1)}=-\frac{2 \sqrt{5}}{7} ,\hspace{.5cm}
\mathbb{F}_{40}^{(B_1)}=\frac{1}{7},\hspace{.5cm}
\mathbb{F}_{44}^{(B_1)}=-\sqrt{\frac{10}{7}},\hspace{.5cm}
{\mathbb{M}}^{(B_1)}=\mathcal{M}_{2,D}^{-1}.
\label{II001B2}
\\
&E:\hspace{.5cm}\mathbb{F}_{00}^{(E)}=1,\hspace{.5cm}
\mathbb{F}_{20}^{(E)}=\frac{\sqrt{5}}{7},\hspace{.5cm}
\mathbb{F}_{40}^{(E)}=-\frac{4}{7},\hspace{.5cm}
{\mathbb{M}}^{(E)}=\mathcal{M}_{2,D}^{-1}.
\label{II001E}
\end{align}
%
\subsubsection{$\mathbf{d}=(1,1,0)$}
\begin{align}
&B_1:\hspace{.5cm}\mathbb{F}_{00}^{(B_1)}=1,\hspace{.5cm}
\mathbb{F}_{20}^{(B_1)}=-\frac{2 \sqrt{5}}{7},\hspace{.5cm}
\mathbb{F}_{40}^{(B_1)}=\frac{1}{7},\hspace{.5cm}
\mathbb{F}_{44}^{(B_1)}=\sqrt{\frac{10}{7}},\hspace{.5cm}
{\mathbb{M}}^{(B_1)}=\mathcal{M}_{2,D}^{-1}.
\label{II110B1}
\\
&B_2:\hspace{.5cm}\mathbb{F}_{00}^{(B_2)}=1,\hspace{.5cm}
\mathbb{F}_{20}^{(B_2)}=\frac{\sqrt{5}}{7},\hspace{.5cm}
\mathbb{F}_{22}^{(B_2)}=-\frac{i \sqrt{30}}{7} ,\hspace{.5cm}
\mathbb{F}_{40}^{(B_2)}=-\frac{4}{7},\hspace{.5cm}
\mathbb{F}_{42}^{(B_2)}=-\frac{2 i \sqrt{10}}{7}  ,
\nonumber\\
&\hspace{1.2cm}
{\mathbb{M}}^{(B_2)}=\mathcal{M}_{2,D}^{-1}.
\label{II110B2}
\\
&B_3:\hspace{.5cm}\mathbb{F}_{00}^{(B_3)}=1,\hspace{.5cm}
\mathbb{F}_{20}^{(B_3)}=\frac{\sqrt{5}}{7},\hspace{.5cm}
\mathbb{F}_{22}^{(B_3)}=\frac{i \sqrt{30}}{7},\hspace{.5cm}
\mathbb{F}_{40}^{(B_3)}=-\frac{4}{7},
\hspace{.5cm}
\mathbb{F}_{42}^{(B_3)}=\frac{2 i \sqrt{10}}{7},
\nonumber\\
&\hspace{1.2cm}
{\mathbb{M}}^{(B_3)}=\mathcal{M}_{2,D}^{-1}.
\label{II110B3}
\\
&A:\hspace{.5cm}
\mathbb{F}_{00}^{(A)}=\textbf{I}_{3},\hspace{.5cm}
\mathbb{F}_{20}^{(A)}=\left(
\begin{array}{ccc}
 0 & 0 & 1 \\
 0 & -\frac{2 \sqrt{5}}{7} & 0 \\
 1 & 0 & \frac{2 \sqrt{5}}{7}
\end{array}
\right),\hspace{.5cm}
\mathbb{F}_{22}^{(A)}=\left(
\begin{array}{ccc}
 0 & \sqrt{2} & 0 \\
 -\sqrt{2} & 0 & \frac{2 \sqrt{10}}{7} \\
 0 & -\frac{2 \sqrt{10}}{7} & 0
\end{array}
\right),
\nonumber\\
&\hspace{1cm}
\mathbb{F}_{40}^{(A)}=\left(
\begin{array}{ccc}
 0 & 0 & 0 \\
 0 & \frac{1}{7} & 0 \\
 0 & 0 & \frac{6}{7}
\end{array}
\right),\hspace{.5cm}
\mathbb{F}_{42}^{(A)}=\left(
\begin{array}{ccc}
 0 & 0 & 0 \\
 0 & 0 & -\frac{\sqrt{30}}{7} \\
 0 & \frac{\sqrt{30}}{7} & 0
\end{array}
\right),\hspace{.5cm}
\mathbb{F}_{44}^{(A)}=\left(
\begin{array}{ccc}
 0 & 0 & 0 \\
 0 & -\sqrt{\frac{10}{7}} & 0 \\
 0 & 0 & 0
\end{array}
\right),
\nonumber\\
&\hspace{1cm}{\mathbb{M}}^{(A)}=\left(
\begin{array}{ccc}
 \mathcal{M}_{0,S}^{-1} & 0 & 0 \\
 0 & \mathcal{M}_{2,D}^{-1} & 0 \\
 0 & 0 & \mathcal{M}_{2,D}^{-1} \\
\end{array}
\right).
\label{II110A}
\end{align}
%

\subsection{Negative parity isosinglet channel}
Given the truncation made on the angular momentum in the master QC, the scattering amplitude matrix for this channel is a $10\times10$ matrix, and is given by
\begin{eqnarray}
\mathcal{M}^{\infty}_{(0,0)} = \left( \begin{array}{cccccccccc}
\mathcal{M}_{1,P}&0\\
0&\mathcal{M}_{3,F}\\
\end{array} \right),
\end{eqnarray}
where each element is still a diagonal matrix in the $M_J$ basis. The following matrices should be used in the QC in Eq. (\ref{QC-simplified}) for this channel. 

\subsubsection{$\mathbf{d}=(0,0,0)$}
\begin{align}
&T_1:\hspace{.5cm}
\mathbb{F}_{00}^{(T_1)}=\textbf{I}_{2},\hspace{.5cm}
\mathbb{F}_{40}^{(T_1)}=\left(
\begin{array}{cc}
 0 & -\frac{4}{\sqrt{21}} \\
 -\frac{4}{\sqrt{21}} & \frac{6}{11} \\
\end{array}
\right),\hspace{.5cm}
\mathbb{F}_{60}^{(T_1)}=\left(
\begin{array}{cc}
 0 & 0 \\
 0 & \frac{100}{33 \sqrt{13}} \\
\end{array}
\right),
\nonumber\\
&\hspace{1.2cm}{\mathbb{M}}^{(T_1)}=
\left(
\begin{array}{cc}
\mathcal{M}_{1,P}^{-1} & 0 \\
 0 & \mathcal{M}_{3,F}^{-1} \\
\end{array}
\right).
\label{III000T1}
\\
&A_2:\hspace{.5cm}
\mathbb{F}_{00}^{(A_2)}=1,\hspace{.5cm}
\mathbb{F}_{40}^{(A_2)}=-\frac{12}{11},\hspace{.5cm}
\mathbb{F}_{60}^{(A_2)}=\frac{80}{11 \sqrt{13}},\hspace{.5cm}
{\mathbb{M}}^{(A_2)}=\mathcal{M}_{3,F}^{-1}.
\label{III000A}
\\
&T_2:\hspace{.5cm}
\mathbb{F}_{00}^{(T_2)}=1,\hspace{.5cm}
\mathbb{F}_{40}^{(T_2)}=-\frac{2}{11},\hspace{.5cm}
\mathbb{F}_{60}^{(T_2)}=-\frac{60}{11 \sqrt{13}},\hspace{.5cm}
{\mathbb{M}}^{(T_2)}=\mathcal{M}_{3,F}^{-1}.
\label{III000T2}
\end{align}
%

\subsubsection{$\mathbf{d}=(0,0,1)$}
\begin{align}
&A_2:\hspace{.5cm}
\mathbb{F}_{00}^{(A_2)}=\textbf{I}_{2},\hspace{.5cm}
\mathbb{F}_{20}^{(A_2)}=\left(
\begin{array}{cc}
 \frac{2}{\sqrt{5}} & 3 \sqrt{\frac{3}{35}} \\
 3 \sqrt{\frac{3}{35}} & \frac{4}{3 \sqrt{5}} \\
\end{array}
\right),\hspace{.5cm}
\mathbb{F}_{40}^{(A_2)}=\left(
\begin{array}{cc}
 0 & \frac{4}{\sqrt{21}} \\
 \frac{4}{\sqrt{21}} & \frac{6}{11} \\
\end{array}
\right),
\nonumber\\
&\hspace{1.25cm}\mathbb{F}_{60}^{(A_2)}=\left(
\begin{array}{cc}
 0 & 0 \\
 0 & \frac{100}{33 \sqrt{13}} \\
\end{array}
\right),\hspace{.5cm}
{\mathbb{M}}^{(A_2)}=\left(
\begin{array}{cc}
\mathcal{M}_{1,P}^{-1} & 0 \\
 0 & \mathcal{M}_{3,F}^{-1} \\
\end{array}
\right).
\label{III001A2}
\\
&B_1:\hspace{.5cm}
\mathbb{F}_{00}^{(B_1)}=1,\hspace{.5cm}
\mathbb{F}_{40}^{(B_1)}=-\frac{7}{11},\hspace{.5cm}
\mathbb{F}_{44}^{(B_1)}=-\frac{\sqrt{70}}{11},\hspace{.5cm}
\mathbb{F}_{60}^{(B_1)}=\frac{10}{11 \sqrt{13}},\hspace{.5cm}
\nonumber\\
&\hspace{1.25cm}
\mathbb{F}_{64}^{(B_1)}=-\frac{10 \sqrt{\frac{14}{13}}}{11},\hspace{.5cm}
{\mathbb{M}}^{(B_1)}=\mathcal{M}_{3,F}^{-1}.
\label{III001B1}
\\
&B_2:\hspace{.5cm}
\mathbb{F}_{00}^{(B_2)}=1,\hspace{.5cm}
\mathbb{F}_{40}^{(B_2)}=-\frac{7}{11},\hspace{.5cm}
\mathbb{F}_{44}^{(B_2)}=\frac{\sqrt{70}}{11},\hspace{.5cm}
\mathbb{F}_{60}^{(B_2)}=\frac{10}{11 \sqrt{13}},\hspace{.5cm}
\nonumber\\
&\hspace{1.2cm}
\mathbb{F}_{64}^{(B_2)}=\frac{10 \sqrt{\frac{14}{13}}}{11},\hspace{.5cm}
{\mathbb{M}}^{(B_2)}=\mathcal{M}_{3,F}^{-1}.
\label{III001B2}
\\
&E:\hspace{.5cm}
\mathbb{F}_{00}^{(E)}=\textbf{I}_{3},\hspace{.5cm}
\mathbb{F}_{20}^{(E)}=\left(
\begin{array}{ccc}
 -\frac{1}{\sqrt{5}} & 3 \sqrt{\frac{2}{35}} & 0 \\
 3 \sqrt{\frac{2}{35}} & \frac{1}{\sqrt{5}} & 0 \\
 0 & 0 & -\frac{\sqrt{5}}{3} \\
\end{array}
\right),\hspace{.5cm}
\mathbb{F}_{40}^{(E)}=\left(
\begin{array}{ccc}
 0 & -\sqrt{\frac{2}{7}} & 0 \\
 -\sqrt{\frac{2}{7}} & \frac{1}{11} & 0 \\
 0 & 0 & \frac{3}{11} \\
\end{array}
\right),
\nonumber\\
&\hspace{1.1cm}\mathbb{F}_{44}^{(E)}=\left(
\begin{array}{ccc}
 0 & 0 & -\frac{2}{\sqrt{3}} \\
 0 & 0 & \frac{\sqrt{42}}{11} \\
 -\frac{2}{\sqrt{3}} & \frac{\sqrt{42}}{11} & 0 \\
\end{array}
\right),\hspace{.5cm}
\mathbb{F}_{60}^{(E)}=
\left(
\begin{array}{ccc}
 0 & 0 & 0 \\
 0 & -\frac{25}{11 \sqrt{13}} & 0 \\
 0 & 0 & -\frac{5}{33 \sqrt{13}} \\
\end{array}
\right),
\nonumber\\
&\hspace{1cm}
\mathbb{F}_{64}^{(E)}=\left(
\begin{array}{ccc}
 0 & 0 & 0 \\
 0 & 0 & -\frac{5}{11}  \sqrt{\frac{70}{39}} \\
 0 & -\frac{5}{11}  \sqrt{\frac{70}{39}} & 0 \\
\end{array}
\right),
\hspace{.5cm}
{\mathbb{M}}^{(E)}=\left(
\begin{array}{ccc}
\mathcal{M}_{1,P}^{-1} & 0 & 0 \\
 0 & \mathcal{M}_{3,F}^{-1} & 0 \\
 0 & 0 & \mathcal{M}_{3,F}^{-1} \\
\end{array}
\right).
\label{III001E}
\end{align}
%

\subsubsection{$\mathbf{d}=(1,1,0)$}
\begin{align}
&B_2:\hspace{.5cm}\mathbb{F}_{00}^{(B_2)}=\textbf{I}_{3},\hspace{.5cm}
\mathbb{F}_{20}^{(B_2)}=\left(
\begin{array}{ccc}
 -\frac{1}{\sqrt{5}} & 3 \sqrt{\frac{2}{35}} & 0 \\
 3 \sqrt{\frac{2}{35}} & \frac{1}{\sqrt{5}} & 0 \\
 0 & 0 & -\frac{\sqrt{5}}{3} \\
\end{array}
\right),\hspace{.5cm}
\mathbb{F}_{22}^{(B_2)}=\left(
\begin{array}{ccc}
 i \sqrt{\frac{6}{5}} & -i \sqrt{\frac{3}{35}} & \frac{3}{\sqrt{7}} \\
 -i \sqrt{\frac{3}{35}} & 2 i \sqrt{\frac{2}{15}} & -\frac{\sqrt{2}}{3} \\
 -\frac{3}{\sqrt{7}} & \frac{\sqrt{2}}{3} & 0 \\
\end{array}
\right),
\nonumber
\\
&\hspace{1.3cm}
\mathbb{F}_{40}^{(B_2)}=\left(
\begin{array}{ccc}
 0 & -\sqrt{\frac{2}{7}} & 0 \\
 -\sqrt{\frac{2}{7}} & \frac{1}{11} & 0 \\
 0 & 0 & \frac{3}{11} \\
\end{array}
\right),\hspace{.5cm}
\mathbb{F}_{42}^{(B_2)}=\left(
\begin{array}{ccc}
 0 & i \sqrt{\frac{5}{7}} & -\frac{1}{\sqrt{21}} \\
 i \sqrt{\frac{5}{7}} & \frac{2 i \sqrt{10}}{11} & \frac{3 \sqrt{6}}{11} \\
 \frac{1}{\sqrt{21}} & -\frac{3 \sqrt{6}}{11} & 0 \\
\end{array}
\right),
\nonumber
\end{align}
\begin{align}
&\hspace{1.3cm}\mathbb{F}_{44}^{(B_2)}=\left(
\begin{array}{ccc}
 0 & 0 & \frac{2 i}{\sqrt{3}} \\
 0 & 0 & -\frac{i \sqrt{42}}{11} \\
 -\frac{2 i}{\sqrt{3}} & \frac{i \sqrt{42}}{11} & 0 \\
\end{array}
\right),\hspace{.5cm}
\mathbb{F}_{60}^{(B_2)}=\left(
\begin{array}{ccc}
 0 & 0 & 0 \\
 0 & -\frac{25}{11 \sqrt{13}} & 0 \\
 0 & 0 & -\frac{5}{33 \sqrt{13}} \\
\end{array}
\right),
\nonumber
\\
&\hspace{1.3cm}
\mathbb{F}_{62}^{(B_2)}=\left(
\begin{array}{ccc}
 0 & 0 & 0 \\
 0 & \frac{10}{11} i \sqrt{\frac{35}{39}} & -\frac{10}{33} \sqrt{\frac{7}{13}} \\
 0 & \frac{10 \sqrt{\frac{7}{13}}}{33} & 0 \\
\end{array}
\right),\hspace{.5cm}
\mathbb{F}_{64}^{(B_2)}=\left(
\begin{array}{ccc}
 0 & 0 & 0 \\
 0 & 0 & \frac{5}{11} i \sqrt{\frac{70}{39}} \\
 0 & -\frac{1}{11} 5 i \sqrt{\frac{70}{39}} & 0 \\
\end{array}
\right),
\nonumber
\\
&\hspace{1.3cm}
\mathbb{F}_{66}^{(B_2)}=\left(
\begin{array}{ccc}
 0 & 0 & 0 \\
 0 & 0 & 0 \\
 0 & 0 & -10 i \sqrt{\frac{7}{429}} \\
\end{array}
\right),\hspace{.5cm}
{\mathbb{M}}^{(B_2)}=\left(
\begin{array}{ccc}
\mathcal{M}_{1,P}^{-1} & 0 & 0 \\
 0 & \mathcal{M}_{3,F}^{-1} & 0 \\
 0 & 0 & \mathcal{M}_{3,F}^{-1} \\
\end{array}
\right).
\label{III110B2}
\\
&B_3:\hspace{.5cm}\mathbb{F}_{00}^{(B_3)}=\textbf{I}_{3},\hspace{.5cm}
\mathbb{F}_{20}^{(B_3)}=\left(
\begin{array}{ccc}
 -\frac{1}{\sqrt{5}} & 0 & 3 \sqrt{\frac{2}{35}} \\
 0 & -\frac{\sqrt{5}}{3} & 0 \\
 3 \sqrt{\frac{2}{35}} & 0 & \frac{1}{\sqrt{5}} \\
\end{array}
\right),\hspace{.5cm}
\mathbb{F}_{22}^{(B_3)}=\left(
\begin{array}{ccc}
 -i \sqrt{\frac{6}{5}} & \frac{3}{\sqrt{7}} & i \sqrt{\frac{3}{35}} \\
 -\frac{3}{\sqrt{7}} & 0 & \frac{\sqrt{2}}{3} \\
 i \sqrt{\frac{3}{35}} & -\frac{\sqrt{2}}{3} & -2 i \sqrt{\frac{2}{15}} \\
\end{array}
\right),
\nonumber
\\
&\hspace{1.3cm}\mathbb{F}_{40}^{(B_3)}=\left(
\begin{array}{ccc}
 0 & 0 & -\sqrt{\frac{2}{7}} \\
 0 & \frac{3}{11} & 0 \\
 -\sqrt{\frac{2}{7}} & 0 & \frac{1}{11} \\
\end{array}
\right),\hspace{.5cm}
\mathbb{F}_{42}^{(B_3)}=\left(
\begin{array}{ccc}
 0 & -\frac{1}{\sqrt{21}} & -i \sqrt{\frac{5}{7}} \\
 \frac{1}{\sqrt{21}} & 0 & -\frac{3 \sqrt{6}}{11} \\
 -i \sqrt{\frac{5}{7}} & \frac{3 \sqrt{6}}{11} & -\frac{i2 \sqrt{10} }{11} \\
\end{array}
\right),
\nonumber
\\
&\hspace{1.3cm}
\mathbb{F}_{44}^{(B_3)}=\left(
\begin{array}{ccc}
 0 & -\frac{2 i}{\sqrt{3}} & 0 \\
 \frac{2 i}{\sqrt{3}} & 0 & - \frac{i \sqrt{42}}{11} \\
 0 & \frac{i \sqrt{42}}{11} & 0 \\
\end{array}
\right),\hspace{.5cm}
\mathbb{F}_{60}^{(B_3)}=\left(
\begin{array}{ccc}
 0 & 0 & 0 \\
 0 & -\frac{5}{33 \sqrt{13}} & 0 \\
 0 & 0 & -\frac{25}{11 \sqrt{13}} \\
\end{array}
\right),
\nonumber
\\
&\hspace{1.3cm}
\mathbb{F}_{62}^{(B_3)}=\left(
\begin{array}{ccc}
 0 & 0 & 0 \\
 0 & 0 & \frac{10 \sqrt{\frac{7}{13}}}{33} \\
 0 & -\frac{10}{33}  \sqrt{\frac{7}{13}} & -\frac{10i}{11}   \sqrt{\frac{35}{39}} \\
\end{array}
\right),\hspace{.5cm}
\mathbb{F}_{64}^{(B_3)}=\left(
\begin{array}{ccc}
 0 & 0 & 0 \\
 0 & 0 & \frac{5}{11} i \sqrt{\frac{70}{39}} \\
 0 & -\frac{1}{11} 5 i \sqrt{\frac{70}{39}} & 0 \\
\end{array}
\right),
\nonumber
\\
&\hspace{1.3cm}
\mathbb{F}_{66}^{(B_3)}=\left(
\begin{array}{ccc}
 0 & 0 & 0 \\
 0 & 10 i \sqrt{\frac{7}{429}} & 0 \\
 0 & 0 & 0 \\
\end{array}
\right),\hspace{.5cm}
{\mathbb{M}}^{(B_3)}=\left(
\begin{array}{ccc}
\mathcal{M}_{1,P}^{-1} & 0 & 0 \\
 0 & \mathcal{M}_{3,F}^{-1} & 0 \\
 0 & 0 & \mathcal{M}_{3,F}^{-1} \\
\end{array}
\right).
\label{III110B3}
\\
&A:\hspace{.5cm}\mathbb{F}_{00}^{(A)}=1,\hspace{.5cm}
\mathbb{F}_{40}^{(A)}=-\frac{7}{11},\hspace{.5cm}
\mathbb{F}_{44}^{(A)}=\frac{\sqrt{70}}{11},\hspace{.5cm}\mathbb{F}_{60}^{(A)}=\frac{10}{11 \sqrt{13}},\hspace{.5cm}
\nonumber\\
&\hspace{1.2cm}
\mathbb{F}_{64}^{(A)}=\frac{10 \sqrt{\frac{14}{13}}}{11},\hspace{.5cm}
{\mathbb{M}}^{(A)}=\mathcal{M}_{3,F}^{-1}.
\label{III110A}
\\
&B_1:\hspace{.5cm}
\mathbb{F}_{00}^{(B_1)}=\textbf{I}_{3},\hspace{.5cm}
\mathbb{F}_{20}^{(B_1)}=\left(
\begin{array}{ccc}
 \frac{2}{\sqrt{5}} & 3 \sqrt{\frac{3}{35}} & 0 \\
 3 \sqrt{\frac{3}{35}} & \frac{4}{3 \sqrt{5}} & 0 \\
 0 & 0 & 0 \\
\end{array}
\right),\hspace{.5cm}
\mathbb{F}_{22}^{(B_1)}=\left(
\begin{array}{ccc}
 0 & 0 & \sqrt{\frac{6}{7}} \\
 0 & 0 & -\frac{2 \sqrt{2}}{3}  \\
 -\sqrt{\frac{6}{7}} & \frac{2 \sqrt{2}}{3} & 0 \\
\end{array}
\right),
\nonumber
\\
&\hspace{1.3cm}
\mathbb{F}_{40}^{(B_1)}=\left(
\begin{array}{ccc}
 0 & \frac{4}{\sqrt{21}} & 0 \\
 \frac{4}{\sqrt{21}} & \frac{6}{11} & 0 \\
 0 & 0 & -\frac{7}{11} \\
\end{array}
\right),\hspace{.5cm}
\mathbb{F}_{42}^{(B_1)}=\left(
\begin{array}{ccc}
 0 & 0 & 2 \sqrt{\frac{2}{7}} \\
 0 & 0 & -\frac{\sqrt{6}}{11} \\
 -2 \sqrt{\frac{2}{7}} & \frac{\sqrt{6}}{11} & 0 \\
\end{array}
\right),\hspace{.5cm}
\mathbb{F}_{44}^{(B_1)}=\left(
\begin{array}{ccc}
 0 & 0 & 0 \\
 0 & 0 & 0 \\
 0 & 0 & -\frac{\sqrt{70}}{11} \\
\end{array}
\right),
\nonumber
\end{align}
\begin{align}
&\hspace{-.7cm}
\mathbb{F}_{60}^{(B_1)}=\left(
\begin{array}{ccc}
 0 & 0 & 0 \\
 0 & \frac{100}{33 \sqrt{13}} & 0 \\
 0 & 0 & \frac{10}{11 \sqrt{13}} \\
\end{array}
\right),\hspace{.5cm}
\mathbb{F}_{62}^{(B_1)}=\left(
\begin{array}{ccc}
 0 & 0 & 0 \\
 0 & 0 & \frac{40 \sqrt{\frac{7}{13}}}{33} \\
 0 & -\frac{40}{33} \sqrt{\frac{7}{13}} & 0 \\
\end{array}
\right),
\nonumber
\\
&\hspace{-.7cm}\mathbb{F}_{64}^{(B_1)}=\left(
\begin{array}{ccc}
 0 & 0 & 0 \\
 0 & 0 & 0 \\
 0 & 0 & -\frac{10}{11} \sqrt{\frac{14}{13}} \\
\end{array}
\right),\hspace{.5cm}
{\mathbb{M}}^{(B_1)}=\left(
\begin{array}{ccc}
\mathcal{M}_{1,P}^{-1} & 0 & 0 \\
 0 & \mathcal{M}_{3,F}^{-1} & 0 \\
 0 & 0 & \mathcal{M}_{3,F}^{-1} \\
\end{array}
\right).
\label{III110B1}
\end{align}
%

\subsection{Negative parity isotriplet channel}
The scattering amplitude in this channel is a $30 \times 30$ matrix with the following elements
 \begin{eqnarray}
\mathcal{M}_{(1,1)}^{\infty}=\left(\begin{array}{cccccc}
\mathcal{M}_{0,P} & 0 & 0 & 0 & 0 & 0\\
0 & \mathcal{M}_{1,P} & 0 & 0 & 0 & 0\\
0 & 0 & \mathcal{M}_{2,P} & \mathcal{M}_{2,PF} & 0 & 0\\
0 & 0 & \mathcal{M}_{2,FP} & \mathcal{M}_{2,F} & 0 & 0\\
0 & 0 & 0 & 0 & \mathcal{M}_{3,F} & 0\\
0 & 0 & 0 & 0 & 0 & \mathcal{M}_{4,F}
\end{array}\right)
 \end{eqnarray}
As is seen, various QCs in this channel will give access to the P- F-waves mixing parameter in the NN negative parity sector. The QCs are obtained via inputing the following matrices in Eq. (\ref{QC-simplified}).

\subsubsection{$\mathbf{d}=(0,0,0)$}
\begin{align}
&T_1\hspace{.5cm}\mathbb{F}_{00}^{(T_1)}=\textbf{I}_{3},\hspace{.5cm}
\mathbb{F}_{40}^{(T_1)}=\left(
\begin{array}{ccc}
 0 & \sqrt{\frac{2}{7}} & \sqrt{\frac{10}{21}} \\
 \sqrt{\frac{2}{7}} & \frac{1}{11} & \frac{\sqrt{15}}{11} \\
 \sqrt{\frac{10}{21}} & \frac{\sqrt{15}}{11} & \frac{3}{11} \\
\end{array}
\right),\hspace{.5cm}
\mathbb{F}_{60}^{(T_1)}=\left(
\begin{array}{ccc}
 0 & 0 & 0 \\
 0 & -\frac{25}{11 \sqrt{13}} & \frac{35 \sqrt{\frac{5}{39}}}{11} \\
 0 & \frac{35 \sqrt{\frac{5}{39}}}{11} & -\frac{5}{33 \sqrt{13}} \\
\end{array}
\right),
\nonumber
\\
&\hspace{.9cm}{\mathbb{M}}^{(T_1)}=\left(
\begin{array}{ccc}
\mathcal{M}_{1,P}^{-1} & 0 & 0 \\
 0 & \mathcal{M}_{3,F}^{-1} & 0 \\
 0 & 0 & \mathcal{M}_{4,F}^{-1} \\
\end{array}
\right).
\label{IV000T1}
\\
&T_2:\hspace{.5cm}\mathbb{F}_{00}^{(T_2)}=\textbf{I}_{4},\hspace{.5cm}
\mathbb{F}_{40}^{(T_2)}=\left(
\begin{array}{cccc}
 0 & \frac{4 \sqrt{\frac{2}{3}}}{7} & \sqrt{\frac{10}{21}} & -\frac{\sqrt{\frac{10}{3}}}{7} \\
 \frac{4 \sqrt{\frac{2}{3}}}{7} & -\frac{4}{21} & \frac{2 \sqrt{\frac{5}{7}}}{3} & \frac{6 \sqrt{5}}{77} \\
 \sqrt{\frac{10}{21}} & \frac{2 \sqrt{\frac{5}{7}}}{3} & -\frac{1}{33} & -\frac{15}{11 \sqrt{7}} \\
 -\frac{\sqrt{\frac{10}{3}}}{7} & \frac{6 \sqrt{5}}{77} & -\frac{15}{11 \sqrt{7}} & -\frac{39}{77} \\
\end{array}
\right),
\nonumber
\end{align}
\begin{align}
&\hspace{1.2cm}
\mathbb{F}_{60}^{(T_2)}=\left(
\begin{array}{cccc}
 0 & 0 & 0 & 0 \\
 0 & 0 & 0 & -\frac{80}{33} \sqrt{\frac{5}{13}} \\
 0 & 0 & \frac{45}{11 \sqrt{13}} & \frac{35 \sqrt{\frac{7}{13}}}{33} \\
 0 & -\frac{80}{33} \sqrt{\frac{5}{13}} & \frac{35 \sqrt{\frac{7}{13}}}{33} & \frac{25}{33 \sqrt{13}} \\
\end{array}
\right),\hspace{.5cm}
{\mathbb{M}}^{(T_2)}= \left(
\begin{array}{cccc}
 \frac{\mathcal{M}_{2,F}}{\det\mathcal{M}_{2}} & -\frac{\mathcal{M}_{2,PF}}{\det\mathcal{M}_{2}} & 0 & 0 \\
 -\frac{\mathcal{M}_{2,PF}}{\det\mathcal{M}_{2}} & \frac{\mathcal{M}_{2,P}}{\det\mathcal{M}_{2}} & 0 & 0 \\
 0 & 0 & \mathcal{M}_{3,F}^{-1} & 0 \\
 0 & 0 & 0 & \mathcal{M}_{4,F}^{-1} \\
\end{array}
\right).
\label{IV000T2}
\\
&A_2:\hspace{.5cm}\mathbb{F}_{00}^{(A_2)}=1,\hspace{.5cm}
\mathbb{F}_{40}^{(A_2)}=-\frac{2}{11},\hspace{.5cm}
\mathbb{F}_{60}^{(A_2)}=-\frac{60}{11 \sqrt{13}}\hspace{.5cm}
{\mathbb{M}}^{(A_2)}=\mathcal{M}_{3,F}^{-1}.
\label{IV000A2}
\\
&A_1:\hspace{.5cm}\mathbb{F}_{00}^{(A_1)}=\textbf{I}_{2},\hspace{.5cm}
\mathbb{F}_{40}^{(A_1)}=\left(
\begin{array}{cc}
 0 & -\frac{4}{\sqrt{21}} \\
 -\frac{4}{\sqrt{21}} & \frac{6}{11} \\
\end{array}
\right), \hspace{.5cm}\mathbb{F}_{60}^{(A_1)}=\left(
\begin{array}{cc}
 0 & 0 \\
 0 & \frac{100}{33 \sqrt{13}} \\
\end{array}
\right),
\nonumber
\\
&\hspace{1.2cm}
{\mathbb{M}}^{(A_1)}= \left(
\begin{array}{cc}
 \mathcal{M}_{0,P}^{-1} & 0 \\
 0 & \mathcal{M}_{4,F}^{-1} \\
\end{array}
\right).
\label{IV000A1}
\\
&E:\hspace{.5cm}
\mathbb{F}_{00}^{(E)}=\textbf{I}_{2},\hspace{.5cm}
\mathbb{F}_{40}^{(E)}=\left(
\begin{array}{ccc}
 0 & -\frac{2 \sqrt{6}}{7}   & -\frac{2 \sqrt{\frac{10}{3}}}{7}   \\
 -\frac{2 \sqrt{6}}{7}   & \frac{2}{7} & \frac{12 \sqrt{5}}{77} \\
 -\frac{2 \sqrt{\frac{10}{3}}}{7}   & \frac{12 \sqrt{5}}{77} & \frac{6}{77} \\
\end{array}
\right),\hspace{.5cm}
\mathbb{F}_{60}^{(E)}=\left(
\begin{array}{ccc}
 0 & 0 & 0 \\
 0 & 0 & \frac{20 \sqrt{\frac{5}{13}}}{11} \\
 0 & \frac{20 \sqrt{\frac{5}{13}}}{11} & -\frac{80}{33 \sqrt{13}} \\
\end{array}
\right),
\nonumber
\\
&\hspace{1.2cm}
{\mathbb{M}}^{(E)}=
\left(
\begin{array}{ccc}
 \frac{\mathcal{M}_{2,F}}{\det\mathcal{M}_{2}} & -\frac{\mathcal{M}_{2,PF}}{\det\mathcal{M}_{2}} & 0 \\
 -\frac{\mathcal{M}_{2,PF}}{\det\mathcal{M}_{2}} & \frac{\mathcal{M}_{2,P}}{\det\mathcal{M}_{2}} & 0 \\
 0 & 0 & \mathcal{M}_{4,F}^{-1} \\
\end{array}
\right).
\label{IV000E}
\end{align}
%

\subsubsection{$\mathbf{d}=(0,0,1)$}
\begin{align}
&A_1:\hspace{.5cm}\mathbb{F}_{00}^{(A_1)}=\textbf{I}_{4},\hspace{.5cm}
\mathbb{F}_{20}^{(A_1)}=\left(
\begin{array}{ccccc}
 0 & -\sqrt{\frac{2}{5}} & \sqrt{\frac{3}{5}} & 0 & 0 \\
 -\sqrt{\frac{2}{5}} & \frac{1}{\sqrt{5}} & -\frac{\sqrt{\frac{6}{5}}}{7} & 0 & \frac{9 \sqrt{\frac{2}{5}}}{7} \\
 \sqrt{\frac{3}{5}} & -\frac{\sqrt{\frac{6}{5}}}{7} & \frac{8}{7 \sqrt{5}} & 0 & -\frac{2}{7 \sqrt{15}} \\
 0 & 0 & 0 & -\frac{\sqrt{5}}{3} & 0 \\
 0 & \frac{9 \sqrt{\frac{2}{5}}}{7} & -\frac{2}{7 \sqrt{15}} & 0 & \frac{5 \sqrt{5}}{21} \\
\end{array}
\right),
\nonumber
\\
&\hspace{1cm}\mathbb{F}_{40}^{(A_1)}=\left(
\begin{array}{ccccc}
 0 & 0 & 0 & 0 & -\frac{2}{3} \\
 0 & 0 & -\frac{2 \sqrt{6}}{7}   & 0 & \frac{5 \sqrt{2}}{21} \\
 0 & -\frac{2 \sqrt{6}}{7}   & \frac{2}{7} & 0 & -\frac{10}{77} \sqrt{3} \\
 0 & 0 & 0 & \frac{3}{11} & 0 \\
 -\frac{2}{3} & \frac{5 \sqrt{2}}{21} & -\frac{10}{77} \sqrt{3} & 0 & \frac{27}{77} \\
\end{array}
\right),\hspace{.5cm}
\mathbb{F}_{44}^{(A_1)}=\left(
\begin{array}{ccccc}
 0 & 0 & 0 & -\frac{2 \sqrt{2}}{3}  & 0 \\
 0 & 0 & 0 & -\frac{2}{3} & 0 \\
 0 & 0 & 0 & \frac{2 \sqrt{6}}{11} & 0 \\
 -\frac{2 \sqrt{2}}{3}  & -\frac{2}{3} & \frac{2 \sqrt{6}}{11} & 0 & \frac{3 \sqrt{2}}{11} \\
 0 & 0 & 0 & \frac{3 \sqrt{2}}{11} & 0 \\
\end{array}
\right),
\nonumber
\\
&\hspace{1cm}
\mathbb{F}_{60}^{(A_1)}=\left(
\begin{array}{ccccc}
 0 & 0 & 0 & 0 & 0 \\
 0 & 0 & 0 & 0 & 0 \\
 0 & 0 & 0 & 0 & -\frac{50}{11 \sqrt{39}} \\
 0 & 0 & 0 & -\frac{5}{33 \sqrt{13}} & 0 \\
 0 & 0 & -\frac{50}{11 \sqrt{39}} & 0 & \frac{25}{33 \sqrt{13}} \\
\end{array}
\right),\hspace{.5cm}
\mathbb{F}_{64}^{(A_1)}=\left(
\begin{array}{ccccc}
 0 & 0 & 0 & 0 & 0 \\
 0 & 0 & 0 & 0 & 0 \\
 0 & 0 & 0 & -\frac{10 }{11} \sqrt{\frac{10}{39}} & 0 \\
 0 & 0 & -\frac{10 }{11} \sqrt{\frac{10}{39}} & 0 & -\frac{5}{11}  \sqrt{\frac{10}{13}} \\
 0 & 0 & 0 & -\frac{5}{11}  \sqrt{\frac{10}{13}} & 0 \\
\end{array}
\right),
\nonumber
\end{align}
\begin{align}
&\hspace{1.2cm}{\mathbb{M}}^{(A_1)}=\left(
\begin{array}{ccccc}
 \mathcal{M}_{0,P}^{-1} & 0 & 0 & 0 & 0 \\
 0 & \frac{\mathcal{M}_{2,F}}{\det\mathcal{M}_{2}} & -\frac{\mathcal{M}_{2,PF}}{\det\mathcal{M}_{2}} & 0 & 0 \\
 0 & -\frac{\mathcal{M}_{2,PF}}{\det\mathcal{M}_{2}} & \frac{\mathcal{M}_{2,P}}{\det\mathcal{M}_{2}} & 0 & 0 \\
 0 & 0 & 0 & \mathcal{M}_{4,F}^{-1} & 0 \\
 0 & 0 & 0 & 0 & \mathcal{M}_{4,F}^{-1} \\
\end{array}
\right).
\label{IV001A1}
\\
&A_2:\hspace{.5cm}\mathbb{F}_{00}^{(A_2)}=\textbf{I}_{3},\hspace{.5cm}
\mathbb{F}_{20}^{(A_2)}=\left(
\begin{array}{ccc}
 -\frac{1}{\sqrt{5}} & \sqrt{\frac{18}{35}} & 0 \\
 \sqrt{\frac{18}{35}} & \frac{1}{\sqrt{5}} & 0 \\
 0 & 0 & -\frac{\sqrt{5}}{3} \\
\end{array}
\right),\hspace{.5cm}
\mathbb{F}_{40}^{(A_2)}=\left(
\begin{array}{ccc}
 0 & -\sqrt{\frac{2}{7}} & 0 \\
 -\sqrt{\frac{2}{7}} & \frac{1}{11} & 0 \\
 0 & 0 & \frac{3}{11} \\
\end{array}
\right),
\nonumber
\\
&\hspace{1.2cm}\mathbb{F}_{44}^{(A_2)}=\left(
\begin{array}{ccc}
 0 & 0 & \frac{2}{\sqrt{3}} \\
 0 & 0 & -\frac{\sqrt{42}}{11} \\
 \frac{2}{\sqrt{3}} & -\frac{\sqrt{42}}{11} & 0 \\
\end{array}
\right),\hspace{.5cm}
\mathbb{F}_{60}^{(A_2)}=\left(
\begin{array}{ccc}
 0 & 0 & 0 \\
 0 & -\frac{25}{11 \sqrt{13}} & 0 \\
 0 & 0 & -\frac{5}{33 \sqrt{13}} \\
\end{array}
\right),
\nonumber
\\
&\hspace{1.2cm}
\mathbb{F}_{64}^{(A_2)}=\left(
\begin{array}{ccc}
 0 & 0 & 0 \\
 0 & 0 & \frac{5 \sqrt{\frac{70}{39}}}{11} \\
 0 & \frac{5 \sqrt{\frac{70}{39}}}{11} & 0 \\
\end{array}
\right),\hspace{.5cm}
{\mathbb{M}}^{(A_2)}=
\left(
\begin{array}{ccc}
\mathcal{M}_{1,P}^{-1} & 0 & 0 \\
 0 & \mathcal{M}_{3,F}^{-1} & 0 \\
 0 & 0 & \mathcal{M}_{4,F}^{-1} \\
\end{array}
\right).
\label{IV001A2}
\\
&B_1:\hspace{.5cm}\mathbb{F}_{00}^{(B_1)}=\textbf{I}_{4},\hspace{.5cm}
\mathbb{F}_{20}^{(B_1)}=\left(
\begin{array}{cccc}
 -\frac{1}{\sqrt{5}} & \frac{\sqrt{\frac{6}{5}}}{7} & -\sqrt{\frac{3}{14}} & \frac{3 \sqrt{\frac{3}{2}}}{7} \\
 \frac{\sqrt{\frac{6}{5}}}{7} & -\frac{8}{7 \sqrt{5}} & -\frac{1}{\sqrt{7}} & -\frac{1}{21} \\
 -\sqrt{\frac{3}{14}} & -\frac{1}{\sqrt{7}} & 0 & -\frac{\sqrt{\frac{5}{7}}}{3} \\
 \frac{3 \sqrt{\frac{3}{2}}}{7} & -\frac{1}{21} & -\frac{\sqrt{\frac{5}{7}}}{3} & \frac{2 \sqrt{5}}{21} \\
\end{array}
\right),
\nonumber
\\
&\hspace{1.2cm}
\mathbb{F}_{40}^{(B_1)}=\left(
\begin{array}{cccc}
 0 & -\frac{\sqrt{\frac{2}{3}}}{7} & \sqrt{\frac{5}{42}} & -\frac{\sqrt{\frac{15}{2}}}{7} \\
 -\frac{\sqrt{\frac{2}{3}}}{7} & \frac{1}{21} & \frac{\sqrt{\frac{5}{7}}}{3} & \frac{9 \sqrt{5}}{77} \\
 \sqrt{\frac{5}{42}} & \frac{\sqrt{\frac{5}{7}}}{3} & -\frac{7}{66} & -\frac{15}{22 \sqrt{7}} \\
 -\frac{\sqrt{\frac{15}{2}}}{7} & \frac{9 \sqrt{5}}{77} & -\frac{15}{22 \sqrt{7}} & -\frac{3}{14} \\
\end{array}
\right),\hspace{.5cm}
\mathbb{F}_{44}^{(B_1)}=\left(
\begin{array}{cccc}
 0 & -\sqrt{\frac{20}{21}} & -\frac{1}{\sqrt{3}} & -\frac{1}{\sqrt{21}} \\
 -\sqrt{\frac{20}{21}} & \frac{\sqrt{\frac{10}{7}}}{3} & -\frac{\sqrt{2}}{3} & \frac{\sqrt{\frac{18}{7}}}{11} \\
 -\frac{1}{\sqrt{3}} & -\frac{\sqrt{2}}{3} & -\frac{\sqrt{\frac{35}{2}}}{33} & \frac{\sqrt{\frac{45}{2}}}{11} \\
 -\frac{1}{\sqrt{21}} & \frac{\sqrt{\frac{18}{7}}}{11} & \frac{\sqrt{\frac{45}{2}}}{11} & \frac{9 \sqrt{\frac{5}{14}}}{11} \\
\end{array}
\right),
\nonumber
\\
&\hspace{1.2cm}\mathbb{F}_{60}^{(B_1)}=\left(
\begin{array}{cccc}
 0 & 0 & 0 & 0 \\
 0 & 0 & 0 & -\frac{10}{33} \sqrt{\frac{5}{13}} \\
 0 & 0 & -\frac{15}{22 \sqrt{13}} & \frac{35 \sqrt{\frac{7}{13}}}{66} \\
 0 & -\frac{10}{33} \sqrt{\frac{5}{13}} & \frac{35 \sqrt{\frac{7}{13}}}{66} & -\frac{5}{6 \sqrt{13}} \\
\end{array}
\right),
\hspace{.5cm}
\mathbb{F}_{64}^{(B_1)}=\left(
\begin{array}{cccc}
 0 & 0 & 0 & 0 \\
 0 & 0 & 0 & -\frac{10}{33} \sqrt{\frac{70}{13}} \\
 0 & 0 & \frac{15 \sqrt{\frac{7}{26}}}{11} & \frac{35}{33 \sqrt{26}} \\
 0 & -\frac{10}{33} \sqrt{\frac{70}{13}} & \frac{35}{33 \sqrt{26}} & \frac{5 \sqrt{\frac{7}{26}}}{11} \\
\end{array}
\right),
\nonumber
\\
&\hspace{1.2cm}
{\mathbb{M}}^{(B_1)}=\left(
\begin{array}{cccc}
 \frac{\mathcal{M}_{2,F}}{\det\mathcal{M}_{2}} & -\frac{\mathcal{M}_{2,PF}}{\det\mathcal{M}_{2}} & 0 & 0 \\
 -\frac{\mathcal{M}_{2,PF}}{\det\mathcal{M}_{2}} & \frac{\mathcal{M}_{2,P}}{\det\mathcal{M}_{2}} & 0 & 0 \\
 0 & 0 & \mathcal{M}_{3,F}^{-1} & 0 \\
 0 & 0 & 0 & \mathcal{M}_{4,F}^{-1} \\
\end{array}
\right).
\label{IV001B1}
\end{align}
\begin{align}
&B_2:\hspace{.5cm}
\mathbb{F}_{00}^{(B_2)}=\textbf{I}_{4},\hspace{.5cm}
\mathbb{F}_{20}^{(B_1)}=\left(
\begin{array}{cccc}
 -\frac{1}{\sqrt{5}} & \frac{\sqrt{\frac{6}{5}}}{7} & -\sqrt{\frac{3}{14}} & \frac{3 \sqrt{\frac{3}{2}}}{7} \\
 \frac{\sqrt{\frac{6}{5}}}{7} & -\frac{8}{7 \sqrt{5}} & -\frac{1}{\sqrt{7}} & -\frac{1}{21} \\
 -\sqrt{\frac{3}{14}} & -\frac{1}{\sqrt{7}} & 0 & -\frac{\sqrt{\frac{5}{7}}}{3} \\
 \frac{3 \sqrt{\frac{3}{2}}}{7} & -\frac{1}{21} & -\frac{\sqrt{\frac{5}{7}}}{3} & \frac{2 \sqrt{5}}{21} \\
\end{array}
\right),
\nonumber
\\
&\hspace{1.2cm}\mathbb{F}_{40}^{(B_2)}=\left(
\begin{array}{cccc}
 0 & -\frac{\sqrt{\frac{2}{3}}}{7} & \sqrt{\frac{5}{42}} & -\frac{\sqrt{\frac{15}{2}}}{7} \\
 -\frac{\sqrt{\frac{2}{3}}}{7} & \frac{1}{21} & \frac{\sqrt{\frac{5}{7}}}{3} & \frac{9 \sqrt{5}}{77} \\
 \sqrt{\frac{5}{42}} & \frac{\sqrt{\frac{5}{7}}}{3} & -\frac{7}{66} & -\frac{15}{22 \sqrt{7}} \\
 -\frac{\sqrt{\frac{15}{2}}}{7} & \frac{9 \sqrt{5}}{77} & -\frac{15}{22 \sqrt{7}} & -\frac{3}{14} \\
\end{array}
\right),
\hspace{.5cm}
\mathbb{F}_{44}^{(B_2)}=\left(
\begin{array}{cccc}
 0 & 2 \sqrt{\frac{5}{21}} & \frac{1}{\sqrt{3}} & \frac{1}{\sqrt{21}} \\
 2 \sqrt{\frac{5}{21}} & -\frac{\sqrt{\frac{10}{7}}}{3} & \frac{\sqrt{2}}{3} & -\frac{3}{11}  \sqrt{\frac{2}{7}} \\
 \frac{1}{\sqrt{3}} & \frac{\sqrt{2}}{3} & \frac{\sqrt{\frac{35}{2}}}{33} & -\frac{3}{11} \sqrt{\frac{5}{2}} \\
 \frac{1}{\sqrt{21}} & -\frac{3}{11}  \sqrt{\frac{2}{7}} & -\frac{3}{11} \sqrt{\frac{5}{2}} & -\frac{9}{11}  \sqrt{\frac{5}{14}} \\
\end{array}
\right),
\nonumber
\\
&\hspace{1.2cm}\mathbb{F}_{60}^{(B_2)}=\left(
\begin{array}{cccc}
 0 & 0 & 0 & 0 \\
 0 & 0 & 0 & -\frac{10}{33} \sqrt{\frac{5}{13}} \\
 0 & 0 & -\frac{15}{22 \sqrt{13}} & \frac{35 \sqrt{\frac{7}{13}}}{66} \\
 0 & -\frac{10}{33} \sqrt{\frac{5}{13}} & \frac{35 \sqrt{\frac{7}{13}}}{66} & -\frac{5}{6 \sqrt{13}} \\
\end{array}
\right),
\hspace{.5cm}
\mathbb{F}_{64}^{(B_2)}=\left(
\begin{array}{cccc}
 0 & 0 & 0 & 0 \\
 0 & 0 & 0 & \frac{10 \sqrt{\frac{70}{13}}}{33} \\
 0 & 0 & -\frac{15}{11} \sqrt{\frac{7}{26}} & -\frac{35}{33 \sqrt{26}} \\
 0 & \frac{10 \sqrt{\frac{70}{13}}}{33} & -\frac{35}{33 \sqrt{26}} & -\frac{5}{11} \sqrt{\frac{7}{26}} \\
\end{array}
\right),
\nonumber
\\
&\hspace{1.1cm}
{\mathbb{M}}^{(B_2)}=\left(
\begin{array}{cccc}
 \frac{\mathcal{M}_{2,F}}{\det\mathcal{M}_{2}} & -\frac{\mathcal{M}_{2,PF}}{\det\mathcal{M}_{2}} & 0 & 0 \\
 -\frac{\mathcal{M}_{2,PF}}{\det\mathcal{M}_{2}} & \frac{\mathcal{M}_{2,P}}{\det\mathcal{M}_{2}} & 0 & 0 \\
 0 & 0 & \mathcal{M}_{3,F}^{-1} & 0 \\
 0 & 0 & 0 & \mathcal{M}_{4,F}^{-1} \\
\end{array}
\right).
\label{IV001B2}
\end{align}
\begin{align}
&E:\hspace{.5cm}
\mathbb{F}_{00}^{(E)}=\textbf{I}_{7},
\hspace{.5cm} \mathbb{F}_{20}^{(E)}=\left(
\begin{array}{ccccccc}
 \frac{1}{2 \sqrt{5}} & \frac{3}{2 \sqrt{5}} & \sqrt{\frac{3}{10}} & 2 \sqrt{\frac{3}{35}} & 0 & 0 & 0 \\
 \frac{3}{2 \sqrt{5}} & \frac{1}{2 \sqrt{5}} & -\frac{\sqrt{\frac{3}{10}}}{7} & \sqrt{\frac{3}{35}} & 0 & \frac{3 \sqrt{3}}{7} & 0 \\
 \sqrt{\frac{3}{10}} & -\frac{\sqrt{\frac{3}{10}}}{7} & \frac{4}{7 \sqrt{5}} & \sqrt{\frac{2}{35}} & 0 & -\frac{\sqrt{2}}{21} & 0 \\
 2 \sqrt{\frac{3}{35}} & \sqrt{\frac{3}{35}} & \sqrt{\frac{2}{35}} & \frac{3}{4 \sqrt{5}} & 0 & \frac{5}{12 \sqrt{7}} & 0 \\
 0 & 0 & 0 & 0 & -\frac{\sqrt{5}}{4} & 0 & -\frac{\sqrt{\frac{5}{3}}}{4} \\
 0 & \frac{3 \sqrt{3}}{7} & -\frac{\sqrt{2}}{21} & \frac{5}{12 \sqrt{7}} & 0 & \frac{17 \sqrt{5}}{84} & 0 \\
 0 & 0 & 0 & 0 & -\frac{\sqrt{\frac{5}{3}}}{4} & 0 & -\frac{\sqrt{5}}{12}
\end{array}
\right),
\nonumber
\end{align}
\begin{align}
&\mathbb{F}_{40}^{(E)}=\left(
\begin{array}{ccccccc}
 0 & 0 & 0 & \frac{\sqrt{\frac{3}{7}}}{2} & 0 & \frac{\sqrt{\frac{5}{3}}}{2} & 0 \\
 0 & 0 & \frac{4 \sqrt{\frac{2}{3}}}{7} & \frac{5}{2 \sqrt{21}} & 0 & \frac{\sqrt{\frac{5}{3}}}{14} & 0 \\
 0 & \frac{4 \sqrt{\frac{2}{3}}}{7} & -\frac{4}{21} & \frac{5}{3 \sqrt{14}} & 0 & -\frac{3 \sqrt{\frac{5}{2}}}{77} & 0 \\
 \frac{\sqrt{\frac{3}{7}}}{2} & \frac{5}{2 \sqrt{21}} & \frac{5}{3 \sqrt{14}} & \frac{1}{66} & 0 & \frac{9 \sqrt{\frac{5}{7}}}{22} & 0 \\
 0 & 0 & 0 & 0 & \frac{1}{22} & 0 & \frac{5 \sqrt{3}}{22} \\
 \frac{\sqrt{\frac{5}{3}}}{2} & \frac{\sqrt{\frac{5}{3}}}{14} & -\frac{3 \sqrt{\frac{5}{2}}}{77} & \frac{9 \sqrt{\frac{5}{7}}}{22} & 0 & \frac{27}{154} & 0 \\
 0 & 0 & 0 & 0 & \frac{5 \sqrt{3}}{22} & 0 & -\frac{9}{22}
\end{array}
\right),
\nonumber\\
&\mathbb{F}_{44}^{(E)}=\left(
\begin{array}{ccccccc}
 0 & 0 & 0 & 0 & \frac{1}{\sqrt{2}} & 0 & \frac{1}{\sqrt{6}} \\
 0 & 0 & 0 & 0 & -\frac{1}{\sqrt{2}} & 0 & -\frac{1}{\sqrt{6}} \\
 0 & 0 & 0 & 0 & -\frac{1}{\sqrt{3}} & 0 & \frac{3}{11} \\
 0 & 0 & 0 & 0 & \frac{\sqrt{\frac{7}{6}}}{11} & 0 & -\frac{3 \sqrt{\frac{7}{2}}}{11} \\
 \frac{1}{\sqrt{2}} & -\frac{1}{\sqrt{2}} & -\frac{1}{\sqrt{3}} & \frac{\sqrt{\frac{7}{6}}}{11} & 0 & \frac{\sqrt{\frac{15}{2}}}{11} & 0 \\
 0 & 0 & 0 & 0 & \frac{\sqrt{\frac{15}{2}}}{11} & 0 & \frac{3 \sqrt{\frac{5}{2}}}{11} \\
 \frac{1}{\sqrt{6}} & -\frac{1}{\sqrt{6}} & \frac{3}{11} & -\frac{3 \sqrt{\frac{7}{2}}}{11} & 0 & \frac{3 \sqrt{\frac{5}{2}}}{11} & 0
\end{array}
\right),
\nonumber\\
&\mathbb{F}_{60}^{(E)}=\left(
\begin{array}{ccccccc}
 0 & 0 & 0 & 0 & 0 & 0 & 0 \\
 0 & 0 & 0 & 0 & 0 & 0 & 0 \\
 0 & 0 & 0 & 0 & 0 & \frac{20 \sqrt{\frac{10}{13}}}{33} & 0 \\
 0 & 0 & 0 & \frac{75}{44 \sqrt{13}} & 0 & \frac{35 \sqrt{\frac{35}{13}}}{132} & 0 \\
 0 & 0 & 0 & 0 & \frac{5}{44 \sqrt{13}} & 0 & -\frac{35}{44 \sqrt{39}} \\
 0 & 0 & \frac{20 \sqrt{\frac{10}{13}}}{33} & \frac{35 \sqrt{\frac{35}{13}}}{132} & 0 & -\frac{5}{132 \sqrt{13}} & 0 \\
 0 & 0 & 0 & 0 & -\frac{35}{44 \sqrt{39}} & 0 & \frac{85}{132 \sqrt{13}}
\end{array}
\right),
\nonumber\\
&\mathbb{F}_{64}^{(E)}=\left(
\begin{array}{ccccccc}
 0 & 0 & 0 & 0 & 0 & 0 & 0 \\
 0 & 0 & 0 & 0 & 0 & 0 & 0 \\
 0 & 0 & 0 & 0 & 0 & 0 & \frac{40 \sqrt{\frac{5}{13}}}{33} \\
 0 & 0 & 0 & 0 & \frac{5 \sqrt{\frac{105}{26}}}{22} & 0 & -\frac{25 \sqrt{\frac{35}{26}}}{66} \\
 0 & 0 & 0 & \frac{5 \sqrt{\frac{105}{26}}}{22} & 0 & -\frac{35 \sqrt{\frac{3}{26}}}{22} & 0 \\
 0 & 0 & 0 & 0 & -\frac{35 \sqrt{\frac{3}{26}}}{22} & 0 & \frac{5}{22 \sqrt{26}} \\
 0 & 0 & \frac{40 \sqrt{\frac{5}{13}}}{33} & -\frac{25 \sqrt{\frac{35}{26}}}{66} & 0 & \frac{5}{22 \sqrt{26}} & 0
\end{array}
\right),
\nonumber
\end{align}
\begin{align}
&{\mathbb{M}}^{(E)}=\left(
\begin{array}{ccccccc}
\mathcal{M}_{1,P}^{-1} & 0 & 0 & 0 & 0 & 0 & 0 \\
 0 & \frac{\mathcal{M}_{2,F}}{\det\mathcal{M}_{2}} & -\frac{\mathcal{M}_{2,PF}}{\det\mathcal{M}_{2}} & 0 & 0 & 0 & 0 \\
 0 & -\frac{\mathcal{M}_{2,PF}}{\det\mathcal{M}_{2}} & \frac{\mathcal{M}_{2,P}}{\det\mathcal{M}_{2}} & 0 & 0 & 0 & 0 \\
 0 & 0 & 0 & \mathcal{M}_{3,F}^{-1} & 0 & 0 & 0 \\
 0 & 0 & 0 & 0 & \mathcal{M}_{3,F}^{-1} & 0 & 0 \\
 0 & 0 & 0 & 0 & 0 & \mathcal{M}_{4,F}^{-1} & 0 \\
 0 & 0 & 0 & 0 & 0 & 0 & \mathcal{M}_{4,F}^{-1} \\
\end{array}
\right).
\label{IV001E}
\end{align}
%

\subsubsection{$\mathbf{d}=(1,1,0)$}
\begin{align}
&A:\hspace{.5cm}\mathbb{F}_{00}^{(A)}=\textbf{I}_{9},
\nonumber\\
&\hspace{1cm}\mathbb{F}_{20}^{(A)}=\left(
\begin{array}{ccccccccc}
 0 & -\sqrt{\frac{2}{5}} & 0 & \sqrt{\frac{3}{5}} & 0 & 0 & 0 & 0 & 0 \\
 -\sqrt{\frac{2}{5}} & \frac{1}{\sqrt{5}} & 0 & -\frac{\sqrt{\frac{6}{5}}}{7} & 0 & 0 & 0 & \frac{9 \sqrt{\frac{2}{5}}}{7} & 0 \\
 0 & 0 & -\frac{1}{\sqrt{5}} & 0 & \frac{\sqrt{\frac{6}{5}}}{7} & -\sqrt{\frac{3}{14}} & \frac{3 \sqrt{\frac{3}{2}}}{7} & 0 & 0 \\
 \sqrt{\frac{3}{5}} & -\frac{\sqrt{\frac{6}{5}}}{7} & 0 & \frac{8}{7 \sqrt{5}} & 0 & 0 & 0 & -\frac{2}{7 \sqrt{15}} & 0 \\
 0 & 0 & \frac{\sqrt{\frac{6}{5}}}{7} & 0 & -\frac{8}{7 \sqrt{5}} & -\frac{1}{\sqrt{7}} & -\frac{1}{21} & 0 & 0 \\
 0 & 0 & -\sqrt{\frac{3}{14}} & 0 & -\frac{1}{\sqrt{7}} & 0 & -\frac{\sqrt{\frac{5}{7}}}{3} & 0 & 0 \\
 0 & 0 & \frac{3 \sqrt{\frac{3}{2}}}{7} & 0 & -\frac{1}{21} & -\frac{\sqrt{\frac{5}{7}}}{3} & \frac{2 \sqrt{5}}{21} & 0 & 0 \\
 0 & \frac{9 \sqrt{\frac{2}{5}}}{7} & 0 & -\frac{2}{7 \sqrt{15}} & 0 & 0 & 0 & \frac{5 \sqrt{5}}{21} & 0 \\
 0 & 0 & 0 & 0 & 0 & 0 & 0 & 0 & -\frac{\sqrt{5}}{3} \\
\end{array}
\right),
\nonumber
\end{align}
\begin{align}
&\nonumber\hspace{1cm}\mathbb{F}_{22}^{(A)}=\left(
\begin{array}{ccccccccc}
 0 & 0 & -\frac{2}{\sqrt{5}} & 0 & \sqrt{\frac{6}{5}} & 0 & 0 & 0 & 0 \\
 0 & 0 & -\sqrt{\frac{2}{5}} & 0 & \frac{2 \sqrt{\frac{3}{5}}}{7} & \sqrt{\frac{3}{7}} & \frac{3 \sqrt{3}}{7} & 0 & 0 \\
 \frac{2}{\sqrt{5}} & \sqrt{\frac{2}{5}} & 0 & -\frac{2}{7} \sqrt{\frac{3}{5}} & 0 & 0 & 0 & -\frac{3}{7 \sqrt{5}} & \frac{3}{\sqrt{7}} \\
 0 & 0 & \frac{2 \sqrt{\frac{3}{5}}}{7} & 0 & -\frac{8}{7}\sqrt{\frac{2}{5}} & \sqrt{\frac{2}{7}} & -\frac{\sqrt{2}}{21} & 0 & 0 \\
 -\sqrt{\frac{6}{5}} & -\frac{2}{7} \sqrt{\frac{3}{5}} & 0 & \frac{8 \sqrt{\frac{2}{5}}}{7} & 0 & 0 & 0 & \frac{\sqrt{\frac{2}{15}}}{21} & -\frac{\sqrt{\frac{2}{21}}}{3} \\
 0 & -\sqrt{\frac{3}{7}} & 0 & -\sqrt{\frac{2}{7}} & 0 & 0 & 0 & \frac{5}{3 \sqrt{42}} & \frac{\sqrt{\frac{5}{6}}}{3} \\
 0 & -\frac{3 \sqrt{3}}{7}   & 0 & \frac{\sqrt{2}}{21} & 0 & 0 & 0 & \frac{5 \sqrt{\frac{3}{2}}}{7} & -\sqrt{\frac{5}{42}} \\
 0 & 0 & \frac{3}{7 \sqrt{5}} & 0 & -\frac{\sqrt{\frac{2}{15}}}{21} & -\frac{5}{3 \sqrt{42}} & -\frac{5}{7}  \sqrt{\frac{3}{2}} & 0 & 0 \\
 0 & 0 & -\frac{3}{\sqrt{7}} & 0 & \frac{\sqrt{\frac{2}{21}}}{3} & -\frac{\sqrt{\frac{5}{6}}}{3} & \sqrt{\frac{5}{42}} & 0 & 0 \\
\end{array}
\right),
\nonumber
\\
&\hspace{1cm}\mathbb{F}_{40}^{(A)}=\left(
\begin{array}{ccccccccc}
 0 & 0 & 0 & 0 & 0 & 0 & 0 & -\frac{2}{3} & 0 \\
 0 & 0 & 0 & -\frac{2 \sqrt{6}}{7}   & 0 & 0 & 0 & \frac{5 \sqrt{2}}{21} & 0 \\
 0 & 0 & 0 & 0 & -\frac{\sqrt{\frac{2}{3}}}{7} & \sqrt{\frac{5}{42}} & -\frac{\sqrt{\frac{15}{2}}}{7} & 0 & 0 \\
 0 & -\frac{2 \sqrt{6}}{7}   & 0 & \frac{2}{7} & 0 & 0 & 0 & -\frac{10}{77} \sqrt{3} & 0 \\
 0 & 0 & -\frac{\sqrt{\frac{2}{3}}}{7} & 0 & \frac{1}{21} & \frac{\sqrt{\frac{5}{7}}}{3} & \frac{9 \sqrt{5}}{77} & 0 & 0 \\
 0 & 0 & \sqrt{\frac{5}{42}} & 0 & \frac{\sqrt{\frac{5}{7}}}{3} & -\frac{7}{66} & -\frac{15}{22 \sqrt{7}} & 0 & 0 \\
 0 & 0 & -\frac{\sqrt{\frac{15}{2}}}{7} & 0 & \frac{9 \sqrt{5}}{77} & -\frac{15}{22 \sqrt{7}} & -\frac{3}{14} & 0 & 0 \\
 -\frac{2}{3} & \frac{5 \sqrt{2}}{21} & 0 & -\frac{10}{77} \sqrt{3} & 0 & 0 & 0 & \frac{27}{77} & 0 \\
 0 & 0 & 0 & 0 & 0 & 0 & 0 & 0 & \frac{3}{11} \\
\end{array}
\right),
\nonumber
\\
&\hspace{1cm}\mathbb{F}_{42}^{(A)}=\left(
\begin{array}{ccccccccc}
 0 & 0 & 0 & 0 & 0 & 0 & -\frac{2 \sqrt{2}}{3}  & 0 & 0 \\
 0 & 0 & 0 & 0 & -\frac{2 \sqrt{5}}{7}  & \frac{2}{\sqrt{7}} & \frac{4}{21} & 0 & 0 \\
 0 & 0 & 0 & \frac{2 \sqrt{5}}{7} & 0 & 0 & 0 & \frac{\sqrt{15}}{7} & -\frac{1}{\sqrt{21}} \\
 0 & 0 & -\frac{2 \sqrt{5}}{7}  & 0 & \frac{\sqrt{\frac{10}{3}}}{7} & 2 \sqrt{\frac{2}{21}} & -\frac{4\sqrt{6}}{77}  & 0 & 0 \\
 0 & \frac{2 \sqrt{5}}{7} & 0 & -\frac{\sqrt{\frac{10}{3}}}{7} & 0 & 0 & 0 & -\frac{9 \sqrt{10}}{77}  & \frac{3 \sqrt{\frac{2}{7}}}{11} \\
 0 & -\frac{2}{\sqrt{7}} & 0 & -2 \sqrt{\frac{2}{21}} & 0 & 0 & 0 & -\frac{15}{11 \sqrt{14}} & -\frac{3}{11} \sqrt{\frac{5}{2}} \\
 \frac{2 \sqrt{2}}{3} & -\frac{4}{21} & 0 & \frac{4 \sqrt{6}}{77} & 0 & 0 & 0 & \frac{3}{7 \sqrt{2}} & \frac{9 \sqrt{\frac{5}{14}}}{11} \\
 0 & 0 & -\frac{\sqrt{15}}{7} & 0 & \frac{9 \sqrt{10}}{77} & \frac{15}{11 \sqrt{14}} & -\frac{3}{7 \sqrt{2}} & 0 & 0 \\
 0 & 0 & \frac{1}{\sqrt{21}} & 0 & -\frac{3}{11}  \sqrt{\frac{2}{7}} & \frac{3 \sqrt{\frac{5}{2}}}{11} & -\frac{9}{11}  \sqrt{\frac{5}{14}} & 0 & 0 \\
\end{array}
\right),
\nonumber
\end{align}
\begin{align}
&\mathbb{F}_{44}^{(A)}=\left(
\begin{array}{ccccccccc}
 0 & 0 & 0 & 0 & 0 & 0 & 0 & 0 & -\frac{2 \sqrt{2}}{3}  \\
 0 & 0 & 0 & 0 & 0 & 0 & 0 & 0 & -\frac{2}{3} \\
 0 & 0 & 0 & 0 & 2 \sqrt{\frac{5}{21}} & \frac{1}{\sqrt{3}} & \frac{1}{\sqrt{21}} & 0 & 0 \\
 0 & 0 & 0 & 0 & 0 & 0 & 0 & 0 & \frac{2 \sqrt{6}}{11} \\
 0 & 0 & 2 \sqrt{\frac{5}{21}} & 0 & -\frac{\sqrt{\frac{10}{7}}}{3} & \frac{\sqrt{2}}{3} & -\frac{3}{11}  \sqrt{\frac{2}{7}} & 0 & 0 \\
 0 & 0 & \frac{1}{\sqrt{3}} & 0 & \frac{\sqrt{2}}{3} & \frac{\sqrt{\frac{35}{2}}}{33} & -\frac{3}{11} \sqrt{\frac{5}{2}} & 0 & 0 \\
 0 & 0 & \frac{1}{\sqrt{21}} & 0 & -\frac{3}{11}  \sqrt{\frac{2}{7}} & -\frac{3}{11} \sqrt{\frac{5}{2}} & -\frac{9}{11}  \sqrt{\frac{5}{14}} & 0 & 0 \\
 0 & 0 & 0 & 0 & 0 & 0 & 0 & 0 & \frac{3 \sqrt{2}}{11} \\
 -\frac{2 \sqrt{2}}{3}  & -\frac{2}{3} & 0 & \frac{2 \sqrt{6}}{11} & 0 & 0 & 0 & \frac{3 \sqrt{2}}{11} & 0 \\
\end{array}
\right),
\nonumber
\\
&\mathbb{F}_{60}^{(A)}=\left(
\begin{array}{ccccccccc}
 0 & 0 & 0 & 0 & 0 & 0 & 0 & 0 & 0 \\
 0 & 0 & 0 & 0 & 0 & 0 & 0 & 0 & 0 \\
 0 & 0 & 0 & 0 & 0 & 0 & 0 & 0 & 0 \\
 0 & 0 & 0 & 0 & 0 & 0 & 0 & -\frac{50}{11 \sqrt{39}} & 0 \\
 0 & 0 & 0 & 0 & 0 & 0 & -\frac{10}{33} \sqrt{\frac{5}{13}} & 0 & 0 \\
 0 & 0 & 0 & 0 & 0 & -\frac{15}{22 \sqrt{13}} & \frac{35 \sqrt{\frac{7}{13}}}{66} & 0 & 0 \\
 0 & 0 & 0 & 0 & -\frac{10}{33} \sqrt{\frac{5}{13}} & \frac{35 \sqrt{\frac{7}{13}}}{66} & -\frac{5}{6 \sqrt{13}} & 0 & 0 \\
 0 & 0 & 0 & -\frac{50}{11 \sqrt{39}} & 0 & 0 & 0 & \frac{25}{33 \sqrt{13}} & 0 \\
 0 & 0 & 0 & 0 & 0 & 0 & 0 & 0 & -\frac{5}{33 \sqrt{13}} \\
\end{array}
\right),
\nonumber
\\
&\mathbb{F}_{62}^{(A)}=\left(
\begin{array}{ccccccccc}
 0 & 0 & 0 & 0 & 0 & 0 & 0 & 0 & 0 \\
 0 & 0 & 0 & 0 & 0 & 0 & 0 & 0 & 0 \\
 0 & 0 & 0 & 0 & 0 & 0 & 0 & 0 & 0 \\
 0 & 0 & 0 & 0 & 0 & 0 & -\frac{40}{33}  \sqrt{\frac{7}{13}} & 0 & 0 \\
 0 & 0 & 0 & 0 & 0 & 0 & 0 & \frac{20 \sqrt{\frac{35}{39}}}{33} & -\frac{10}{33 \sqrt{39}} \\
 0 & 0 & 0 & 0 & 0 & 0 & 0 & -\frac{140}{33 \sqrt{39}} & \frac{5 \sqrt{\frac{35}{39}}}{33} \\
 0 & 0 & 0 & \frac{40 \sqrt{\frac{7}{13}}}{33} & 0 & 0 & 0 & 0 & -\frac{5}{11}  \sqrt{\frac{5}{39}} \\
 0 & 0 & 0 & 0 & -\frac{20}{33} \sqrt{\frac{35}{39}} & \frac{140}{33 \sqrt{39}} & 0 & 0 & 0 \\
 0 & 0 & 0 & 0 & \frac{10}{33 \sqrt{39}} & -\frac{5}{33}  \sqrt{\frac{35}{39}} & \frac{5 \sqrt{\frac{5}{39}}}{11} & 0 & 0 \\
\end{array}
\right),
\nonumber
\\
&\mathbb{F}_{64}^{(A)}=\left(
\begin{array}{ccccccccc}
 0 & 0 & 0 & 0 & 0 & 0 & 0 & 0 & 0 \\
 0 & 0 & 0 & 0 & 0 & 0 & 0 & 0 & 0 \\
 0 & 0 & 0 & 0 & 0 & 0 & 0 & 0 & 0 \\
 0 & 0 & 0 & 0 & 0 & 0 & 0 & 0 & -\frac{10 }{11} \sqrt{\frac{10}{39}} \\
 0 & 0 & 0 & 0 & 0 & 0 & \frac{10 \sqrt{\frac{70}{13}}}{33} & 0 & 0 \\
 0 & 0 & 0 & 0 & 0 & -\frac{15}{11} \sqrt{\frac{7}{26}} & -\frac{35}{33 \sqrt{26}} & 0 & 0 \\
 0 & 0 & 0 & 0 & \frac{10 \sqrt{\frac{70}{13}}}{33} & -\frac{35}{33 \sqrt{26}} & -\frac{5}{11} \sqrt{\frac{7}{26}} & 0 & 0 \\
 0 & 0 & 0 & 0 & 0 & 0 & 0 & 0 & -\frac{5}{11}  \sqrt{\frac{10}{13}} \\
 0 & 0 & 0 & -\frac{10 }{11} \sqrt{\frac{10}{39}} & 0 & 0 & 0 & -\frac{5}{11}  \sqrt{\frac{10}{13}} & 0 \\
\end{array}
\right),
\nonumber
\end{align}
\begin{align}
&\hspace{1.1cm}\mathbb{F}_{66}^{(A)}=\left(
\begin{array}{ccccccccc}
 0 & 0 & 0 & 0 & 0 & 0 & 0 & 0 & 0 \\
 0 & 0 & 0 & 0 & 0 & 0 & 0 & 0 & 0 \\
 0 & 0 & 0 & 0 & 0 & 0 & 0 & 0 & 0 \\
 0 & 0 & 0 & 0 & 0 & 0 & 0 & 0 & 0 \\
 0 & 0 & 0 & 0 & 0 & 0 & 0 & 0 & 10 \sqrt{\frac{5}{429}} \\
 0 & 0 & 0 & 0 & 0 & 0 & 0 & 0 & 5 \sqrt{\frac{7}{429}} \\
 0 & 0 & 0 & 0 & 0 & 0 & 0 & 0 & \frac{5}{\sqrt{429}} \\
 0 & 0 & 0 & 0 & 0 & 0 & 0 & 0 & 0 \\
 0 & 0 & 0 & 0 & -10 \sqrt{\frac{5}{429}} & -5 \sqrt{\frac{7}{429}} & -\frac{5}{\sqrt{429}} & 0 & 0 \\
\end{array}
\right),
\nonumber
\\
&\hspace{1cm}{\mathbb{M}}^{(A)}=
\left(
\begin{array}{ccccccccc}
 \mathcal{M}_{0,P}^{-1} & 0 & 0 & 0 & 0 & 0 & 0 & 0 & 0 \\
 0 & \frac{\mathcal{M}_{2,F}}{\det\mathcal{M}_{2}} & 0 & -\frac{\mathcal{M}_{2,PF}}{\det\mathcal{M}_{2}} & 0 & 0 & 0 & 0 & 0 \\
 0 & 0 & \frac{\mathcal{M}_{2,F}}{\det\mathcal{M}_{2}} & 0 & -\frac{\mathcal{M}_{2,PF}}{\det\mathcal{M}_{2}} & 0 & 0 & 0 & 0 \\
 0 & -\frac{\mathcal{M}_{2,PF}}{\det\mathcal{M}_{2}} & 0 & \frac{\mathcal{M}_{2,P}}{\det\mathcal{M}_{2}} & 0 & 0 & 0 & 0 & 0 \\
 0 & 0 & -\frac{\mathcal{M}_{2,PF}}{\det\mathcal{M}_{2}} & 0 & \frac{\mathcal{M}_{2,P}}{\det\mathcal{M}_{2}} & 0 & 0 & 0 & 0 \\
 0 & 0 & 0 & 0 & 0 & \mathcal{M}_{3,F}^{-1} & 0 & 0 & 0 \\
 0 & 0 & 0 & 0 & 0 & 0 & \mathcal{M}_{4,F}^{-1} & 0 & 0 \\
 0 & 0 & 0 & 0 & 0 & 0 & 0 & \mathcal{M}_{4,F}^{-1} & 0 \\
 0 & 0 & 0 & 0 & 0 & 0 & 0 & 0 & \mathcal{M}_{4,F}^{-1} \\
\end{array}
\right).
\label{IV110A}
\\
&B_1:\hspace{.5cm}\mathbb{F}_{00}^{(B_1)}=\textbf{I}_{7},\hspace{.5cm}
\mathbb{F}_{20}^{(B_1)}=\left(
\begin{array}{ccccccc}
 -\frac{1}{\sqrt{5}} & 0 & 0 & 3 \sqrt{\frac{2}{35}} & 0 & 0 & 0 \\
 0 & -\frac{1}{\sqrt{5}} & \frac{\sqrt{\frac{6}{5}}}{7} & 0 & -\sqrt{\frac{3}{14}} & 0 & \frac{3 \sqrt{\frac{3}{2}}}{7} \\
 0 & \frac{\sqrt{\frac{6}{5}}}{7} & -\frac{8}{7 \sqrt{5}} & 0 & -\frac{1}{\sqrt{7}} & 0 & -\frac{1}{21} \\
 3 \sqrt{\frac{2}{35}} & 0 & 0 & \frac{1}{\sqrt{5}} & 0 & 0 & 0 \\
 0 & -\sqrt{\frac{3}{14}} & -\frac{1}{\sqrt{7}} & 0 & 0 & 0 & -\frac{\sqrt{\frac{5}{7}}}{3} \\
 0 & 0 & 0 & 0 & 0 & -\frac{\sqrt{5}}{3} & 0 \\
 0 & \frac{3 \sqrt{\frac{3}{2}}}{7} & -\frac{1}{21} & 0 & -\frac{\sqrt{\frac{5}{7}}}{3} & 0 & \frac{2 \sqrt{5}}{21} \\
\end{array}
\right),
\nonumber
\\
&\hspace{1cm}\mathbb{F}_{22}^{(B_1)}=\left(
\begin{array}{ccccccc}
 0 & \sqrt{\frac{6}{5}} & \frac{2}{\sqrt{5}} & 0 & \frac{2}{\sqrt{7}} & 0 & 0 \\
 -\sqrt{\frac{6}{5}} & 0 & 0 & \sqrt{\frac{3}{35}} & 0 & \frac{3}{\sqrt{7}} & 0 \\
 -\frac{2}{\sqrt{5}} & 0 & 0 & \sqrt{\frac{2}{35}} & 0 & -\frac{\sqrt{\frac{2}{21}}}{3} & 0 \\
 0 & -\sqrt{\frac{3}{35}} & -\sqrt{\frac{2}{35}} & 0 & -\frac{1}{\sqrt{2}} & 0 & \frac{5}{3 \sqrt{14}} \\
 -\frac{2}{\sqrt{7}} & 0 & 0 & \frac{1}{\sqrt{2}} & 0 & \frac{\sqrt{\frac{5}{6}}}{3} & 0 \\
 0 & -\frac{3}{\sqrt{7}} & \frac{\sqrt{\frac{2}{21}}}{3} & 0 & -\frac{\sqrt{\frac{5}{6}}}{3} & 0 & \sqrt{\frac{5}{42}} \\
 0 & 0 & 0 & -\frac{5}{3 \sqrt{14}} & 0 & -\sqrt{\frac{5}{42}} & 0 \\
\end{array}
\right),
\nonumber
\end{align}
\begin{align}
&\hspace{.5cm}
\mathbb{F}_{40}^{(B_1)}=\left(
\begin{array}{ccccccc}
 0 & 0 & 0 & -\sqrt{\frac{2}{7}} & 0 & 0 & 0 \\
 0 & 0 & -\frac{\sqrt{\frac{2}{3}}}{7} & 0 & \sqrt{\frac{5}{42}} & 0 & -\frac{\sqrt{\frac{15}{2}}}{7} \\
 0 & -\frac{\sqrt{\frac{2}{3}}}{7} & \frac{1}{21} & 0 & \frac{\sqrt{\frac{5}{7}}}{3} & 0 & \frac{9 \sqrt{5}}{77} \\
 -\sqrt{\frac{2}{7}} & 0 & 0 & \frac{1}{11} & 0 & 0 & 0 \\
 0 & \sqrt{\frac{5}{42}} & \frac{\sqrt{\frac{5}{7}}}{3} & 0 & -\frac{7}{66} & 0 & -\frac{15}{22 \sqrt{7}} \\
 0 & 0 & 0 & 0 & 0 & \frac{3}{11} & 0 \\
 0 & -\frac{\sqrt{\frac{15}{2}}}{7} & \frac{9 \sqrt{5}}{77} & 0 & -\frac{15}{22 \sqrt{7}} & 0 & -\frac{3}{14} \\
\end{array}
\right),
\nonumber
\\
&\hspace{.5cm}\mathbb{F}_{42}^{(B_1)}=\left(
\begin{array}{ccccccc}
 0 & 0 & 0 & 0 & -\sqrt{\frac{3}{7}} & 0 & \frac{1}{\sqrt{3}} \\
 0 & 0 & 0 & -\sqrt{\frac{5}{7}} & 0 & -\frac{1}{\sqrt{21}} & 0 \\
 0 & 0 & 0 & -\sqrt{\frac{10}{21}} & 0 & \frac{3 \sqrt{\frac{2}{7}}}{11} & 0 \\
 0 & \sqrt{\frac{5}{7}} & \sqrt{\frac{10}{21}} & 0 & -\frac{1}{11 \sqrt{6}} & 0 & \frac{13 \sqrt{\frac{3}{14}}}{11} \\
 \sqrt{\frac{3}{7}} & 0 & 0 & \frac{1}{11 \sqrt{6}} & 0 & -\frac{3}{11} \sqrt{\frac{5}{2}} & 0 \\
 0 & \frac{1}{\sqrt{21}} & -\frac{3}{11}  \sqrt{\frac{2}{7}} & 0 & \frac{3 \sqrt{\frac{5}{2}}}{11} & 0 & -\frac{9}{11}  \sqrt{\frac{5}{14}} \\
 -\frac{1}{\sqrt{3}} & 0 & 0 & -\frac{13}{11} \sqrt{\frac{3}{14}} & 0 & \frac{9 \sqrt{\frac{5}{14}}}{11} & 0 \\
\end{array}
\right),
\nonumber
\\
&\hspace{.5cm}\mathbb{F}_{44}^{(B_1)}=\left(
\begin{array}{ccccccc}
 0 & 0 & 0 & 0 & 0 & \frac{2}{\sqrt{3}} & 0 \\
 0 & 0 & -2 \sqrt{\frac{5}{21}} & 0 & -\frac{1}{\sqrt{3}} & 0 & -\frac{1}{\sqrt{21}} \\
 0 & -2 \sqrt{\frac{5}{21}} & \frac{\sqrt{\frac{10}{7}}}{3} & 0 & -\frac{\sqrt{2}}{3} & 0 & \frac{3 \sqrt{\frac{2}{7}}}{11} \\
 0 & 0 & 0 & 0 & 0 & -\frac{\sqrt{42}}{11} & 0 \\
 0 & -\frac{1}{\sqrt{3}} & -\frac{\sqrt{2}}{3} & 0 & -\frac{\sqrt{\frac{35}{2}}}{33} & 0 & \frac{3 \sqrt{\frac{5}{2}}}{11} \\
 \frac{2}{\sqrt{3}} & 0 & 0 & -\frac{\sqrt{42}}{11} & 0 & 0 & 0 \\
 0 & -\frac{1}{\sqrt{21}} & \frac{3 \sqrt{\frac{2}{7}}}{11} & 0 & \frac{3 \sqrt{\frac{5}{2}}}{11} & 0 & \frac{9 \sqrt{\frac{5}{14}}}{11} \\
\end{array}
\right),
\nonumber
\\
&\hspace{.5cm}\mathbb{F}_{60}^{(B_1)}=\left(
\begin{array}{ccccccc}
 0 & 0 & 0 & 0 & 0 & 0 & 0 \\
 0 & 0 & 0 & 0 & 0 & 0 & 0 \\
 0 & 0 & 0 & 0 & 0 & 0 & -\frac{10}{33} \sqrt{\frac{5}{13}} \\
 0 & 0 & 0 & -\frac{25}{11 \sqrt{13}} & 0 & 0 & 0 \\
 0 & 0 & 0 & 0 & -\frac{15}{22 \sqrt{13}} & 0 & \frac{35 \sqrt{\frac{7}{13}}}{66} \\
 0 & 0 & 0 & 0 & 0 & -\frac{5}{33 \sqrt{13}} & 0 \\
 0 & 0 & -\frac{10}{33} \sqrt{\frac{5}{13}} & 0 & \frac{35 \sqrt{\frac{7}{13}}}{66} & 0 & -\frac{5}{6 \sqrt{13}} \\
\end{array}
\right),
\nonumber
\end{align}
\begin{align}
&\hspace{1cm}\mathbb{F}_{62}^{(B_1)}=\left(
\begin{array}{ccccccc}
 0 & 0 & 0 & 0 & 0 & 0 & 0 \\
 0 & 0 & 0 & 0 & 0 & 0 & 0 \\
 0 & 0 & 0 & 0 & 0 & -\frac{10}{33 \sqrt{39}} & 0 \\
 0 & 0 & 0 & 0 & -\frac{10}{11} \sqrt{\frac{7}{13}} & 0 & \frac{70}{33 \sqrt{13}} \\
 0 & 0 & 0 & \frac{10 \sqrt{\frac{7}{13}}}{11} & 0 & \frac{5 \sqrt{\frac{35}{39}}}{33} & 0 \\
 0 & 0 & \frac{10}{33 \sqrt{39}} & 0 & -\frac{5}{33}  \sqrt{\frac{35}{39}} & 0 & \frac{5 \sqrt{\frac{5}{39}}}{11} \\
 0 & 0 & 0 & -\frac{70}{33 \sqrt{13}} & 0 & -\frac{5}{11}  \sqrt{\frac{5}{39}} & 0 \\
\end{array}
\right),
\nonumber
\\
&\hspace{1cm}\mathbb{F}_{64}^{(B_1)}=\left(
\begin{array}{ccccccc}
 0 & 0 & 0 & 0 & 0 & 0 & 0 \\
 0 & 0 & 0 & 0 & 0 & 0 & 0 \\
 0 & 0 & 0 & 0 & 0 & 0 & -\frac{10}{33} \sqrt{\frac{70}{13}} \\
 0 & 0 & 0 & 0 & 0 & \frac{5 \sqrt{\frac{70}{39}}}{11} & 0 \\
 0 & 0 & 0 & 0 & \frac{15 \sqrt{\frac{7}{26}}}{11} & 0 & \frac{35}{33 \sqrt{26}} \\
 0 & 0 & 0 & \frac{5 \sqrt{\frac{70}{39}}}{11} & 0 & 0 & 0 \\
 0 & 0 & -\frac{10}{33} \sqrt{\frac{70}{13}} & 0 & \frac{35}{33 \sqrt{26}} & 0 & \frac{5 \sqrt{\frac{7}{26}}}{11} \\
\end{array}
\right),
\nonumber\
\\
&\hspace{1cm}\mathbb{F}_{66}^{(B_1)}=\left(
\begin{array}{ccccccc}
 0 & 0 & 0 & 0 & 0 & 0 & 0 \\
 0 & 0 & 0 & 0 & 0 & 0 & 0 \\
 0 & 0 & 0 & 0 & 0 & -10 \sqrt{\frac{5}{429}} & 0 \\
 0 & 0 & 0 & 0 & 0 & 0 & 0 \\
 0 & 0 & 0 & 0 & 0 & -5 \sqrt{\frac{7}{429}} & 0 \\
 0 & 0 & 10 \sqrt{\frac{5}{429}} & 0 & 5 \sqrt{\frac{7}{429}} & 0 & \frac{5}{\sqrt{429}} \\
 0 & 0 & 0 & 0 & 0 & -\frac{5}{\sqrt{429}} & 0 \\
\end{array}
\right),
\nonumber
\\
&\hspace{1cm}{\mathbb{M}}^{(B_1)}=
\left(
\begin{array}{ccccccc}
\mathcal{M}_{1,P}^{-1} & 0 & 0 & 0 & 0 & 0 & 0 \\
 0 & \frac{\mathcal{M}_{2,F}}{\det\mathcal{M}_{2}} & -\frac{\mathcal{M}_{2,PF}}{\det\mathcal{M}_{2}} & 0 & 0 & 0 & 0 \\
 0 & -\frac{\mathcal{M}_{2,PF}}{\det\mathcal{M}_{2}} & \frac{\mathcal{M}_{2,P}}{\det\mathcal{M}_{2}} & 0 & 0 & 0 & 0 \\
 0 & 0 & 0 & \mathcal{M}_{3,F}^{-1} & 0 & 0 & 0 \\
 0 & 0 & 0 & 0 & \mathcal{M}_{3,F}^{-1} & 0 & 0 \\
 0 & 0 & 0 & 0 & 0 & \mathcal{M}_{4,F}^{-1} & 0 \\
 0 & 0 & 0 & 0 & 0 & 0 & \mathcal{M}_{4,F}^{-1} \\
\end{array}
\right).
\label{IV110B1}
\\
&B_2:\hspace{.5cm}\mathbb{F}_{00}^{(B_2)}=\textbf{I}_{7},\hspace{.5cm}
\mathbb{F}_{20}^{(B_2)}=\left(
\begin{array}{ccccccc}
 \frac{1}{2 \sqrt{5}} & -\frac{3}{2 \sqrt{5}} & -\sqrt{\frac{3}{10}} & 2 \sqrt{\frac{3}{35}} & 0 & 0 & 0 \\
 -\frac{3}{2 \sqrt{5}} & \frac{1}{2 \sqrt{5}} & -\frac{\sqrt{\frac{3}{10}}}{7} & -\sqrt{\frac{3}{35}} & 0 & 0 & \frac{3 \sqrt{3}}{7} \\
 -\sqrt{\frac{3}{10}} & -\frac{\sqrt{\frac{3}{10}}}{7} & \frac{4}{7 \sqrt{5}} & -\sqrt{\frac{2}{35}} & 0 & 0 & -\frac{\sqrt{2}}{21} \\
 2 \sqrt{\frac{3}{35}} & -\sqrt{\frac{3}{35}} & -\sqrt{\frac{2}{35}} & \frac{3}{4 \sqrt{5}} & 0 & 0 & -\frac{5}{12 \sqrt{7}} \\
 0 & 0 & 0 & 0 & -\frac{\sqrt{5}}{4} & -\frac{\sqrt{\frac{5}{3}}}{4} & 0 \\
 0 & 0 & 0 & 0 & -\frac{\sqrt{\frac{5}{3}}}{4} & -\frac{\sqrt{5}}{12} & 0 \\
 0 & \frac{3 \sqrt{3}}{7} & -\frac{\sqrt{2}}{21} & -\frac{5}{12 \sqrt{7}} & 0 & 0 & \frac{17 \sqrt{5}}{84} \\
\end{array}
\right),
\nonumber
\end{align}
\begin{align}
&\hspace{.5cm}\mathbb{F}_{22}^{(B_2)}=\left(
\begin{array}{ccccccc}
 -i \sqrt{\frac{3}{10}} & -i \sqrt{\frac{3}{10}} & -\frac{i}{\sqrt{5}} & -i \sqrt{\frac{2}{35}} & \sqrt{\frac{6}{7}} & 0 & 0 \\
 -i \sqrt{\frac{3}{10}} & -i \sqrt{\frac{3}{10}} & \frac{3 i}{7 \sqrt{5}} & \frac{3 i}{\sqrt{70}} & \sqrt{\frac{3}{14}} & \frac{3}{\sqrt{14}} & \frac{3 i}{7 \sqrt{2}} \\
 -\frac{i}{\sqrt{5}} & \frac{3 i}{7 \sqrt{5}} & \frac{-4i}{7}  \sqrt{\frac{6}{5}} & i \sqrt{\frac{3}{35}} & \frac{1}{\sqrt{7}} & -\frac{1}{3 \sqrt{21}} & -\frac{i}{21 \sqrt{3}} \\
 -i \sqrt{\frac{2}{35}} & \frac{3 i}{\sqrt{70}} & i \sqrt{\frac{3}{35}} & i \sqrt{\frac{3}{10}} & -\frac{1}{2 \sqrt{2}} & \frac{5}{6 \sqrt{6}} & -\frac{5 i}{3 \sqrt{42}} \\
 -\sqrt{\frac{6}{7}} & -\sqrt{\frac{3}{14}} & -\frac{1}{\sqrt{7}} & \frac{1}{2 \sqrt{2}} & 0 & 0 & \frac{\sqrt{\frac{5}{14}}}{6} \\
 0 & -\frac{3}{\sqrt{14}} & \frac{1}{3 \sqrt{21}} & -\frac{5}{6 \sqrt{6}} & 0 & 0 & \frac{\sqrt{\frac{15}{14}}}{2} \\
 0 & \frac{3 i}{7 \sqrt{2}} & -\frac{i}{21 \sqrt{3}} & -\frac{5 i}{3 \sqrt{42}} & -\frac{\sqrt{\frac{5}{14}}}{6} & -\frac{\sqrt{\frac{15}{14}}}{2} & \frac{-5i}{7}  \sqrt{\frac{5}{6}} \\
\end{array}
\right),
\nonumber
\\
&\hspace{.5cm}\mathbb{F}_{40}^{(B_2)}=\left(
\begin{array}{ccccccc}
 0 & 0 & 0 & \frac{\sqrt{\frac{3}{7}}}{2} & 0 & 0 & -\frac{\sqrt{\frac{5}{3}}}{2} \\
 0 & 0 & \frac{4 \sqrt{\frac{2}{3}}}{7} & -\frac{5}{2 \sqrt{21}} & 0 & 0 & \frac{\sqrt{\frac{5}{3}}}{14} \\
 0 & \frac{4 \sqrt{\frac{2}{3}}}{7} & -\frac{4}{21} & -\frac{5}{3 \sqrt{14}} & 0 & 0 & -\frac{3}{77}  \sqrt{\frac{5}{2}} \\
 \frac{\sqrt{\frac{3}{7}}}{2} & -\frac{5}{2 \sqrt{21}} & -\frac{5}{3 \sqrt{14}} & \frac{1}{66} & 0 & 0 & -\frac{9}{22} \sqrt{\frac{5}{7}} \\
 0 & 0 & 0 & 0 & \frac{1}{22} & \frac{5 \sqrt{3}}{22} & 0 \\
 0 & 0 & 0 & 0 & \frac{5 \sqrt{3}}{22} & -\frac{9}{22} & 0 \\
 -\frac{\sqrt{\frac{5}{3}}}{2} & \frac{\sqrt{\frac{5}{3}}}{14} & -\frac{3}{77}  \sqrt{\frac{5}{2}} & -\frac{9}{22} \sqrt{\frac{5}{7}} & 0 & 0 & \frac{27}{154} \\
\end{array}
\right),
\nonumber
\\
&\hspace{.5cm}\mathbb{F}_{42}^{(B_2)}=\left(
\begin{array}{ccccccc}
 0 & 0 & 0 & -\frac{1}{2} i \sqrt{\frac{15}{14}} & \frac{1}{2 \sqrt{14}} & -\frac{\sqrt{\frac{7}{6}}}{2} & -\frac{1}{2} i \sqrt{\frac{3}{2}} \\
 0 & 0 & \frac{4}{7} i \sqrt{\frac{5}{3}} & \frac{1}{2} i \sqrt{\frac{5}{42}} & -\frac{3}{2 \sqrt{14}} & \frac{5}{2 \sqrt{42}} & \frac{-3}{14} i \sqrt{\frac{3}{2}} \\
 0 & \frac{4}{7} i \sqrt{\frac{5}{3}} & -i\frac{2  \sqrt{10}}{21}  & \frac{1}{6} i \sqrt{\frac{5}{7}} & -\frac{\sqrt{\frac{3}{7}}}{2} & -\frac{15}{22 \sqrt{7}} & \frac{27 i}{154} \\
 -\frac{1}{2} i \sqrt{\frac{15}{14}} & \frac{1}{2} i \sqrt{\frac{5}{42}} & \frac{1}{6} i \sqrt{\frac{5}{7}} & \frac{i \sqrt{10}}{33} & \frac{\sqrt{\frac{3}{2}}}{11} & \frac{3}{11 \sqrt{2}} & \frac{-i3}{11}  \sqrt{\frac{2}{7}} \\
 -\frac{1}{2 \sqrt{14}} & \frac{3}{2 \sqrt{14}} & \frac{\sqrt{\frac{3}{7}}}{2} & -\frac{\sqrt{\frac{3}{2}}}{11} & 0 & 0 & -\frac{5}{11} \sqrt{\frac{15}{14}} \\
 \frac{\sqrt{\frac{7}{6}}}{2} & -\frac{5}{2 \sqrt{42}} & \frac{15}{22 \sqrt{7}} & -\frac{3}{11 \sqrt{2}} & 0 & 0 & -\frac{3}{11} \sqrt{\frac{5}{14}} \\
 -\frac{1}{2} i \sqrt{\frac{3}{2}} & \frac{-3}{14} i \sqrt{\frac{3}{2}} & \frac{27 i}{154} & \frac{-3i}{11}  \sqrt{\frac{2}{7}} & \frac{5 \sqrt{\frac{15}{14}}}{11} & \frac{3 \sqrt{\frac{5}{14}}}{11} & -i\frac{9}{77}  \sqrt{10} \\
\end{array}
\right),
\nonumber
\\
&\hspace{.5cm}\mathbb{F}_{44}^{(B_2)}=\left(
\begin{array}{ccccccc}
 0 & 0 & 0 & 0 & -\frac{i}{\sqrt{2}} & -\frac{i}{\sqrt{6}} & 0 \\
 0 & 0 & 0 & 0 & -\frac{i}{\sqrt{2}} & -\frac{i}{\sqrt{6}} & 0 \\
 0 & 0 & 0 & 0 & -\frac{i}{\sqrt{3}} & \frac{3 i}{11} & 0 \\
 0 & 0 & 0 & 0 & -\frac{1}{11} i \sqrt{\frac{7}{6}} & \frac{3}{11} i \sqrt{\frac{7}{2}} & 0 \\
 \frac{i}{\sqrt{2}} & \frac{i}{\sqrt{2}} & \frac{i}{\sqrt{3}} & \frac{1}{11} i \sqrt{\frac{7}{6}} & 0 & 0 & -\frac{1}{11} i \sqrt{\frac{15}{2}} \\
 \frac{i}{\sqrt{6}} & \frac{i}{\sqrt{6}} & -\frac{3 i}{11} & \frac{-3i}{11}  \sqrt{\frac{7}{2}} & 0 & 0 & \frac{-3i}{11}  \sqrt{\frac{5}{2}} \\
 0 & 0 & 0 & 0 & \frac{1}{11} i \sqrt{\frac{15}{2}} & \frac{3}{11} i \sqrt{\frac{5}{2}} & 0 \\
\end{array}
\right),
\nonumber
\end{align}
\begin{align}
&\mathbb{F}_{60}^{(B_2)}=\left(
\begin{array}{ccccccc}
 0 & 0 & 0 & 0 & 0 & 0 & 0 \\
 0 & 0 & 0 & 0 & 0 & 0 & 0 \\
 0 & 0 & 0 & 0 & 0 & 0 & \frac{20 \sqrt{\frac{10}{13}}}{33} \\
 0 & 0 & 0 & \frac{75}{44 \sqrt{13}} & 0 & 0 & -\frac{35}{132} \sqrt{\frac{35}{13}} \\
 0 & 0 & 0 & 0 & \frac{5}{44 \sqrt{13}} & -\frac{35}{44 \sqrt{39}} & 0 \\
 0 & 0 & 0 & 0 & -\frac{35}{44 \sqrt{39}} & \frac{85}{132 \sqrt{13}} & 0 \\
 0 & 0 & \frac{20 \sqrt{\frac{10}{13}}}{33} & -\frac{35}{132} \sqrt{\frac{35}{13}} & 0 & 0 & -\frac{5}{132 \sqrt{13}} \\
\end{array}
\right),
\nonumber
\\
&\mathbb{F}_{62}^{(B_2)}=\left(
\begin{array}{ccccccc}
 0 & 0 & 0 & 0 & 0 & 0 & 0 \\
 0 & 0 & 0 & 0 & 0 & 0 & 0 \\
 0 & 0 & 0 & 0 & 0 & \frac{40 \sqrt{\frac{2}{39}}}{33} & \frac{40}{33} i \sqrt{\frac{14}{39}} \\
 0 & 0 & 0 & \frac{-5i}{22}  \sqrt{\frac{105}{13}} & \frac{5 \sqrt{\frac{7}{13}}}{22} & -\frac{5}{66} \sqrt{\frac{91}{3}} & -\frac{35 i}{66 \sqrt{39}} \\
 0 & 0 & 0 & -\frac{5}{22}  \sqrt{\frac{7}{13}} & 0 & 0 & \frac{35 \sqrt{\frac{5}{13}}}{66} \\
 0 & 0 & -\frac{40}{33} \sqrt{\frac{2}{39}} & \frac{5 \sqrt{\frac{91}{3}}}{66} & 0 & 0 & -\frac{5}{22} \sqrt{\frac{15}{13}} \\
 0 & 0 & \frac{40}{33} i \sqrt{\frac{14}{39}} & -\frac{35 i}{66 \sqrt{39}} & -\frac{35}{66} \sqrt{\frac{5}{13}} & \frac{5 \sqrt{\frac{15}{13}}}{22} & \frac{-5i}{22}  \sqrt{\frac{35}{39}} \\
\end{array}
\right),
\nonumber
\\
&\mathbb{F}_{64}^{(B_2)}=\left(
\begin{array}{ccccccc}
 0 & 0 & 0 & 0 & 0 & 0 & 0 \\
 0 & 0 & 0 & 0 & 0 & 0 & 0 \\
 0 & 0 & 0 & 0 & 0 & \frac{40}{33} i \sqrt{\frac{5}{13}} & 0 \\
 0 & 0 & 0 & 0 & \frac{-5i}{22}  \sqrt{\frac{105}{26}} & \frac{25}{66} i \sqrt{\frac{35}{26}} & 0 \\
 0 & 0 & 0 & \frac{5}{22} i \sqrt{\frac{105}{26}} & 0 & 0 & \frac{35}{22} i \sqrt{\frac{3}{26}} \\
 0 & 0 & -i\frac{40}{33}  \sqrt{\frac{5}{13}} & \frac{-25i}{66}  \sqrt{\frac{35}{26}} & 0 & 0 & -\frac{5 i}{22 \sqrt{26}} \\
 0 & 0 & 0 & 0 & \frac{-35i}{22} \sqrt{\frac{3}{26}} & \frac{5 i}{22 \sqrt{26}} & 0 \\
\end{array}
\right),
\nonumber
\\
&\mathbb{F}_{66}^{(B_2)}=\left(
\begin{array}{ccccccc}
 0 & 0 & 0 & 0 & 0 & 0 & 0 \\
 0 & 0 & 0 & 0 & 0 & 0 & 0 \\
 0 & 0 & 0 & 0 & 0 & 0 & 0 \\
 0 & 0 & 0 & 0 & 0 & 0 & 0 \\
 0 & 0 & 0 & 0 & \frac{5}{2} i \sqrt{\frac{21}{143}} & \frac{5}{2} i \sqrt{\frac{7}{143}} & 0 \\
 0 & 0 & 0 & 0 & \frac{5}{2} i \sqrt{\frac{7}{143}} & \frac{5}{2} i \sqrt{\frac{7}{429}} & 0 \\
 0 & 0 & 0 & 0 & 0 & 0 & 0 \\
\end{array}
\right),
\nonumber
\\
&{\mathbb{M}}^{(B_2)}=\left(
\begin{array}{ccccccc}
\mathcal{M}_{1,P}^{-1} & 0 & 0 & 0 & 0 & 0 & 0 \\
 0 & \frac{\mathcal{M}_{2,F}}{\det\mathcal{M}_{2}} & -\frac{\mathcal{M}_{2,PF}}{\det\mathcal{M}_{2}} & 0 & 0 & 0 & 0 \\
 0 & -\frac{\mathcal{M}_{2,PF}}{\det\mathcal{M}_{2}} & \frac{\mathcal{M}_{2,P}}{\det\mathcal{M}_{2}} & 0 & 0 & 0 & 0 \\
 0 & 0 & 0 & \mathcal{M}_{3,F}^{-1} & 0 & 0 & 0 \\
 0 & 0 & 0 & 0 & \mathcal{M}_{3,F}^{-1} & 0 & 0 \\
 0 & 0 & 0 & 0 & 0 & \mathcal{M}_{4,F}^{-1} & 0 \\
 0 & 0 & 0 & 0 & 0 & 0 & \mathcal{M}_{4,F}^{-1} \\
\end{array}
\right).
\label{IV110B2}
\end{align}
\begin{align}
&B_3:\hspace{.5cm}\mathbb{F}_{00}^{(B_3)}=\textbf{I}_{7},\hspace{.5cm}
\mathbb{F}_{20}^{(B_3)}=\left(
\begin{array}{ccccccc}
 \frac{1}{2 \sqrt{5}} & -\frac{3}{2 \sqrt{5}} & -\sqrt{\frac{3}{10}} & 0 & 2 \sqrt{\frac{3}{35}} & 0 & 0 \\
 -\frac{3}{2 \sqrt{5}} & \frac{1}{2 \sqrt{5}} & -\frac{\sqrt{\frac{3}{10}}}{7} & 0 & -\sqrt{\frac{3}{35}} & \frac{3 \sqrt{3}}{7} & 0 \\
 -\sqrt{\frac{3}{10}} & -\frac{\sqrt{\frac{3}{10}}}{7} & \frac{4}{7 \sqrt{5}} & 0 & -\sqrt{\frac{2}{35}} & -\frac{\sqrt{2}}{21} & 0 \\
 0 & 0 & 0 & -\frac{\sqrt{5}}{4} & 0 & 0 & -\frac{\sqrt{\frac{5}{3}}}{4} \\
 2 \sqrt{\frac{3}{35}} & -\sqrt{\frac{3}{35}} & -\sqrt{\frac{2}{35}} & 0 & \frac{3}{4 \sqrt{5}} & -\frac{5}{12 \sqrt{7}} & 0 \\
 0 & \frac{3 \sqrt{3}}{7} & -\frac{\sqrt{2}}{21} & 0 & -\frac{5}{12 \sqrt{7}} & \frac{17 \sqrt{5}}{84} & 0 \\
 0 & 0 & 0 & -\frac{\sqrt{\frac{5}{3}}}{4} & 0 & 0 & -\frac{\sqrt{5}}{12} \\
\end{array}%
\right).
\label{IV110B3}
\end{align}
%

 
\bibliography{bibi}

\end{document}